ПЕРМСКИЙ ГОСУДАРСТВЕННЫЙ УНИВЕРСИТЕТ

*На правах рукописи*

Голдобин Денис Сергеевич

# ПАРАМЕТРИЧЕСКОЕ ВОЗБУЖДЕНИЕ, ЛОКАЛИЗАЦИЯ И СИНХРОНИЗАЦИЯ В РАСПРЕДЕЛЕННЫХ НЕЛИНЕЙНЫХ СИСТЕМАХ ГИДРОДИНАМИЧЕСКОГО ТИПА

01.02.05 – Механика жидкости, газа и плазмы

Диссертация на соискание ученой степени
кандидата физико-математических наук

Научный руководитель:
д.ф.-м.н., профессор Д.В.Любимов

# ОГЛАВЛЕНИЕ

# Введение

В данной работе явления параметрического возбуждения и локализации в распределенных нелинейных системах изучаются на примере тепловой конвекции в пористой среде. Тепловая конвекция в пористых средах представляет интерес как в связи с прикладными задачами, связанными с технологическими (например, охлаждение реакторов, фильтрация) и природными процессами (например, течения грунтовых вод), так и с точки зрения математической физики (например, явление косимметрии [1,2,3]).

В задачах о течении жидкости через пористую среду необходимо иметь дело со сложной пористой структурой среды и тем, как эта структура влияет на процессы тепло- и массопереноса (в том числе, диффузии в случае смесей). В обощем случае механизмы этих процессов зависят как от деформируемости твердого скелета, так и от его геометрии, которая может быть как упорядоченной, так и неупорядоченной (подробное освещение как этого, так и многих других вопросов можно найти, например, в книге [4]). Из двух классов неупорядоченных сред: (1) микроскопически неупорядоченных, но макроскопически однородных и (2) микроскопически неупорядоченных и макроскопически неоднородных – в задачах, представленных в данной диссертации, рассматриваются среды первого типа, причем скелет будет также полагаться макроскопически изотропным и недеформируемым.

**Закон Дарси.** При протекании однофазной жидкости через макроскопически однородную изотропную пористую среду во многих реалистичных ситуациях инерционные эффекты несущественны и справедлив закон Дарси:

$$\left\langle \vec{v} \right\rangle = -\frac{K}{\eta}\left(\nabla P - \rho\vec{g}\right), \tag{I.1}$$

где $\left\langle \vec{v} \right\rangle$, $\eta$ и $\rho$ – скорость фильтрации, динамическая вязкость и плотность жидкости, соответственно, $K$ – проницаемость, $P$ – давление и $\vec{g}$ – поле массовых сил. Одномерная версия этого уравнения была открыта Дарси экспериментально в 1856 году. Один из примеров строгого теоретического вывода этого закона можно найти в работе [5], где окончательный результат, приведенный в тензорном виде:

$$\langle \vec{v} \rangle = -\frac{1}{\eta} \overline{\overline{K}} \cdot \left( \nabla P - \rho \vec{g} \right), \tag{I.2}$$

– для случая макроскопически изотропной среды принимает вид (I.1).

Использование уравнения (I.1) в качестве закона сохранения импульса понижает порядок дифференциальных уравнений относительно производных по пространственным координатам по сравнению со случаем течений в однородной вязкой жидкости (без твердого скелета). Это находит свое отражение, в том числе, и в изменении граничных условий для поля скорости (например, см. [4]), когда, кроме прочего, условие прилипания на твердой границе теряет смысл, поскольку жидкость точно также прилипает и к стенкам твердого скелета во всем объеме пористой среды.

**Тепловая конвекция в слое пористой среды.** В диссертации рассматриваются задачи, касающиеся тепловой (либо термоконцентрационной) конвекции в горизонтальном слое пористой среды при вертикальном градиенте температуры в основном состоянии, и настоящий обзор не касается смежных задач (например, о тепловой конвекции в наклонных слоях).

Впервые задача о возникновении тепловой конвекции в слое пористой среды рассматривалась в работе [6], где, фактически, полагались непроницаемые изотермические границы. Затем в работе [7] был рассмотрен другой вариант граничных условий, соответствующий тому, что поверх насыщенного жидкостью пористого слоя налита свободная жидкость. В этом случае верхняя граница становится проницаемой для жидкости, а горизонтальная компонента гра-диента давления на верхней границе обращается в ноль. Такое "смягчение" граничных условий ожидаемо приводит к понижению порога возбуждения тепловой конвекции и увеличению длины волны критических возмущений.

Для проверки теоретических предсказаний работы [6] было предпринято первое экспериментальное исследование возникновения конвекции в горизонтальном пористом слое [8]. В опытах использовались слои песка, насыщенного различными жидкостями: водой, растворами глицерина, четыреххлористым углеродом. Результаты этих экспериментов отличались от теоретических предсказаний на порядок, хотя качественно зависимость критического градиента температуры от параметров воспроизводилась. Это расхождение было связано с указанной авторами существенной зависимостью вязкости от температуры (разности температур в слое были очень велики), а также некорректным определением температуропроводности системы. Попытки учета зависимости вязкости от температуры и нелинейности профиля температуры привели к некоторому улучшению соответствия теории и эксперимента [9,10], но определение температуропроводности исправлено не было.

В итоге, более аккуратное экспериментальное исследование данного явления (при корректном определении температуропроводности) было проведено в работе [11] и дало результаты, хорошо согласующиеся с теорией.

Здесь следует отметить, что задача о возникновении тепловой конвекции в пористой среде при вертикальном градиенте температуры в основном состоянии существенно зависит от геометрии области: для слоя и горизонтального цилиндра задачи о возбуждении двухмерных конвективных движений при изотермических непроницаемых границах существенно отличаются. Дело в том, что во втором случае в системе имеется косимметрия [1,3], в результате чего при потере устойчивости состояния механического равновесия возникает не конечное количество стационарных решений, а непрерывное однопараметрическое множество. Несмотря на то, что нарушение строгой вертикальности гради-ента температуры в основном состоянии, однородности системы вдоль одного из горизонтальных направлений, неидеальная изотермичность границ и некоторые другие осложняющие факторы должны в реальных системах снимать вырождение, связанное с косимметрией, и разрушать множество стационарных решений, возникновение этого множества удалось пронаблюдать экспериментально [2]. Поведение системы при слабых нарушениях косимметрии также рассмотрено, например, в работах [2,12].

**Термоконцентрационная конвекция в слое пористой среды.** При конвекции бинарных смесей в пористой среде как и для однокомпонентной жидкости, возможна монотонная неустойчивость стратифицированного состояния, но, дополнительно к тому, конкуренция теплового и концентрационного механизмов неустойчивости может приводить к возникновению колебательной неустойчивости [13,14].

Эффект Соре (термодиффузия, [15]), особенно важный в газах, способен существенным образом влиять на термоконцентрационную конвекцию. В связи с этим имеет смысл выделить две группы задач: (1) задачи о конвекции при двойной диффузии (double diffusion), в которых не учитывается термодиффузия, т.е. влияние составляющей потока концентрации, индуцированной неоднородностью нагрева; (2) задачи с учетом эффекта Соре.

Ранние исследования конвекции при двойной диффузии в пористой среде были сосредоточены главным образом на задаче о возбуждении конвекции в горизонтальном слое. В работах [13,14,16,17,18,19] для исследования возникновения термохалиновой (thermohaline) конвекции производился линейный анализ устойчивости стратифицированного состояния. Критерии возникновения монотонного и колебательного движений были получены авторами этих работ в различных вариантах постановки задачи: для различных граничных условий и моделей жидкости.

Конечно-амплитудная термохалиновая конвекция исследовалась в работе [20], где для предсказания зависимости чисел Нуссельта и Шервуда от управляющих параметров использовались усеченные разложения в ряды Фурье. Комбинированное аналитическое и численное исследование массопереноса конвекцией Бенара при больших числах Рэлея представлено в работе [21], где суммарный массоперенос предсказывался авторами в терминах трех различных законов скейлинга. Работа [22] посвящена исследованию пальцеобразования при двухмерной конвекции, связанной с двойной диффузией, в горизонтальном слое с заданным пространственным периодом структуры (приняты периодические боковые граничные условия). На плоскости теплового и концентрационного чисел Рэлея были найдены границы устойчивости, разделяющие области параметров с различными типами конвективного движения (стационарное пальцеобразование, периодическая конвекция, нерегулярная конвекция). Недавно, обусловленная двойной диффузией конвекция в горизонтальном слое пористой среды исследовалась в работе [23] при вертикальном градиенте температуры и концентрации в основном состоянии. Неустойчивости к монотонным и колебательным возмущениям были исследованы аналитически в рамках линейной и нелинейной теорий возмущений, и была предсказана возможность жесткого возбуждения конвективных движений.

В задачах о термоконцентрационной конвекции с учетом эффекта Соре градиенты концентрации компонент в стратифицированном состоянии не навязываются особыми граничными условиями для концентрации, как при двойной диффузии, а обычно являются следствием навязанного градиента температуры и эффекта термодиффузии, в отсутствии которого концентрация была бы однородной. В работах [24,25] исследовано влияние сил концентрационной плавучести, вызванных эффектом Соре, на конвективную неустойчивость жидкой смеси в пористой среде при подогреве, обеспечиваемом изотермическими граничными условиями. Были получены результаты, касающиеся возбуждения конвективных движений и нелинейных режимов конвекции. Было показано, что количество независимых параметров, определяющих поведение системы, может быть сокращено: достаточно числа Рэлея и отношения плавучести. Было показано, что в зависимости от величины и знака отношения плавучести потеря устойчивости стратифицированного состояния может происходить в первую очередь либо монотонно, либо колебательно. Та же задача была рассмотрена в работе [18] с учетом не только эффекта Соре, но и термодинамически сопряженного ему диффузионного термоэффекта Дюфора (поток тепла индуцируется неоднородностью концентрации). Влияние эффекта Соре на линейную устойчивость жидкой смеси в пористой среде при периодически меняющемся со временем градиенте температуры было исследовано в работе [26]. Авторами определялся порог возбуждения двухмерной конвекции как для колебательной, так и для монотонной мод. Позже возбуждение конвекции, вызванной эффектом Соре, в бесконечном горизонтальном слое, подогреваемом изотермически снизу и сверху, было рассмотрено в работе [27]. Порог линейной устойчивости определялся в терминах отношения плавучести, числа Льюиса и нормированной пористости среды. Аналитически и численно было получено, что в зависимости от отношения плавучести состояние механического равновесия теряет устойчивость либо через вилочную бифуркацию, либо через бифуркацию Хопфа.

Наиболее близкой задачам первой и второй глав оказывается работа [28], где аналитически и численно исследуется конвекция бинарной смеси в горизонтальном слое пористой среды. В этой работе полагаются заданный тепловой поток через горизонтальные границы и адиабатические непроницаемые боковые границы (рассматриваются большие аспектные соотношения, делающие эту задачу эквивалентной задаче о конвекции в слое). Рассматриваются два способа создания неоднородностей концентрации: (1) навязывание постоянного потока примеси через горизонтальные границы в случае двойной диффузии или (2) за счет неоднородностей поля температуры при учете эффекта Соре. Рассмотрена линейная устойчивость стратифицированного состояния в слое. Оп-ределены пороги возбуждения конечно-амплитудных стационарных и колебательных конвективных течений. В приближении плоскопараллельного течения найдены аналитические решения уравнений в стационарном случае (исходно рассматриваемая система ограничена в горизонтальном направлении, и плоскопараллельное течение реализуется только вдали от боковых границ). На основе этих аналитических результатов оцениваются критические значения числа Рэлея для мягкого и жесткого возбуждения конвекции. Для установления точки бифуркации Хопфа исследуется линейная устойчивость плоскопараллельных течений. Представленные в работе результаты численного интегрирования полной нелинейной системы уравнений конвекции бинарной смеси хорошо согласуются с аналитическими результатами.

В контексте исследований, проведенных в первой и второй главах диссертации, следует отметить, что в работе [96] для длинноволновых возмущений была выявлена пропорциональность характерных времен эволюции квадрату длины волны возмущений. В свете этого становится очевидным, что при рассмотрении плоскопараллельных течений, соответствующих бесконечной длине волны, невозможно отличить решения, являющиеся предельным случаем колебательных возмущений, от решений, соответствующих монотонным возмущениям. Это накладывает существенные ограничения на информативность результатов, приведенных в работе [28] в связи с плоскопараллельным приближением.

Кроме того, в работе [28] задача о линейной устойчивости состояния механического равновесия решена до конца лишь для случая двойной диффузии, но не для случая конвекции, вызнаной эффектом Соре.

Для полноты представления о положении исследований в данной области имеет смысл упомянуть работы [29,30,31,32,33], независимо появившиеся после подготовки работы [96]. В работе [29] сопоставляются результаты исследований, аналогичных [28,30] ([30] можно рассматривать как продолжение [28], покрываемое работой [29]), для конвекции в отсутствие твердого скелета и в пористой среде, проводится до конца анализ линейной устойчивости состояния механического равновесия при эффекте Соре, незавершенный в [28], и отмечается (как и в [96]) рост характерных времен эволюции возмущений по мере роста их длины волны. Однако, для аналитического исследования в этой и остальных процитированных здесь работах по-прежнему используется приближение плоскопараллельного течения. В работах [31,32] по сравнению с [28,30] рассматривается случай неплотной упаковки пористого скелета; при этом в [31] верхняя граница слоя полагается свободной, а в [32] – непроницаемой. В работе [33] проводится сравнительное исследование влияния эффекта Соре на течения и процессы тепло- и массопереноса при различных граничных условиях для течений: свободных и непроницаемых границах.

**Длинноволновое приближение.** При фиксированном тепловом потоке через границы горизонтального слоя оказывается возможна длинноволновая конвективная неустойчивость состояния механического равновесия как в случае пористой среды, так и в однородной жидкости (см., например, [4,34,35]). В некоторых обстоятельствах длинноволновые возмущения могут быть критическими. Причем длинноволновая неустойчивость сохраняется и при термоконцентрационной конвекции, когда при фиксированном тепловом потоке дополнительно либо навязывается заданный поток примеси при двойной диффузии, либо непроницаемые для примеси граничные условия при термоконцентрационной конвекции, вызванной эффектом Соре (как это, например, было показано в работах [28,29,30,31,32,33,96]).

Помимо приближения плоскопараллельного течения, об использовании которого в цитируемых здесь работах уже упоминалось, для описания длинноволновой конвекции можно строить приближенные теории, учитывающие конечность длины волны возмущений и позволяющие отследить эволюцию длинноволновых структур. В работе [36] исследуется отбор структур при околокри-тической длинноволновой тепловой конвекции в слое однородной жидкости; подобные же уравнения описывают конвекцию в слое турбулентной жидкости [37].

В первой и второй главах данной диссертации выводятся и используются уравнения длинноволновой термоконцентрационной конвекции, вызванной эффектом Соре, в слое пористой среды при заданном вертикальном потоке тепла через слой [96,97]. Причем, в отличие от, например, работ [36,37], используемые уравнения справедливы при произвольной, а не малой надкритичности. Для подтверждения обоснованности использования длинноволнового приближения в первой главе исследуется линейная устойчивость стратифицированного состояния по отношению к возмущениям произвольной длины волны. В результате, для монотонной моды доказывается строго, а для колебательной демонстрируется, что длинноволновые возмущения всегда являются наиболее опасными.

**Параметрическое возбуждение термоконцентрационной конвекции.** В связи с колебательной модой термоконцентрационной неустойчивости, вызванной эффектом Соре, представляет интерес параметрическое возбуждение конвекции. Причем, поскольку характерные времена эволюции возмущений пропорциональны квадрату длины волны, для длинноволновых возмущений резонансными оказываются медленные частоты, которые и рассматриваются в первой главе диссертации.

Вибрационное воздействие на гидродинамические системы (включая термоконвективные) имеет самые разнообразные приложения: например, вибрационный контроль процессов тепло- и массопереноса в теплообменниках, смесителях, сепараторах минеральных веществ и системах для выращивания кристаллов. Эти приложения обусловили большой теоретический интерес к задачам о тепловой конвекции при периодической модуляции параметров во времени (например, вибрациях) и множество работ. Общее представление об ис-следованиях в данной области можно составить на основе, в первую очередь, монографии [38], а основные ранние исследования такого типа отражены, например, в обзоре [39] и монографиях [34,35].

Среди более поздних и близких к задачам, представленным в данной диссертации, работ следует остановиться на работах, посвященных исследованию влияния модуляции параметров на термоконцентрационную конвекцию в горизонтальном слое [40,41,42,43]. В работе [40] исследуется возникновение термоконцентрационной конвекции, вызванной двойной диффузией, в слое пористой среды, температура верхней границы которого периодически меняется во времени. С использованием теории Флоке показано, что модуляция может понижать критическое значение теплового числа Рэлея. Однако, ни при каких из рассматривавшихся значений параметров модуляция нагрева не приводила к стабилизации стратифицированного состояния.

В работе [41] рассматривается влияние высокочастотных вертикальных вибраций на конвекцию, обусловленную двойной диффузией, в двухмерной прямоугольной полости с фиксированными разностями температуры и концентрации между горизонтальными границами. Определены границы линейной устойчивости стратифицированного состояния по отношению к монотонным и колебательным возмущениям. В отличие от модуляции нагрева [40], модуляция силы тяжести может как понижать порог возбуждения конвективных движений при одних значениях параметров, так и повышать его при других. Также проанализировано влияние вибраций на структуру течений вблизи бифуркации. Определено, в каких диапазонах параметров возбуждение конвекции происходит жестко, а в каких – мягко. Для проверки и иллюстрации аналитических результатов было произведено прямое численное моделирование.

Авторы работы [42] вновь возвращаются к линейному анализу устойчивости стратифицированного состояния в задаче о периодической модуляции температуры границы пористой среды, насыщенной бинарной смесью. Однако, в этой работе рассматривается анизотропная пористая среда и учитываются инерционные слагаемые в законе сохранения импульса. При малых амплитудах и умеренных частотах модуляции критические число Рэлея и волновой вектор вычисляются аналитически. Обнаружено, что соответствующим выбором частоты модуляции можно как понизить, так и повысить порог линейной устойчивости системы.

Единственной работой (помимо [96]), в которой рассматривается термоконцентрационная конвекция, вызванная эффектом Соре, в пористой среде при модуляции параметров, является работа [43]. В этой работе вибрационное воздействие полагается высокочастотным. Аналитическое и численное исследования выявили, что вертикальные вибрации оказывают стабилизирующее воздействие как на монотонную, так и на колебательную моды неустойчивости стратифицированного состояния, а горизонтальные – дестабилизирующее.

Таким образом, вопрос о резонансном параметрическом возбуждении термоконцентрационной конвекции, вызванной эффектом Соре, в пористой среде, не освещен в доступной литературе.

**Источниковые решения.** В связи с тепловой конвекцией особый интерес вызывает развитие конвективных течений от локализованных источников тепла. Для однородной жидкости в различных вариантах постановки задачи рассматривался конвективный факел как от точечного источника в бесконечной среде, так и от горизонтального линейного. Среди недавних работ, представляющих интересные результаты и краткие обзоры состояния исследований соответствующих проблем, можно отметить работы [44,45], где рассматривается конвективный факел при малых числах Грасгофа около точечного/сферического и линейного/цилиндрического источников тепла, соответственно, и работу [46], где рассматривается линейный источник тепла при числах Рэлея в диапазоне $10^4$–$10^8$.

Исследования конвективного факела от точечного или линейного источника тепла в неограниченном массиве пористой среды впервые было представлено в работе [47]. В приближении пограничного слоя были получены самоподобные стационарные решения, предполагавшие достаточно большие значения числа Рэлея. Вопрос об обоснованности погранслойного приближения для задачи такого типа был подвергнут ревизии в [48]. Позже, в работе [49] было продемонстрировано, что, несмотря на критику, приведенную в [48], погранслойное приближение может быть использовано для исследования термоконвективных течений от точечного источника тепла в пористой среде. Самоподобные решения задачи об осесимметричном конвективном факеле в пористой среде, полученные в погранслойном приближении, представлены также в [50], а конвекция от горизонтального линейного источника была дополнительно аналитически исследована в работе [51]. В связи с тем, что при больших числах Рейнольдса, связанных с масштабом пор, приближение Дарси становится недействительным, в работе [52] для описания конвекции от точечного источника тепла использована модель Форхаймера. В работе [53] рассматривается как свободная, так и вынужденная конвекция в окрестности линейного источника тепла или нагреваемого цилиндра.

Единственной работой, посвященной термоконцентрационной конвекции от источника тепла в пористой среде, является работа [54], в которой рассматривается двойная диффузия вокруг горизонтального линейного источника в неограниченном объеме.

В работе [97] рассматриваются источниковые течения другого типа – это течения от локализованного источника тепла или примеси в тонком слое пористой среды, а не в неограниченном пространстве. Причем, эффективно, благодаря конвективной неустойчивости, процессы переноса, исследуемые в [97], могут рассматриваться как процессы в активной среде, в то время, как в приведенных выше примерах среда всегда диссипативная. Соответствующему иссле-дованию посвящена вторая глава диссертации. Можно также отметить, что оказывающиеся возможными режимы течения, при которых области длинноволнового течения вблизи источника и на бесконечности разделены кольцами коротковолнового переходного течения, не являются уникальными в своем роде. Другим примером такого рода локализованных переходных режимов течения является гидравлический скачок, возникающий при растекании струи жидкости по горизонтальной поверхности [55].

**Локализация течений в горизонтальном слое при случайно неоднородном нагреве.** Впервые локализация решений при случайной пространственной модуляции параметров распределенной системы рассматривалась в квантово-механических системах. А именно, Андерсон (Anderson) в работе [56] рассматривал теоретическую модель, описывавшую такие проблемы, как диффузия спинов, наблюдавшаяся в экспериментах [57] (в меньшей степени, теория Андерсона отвечала исследованиям [58,59]), или проводимость (последняя проблема затрагивается в контексте возможности возникновения связанных состояний электронов) в полупроводниках при наличии случайно распределенных примесей. В связи с этим, за эффектами локализации в системах со случайной модуляцией параметров закрепилось название "локализация Андерсона".

Классическим вариантом математической постановки задачи о локализации Андерсона является пространственно-дискретное уравнение Шредингера для частицы в случайном потенциале:

$$-\left(\Delta_{\mathrm{d}}\psi\right)_{\vec{j}} + U_{\vec{j}}\psi_{\vec{j}} = E\psi_{\vec{j}}, \tag{I.3}$$

где $\vec{j}$ – вектор, компонентами которого являются целые числа, $\psi_{\vec{j}}$ – амплитуда волновой функции электрона в окрестности $\vec{j}$-го узла кристаллической решетки, $\Delta_{\mathrm{d}}$ – дискретный оператор Лапласа в пространстве соответствующей раз-мерности, $U_{\vec{j}}$ – потенциал кристаллической решетки, полагается случайной гауссовской величиной, значения в соседних узлах независимы, $E$ – энергия данного состояния электрона. Аккуратное математическое исследование локализации решений задачи (I.3) может быть найдено в работе [60]. Среди результатов этих работ следует отметить, что для гуссовского шума в одномерном случае оказываются локализованы состояния с любой энергией, в то время как в двух- и трехмерных случаях возможны как локализованные, так и нелокализованные состояния.

Другим классическим вариантом постановки задачи является стационарное уравнение Шредингера в непрерывном пространстве, но с потенциалом в виде $\delta$-коррелированного шума. Подробное исследование такой задачи представлено в монографии [61] (а в монографии [62], например, можно найти исследование аналогичной, до некоторой степени, по духу своей математической постановки задачи о распространении света в случайно неоднородной среде). Для описания свойств локализации решений в одномерном случае в [61] рассматривается показатель экспоненциального роста (показатель локализации), который фактически является показателем Ляпунова, определяющим скорость асимптотического (на больших расстояниях) экспоненциального роста возмущений решения в стохастическом дифференциальном уравнении по мере движения в пространстве при заданной реализации шума.

Задача об отыскании показателя локализации (показателя Ляпунова) даже для стационарного уравнения Шредингера с потенциалом в виде белого ($\delta$-коррелированного) гауссовского шума требует решения уравнения Фоккера–Планка для логарифмической производной волновой функции и оказывается достаточно громоздкой. Если линейное дифференциальное стохастическое уравнение имеет порядок выше второго, то уравнение Фоккера–Планка следует решать в многомерном пространстве (логарифмическая производная и некоторые другие характеристики состояния системы) – аналитическое решение этой задачи в обобщенном случае невозможно. Однако, показатель Ляпунова может быть оценен, при помощи показателей роста моментов полей, осредненных по реализациям шума, подобно тому, как это было сделано в работе [63]. Для аналитического вычисления последних можно использовать результаты монографии [62] для линейных дифференциальных уравнений с белым шумом в коэффициентах.

Для полноты представления о состоянии исследований в данной области можно отметить, что локализация Андерсона возможна не только при случайной модуляции параметров, но и при квазипериодической, как это было показано, например, в работах [64,65].

Вслед за квантово-механическими системами, локализация Андерсона наблюдалась и исследовалась в других областях. Например, в обзорной работе [66], где представлен обширный разбор акустических проблем, аналогичных проблемам физики конденсированного состояния вещества, отдельное внимание уделяется локализации Андерсона в акустических системах. В работе [67] представлены простейшие варианты экспериментов по одномерной и многомерной локализации Андерсона (дискретного случая) в акустических системах. В теоретических работах [68,69] исследуются более сложные варианты системы, использованной в [67] для наблюдения одномерной локализации Андерсона: добавляется беспорядок в силе связи между элементами. В работе [70] рассматривается многомерная локализация Андерсона в непрерывной акустической системе (локализация ультразвуковых волн).

При этом упоминания задач о пространственной локализации конвективных течений на данный момент в литературе не встречается, хотя задачи такого типа могут представлять интерес, например, по причине того, что в реальных системах невозможно полностью избежать случайных пространственных неоднородностей. В этой связи, в третьей главе диссертации рассматривается тепловая конвекция в тонком горизонтальном слое пористой среды со случайной стационарной пространственной неоднородностью нагрева [98], обеспечиваемого посредством заданного потока тепла через слой. Как было отмечено выше, в таких условиях в системе при однородном нагреве возможна длинноволновая неустойчивость, и система рассматривается именно в длинноволновом приближении. Представлен вывод уравнений длинноволновой конвекции вблизи порога устойчивости при стационарной неоднородности нагрева (другой пример уравнений длинноволновой конвекции при пространственно неоднородном нагреве можно найти в работе [72]).

Ранее, в работе [73], исследовалось влияние случайной неоднородности локальной надкритичности на порог устойчивости в одномерном уравнении Ландау–Гинзбурга, но свойства локализации решений в линеаризованном варианте этой системы не отличаются от таковых в стационарном уравнении Шредингера с потенциалом в виде белого гауссовского шума, описанных в [61]. Гидродинамическая система, фигурирующая в третьей главе, описывается уравнением, линеаризация которого может быть сведена к упомянутому уравнению Шредингера лишь в отсутствии горизонтального прокачивания жидкости через слой. Кроме того, есть существенная разница в наблюдаемости эффектов, связанных с формальными свойствами уравнений, описывающих разные по своей природе системы.

В свете всего вышеперечисленного, третья глава диссертации была посвящена анализу вопроса об интерпретации локализации формальных решений линеаризованного варианта уравнений, описывающих тепловую конвекцию в тонком слое при случайно неоднородном нагреве, исследованию влияния на эти свойства прокачивания жидкости и наблюдению проявления этих свойств в исходной нелинейной системе.

**Синхронизация нелинейных систем общим шумом.** В теории стохастических процессов широко распространены ситуации, требующие осреднения по реализациям шума. При этом имеется два классических круга задач, связан-ных с осреднением вдоль траектории системы при заданной реализации шума. Количественное описание явлений, наблюдаемых в этих задачах, в той или иной степени фактически связано с вычислением показателей Ляпунова. Для фигурирующих в таких задачах стохастических систем показатели Ляпунова определяются в том смысле, что они описывают эволюцию малых возмущений траектории системы при заданной реализации шума, т.е. описывают устойчивость отклика не к возмущениям шума, а к возмущениям траектории системы.

Первый круг задач связан с явлениями локализации в распределенных системах со случайной пространственной модуляцией параметров – задача такого типа рассматривается в третьей главе. Второй круг задач упомянутого типа связан с синхронизацией нелинейных систем общим шумом, и некоторым общим результатам относительно этого явления посвящена четвертая глава диссертации.

Основным эффектом при воздействии шума на периодические автоколебания является появление диффузии фазы: автоколебания становятся неидеальными [74,75]. Однако шум может играть и упорядочивающую роль, в частности синхронизовать автоколебания. Если на две одинаковые (или слабо отличающиеся) автоколебательные системы действует общий шум, то их состояния могут под действием этого шума синхронизоваться. Этот эффект определяется знаком максимального показателя Ляпунова, при периодических автоколебаниях он отвечает направлению вдоль предельного цикла. Для автономных систем он нулевой, и синхронизации в описанном выше смысле нет. Под действием шума показатель Ляпунова может стать отрицательным, что означает синхронизацию.

Эффекты, связанные с проявлением синхронизации нелинейных систем под действием общего внешнего шума, в различных областях науки фигурируют под разными названиями. В нейрофизиологии используется понятие "надежности" ("reliability") нейронов [76,77,78,79], подразумевающее способность нейрона давать идентичный выходной сигнал при повторных подачах одной и той же предварительно записанной реализации случайного входного сигнала. В недавних экспериментах с Nd:YAG (неодим:иттрий гранат алюминия) лазерами [80,81] аналогичное свойство упоминалось как "устойчивость" ("consistency"). Если для воздействия используется не стохастический шум, а хаотический сигнал, говорится об обобщенной синхронизации (generalized synchronization) [82, 81].

Впервые задача о синхронизации идентичных систем общим внешним шумом была рассмотрена в работах [83,84], где рассматривался слабонелинейный квазигармонический автогенератор с шумом в виде случайной последовательности импульсов. Было установлено, что слабый шум синхронизует колебания (показатель Ляпунова отрицателен), а достаточно сильный – десинхронизует (показатель Ляпунова положителен). Позднее в таком смысле синхронизация нелинейных систем общим внешним шумом рассматривалась в книге [85] и работах [86,87,88,89,90].

В работе [99] и появившихся следом, независимо друг от друга, работах [91,100] авторы обращаются к автоколебательным динамическим системам общего положения с белым гауссовским шумом и исследуют их в рамках фазового приближения [92]. В рамках этого приближения оказываются возможными только отрицательные значения показателя Ляпунова, в то время, как в работах [83,84,93] демонстрируется возможность возникновения положительных показателей Ляпунова для стохастических систем. Действительно, в автоколебательных (в отсутствие шума) системах при умеренных интенсивностях шума в некоторых случаях также наблюдаются положительные значения показателя Ляпунова как для белого гауссовского шума [101,103], так и для телеграфного [102].

## Общая характеристика работы

**Содержание и структура работы.** Диссертация состоит из введения, четырех глав и заключения. В ведении представлен обзор литературы, содержание и основные цели данной работы.

В первой главе исследуется термоконцентрационная конвекция бинарной смеси в подогреваемом снизу горизонтальном слое пористой среды при наличии модуляции поля тяжести. Для случая фиксированного теплового потока через границы с учетом эффекта Соре выводятся уравнения нелинейной длинноволновой конвекции. Исследуется линейная устойчивость состояния механического равновесия при статическом поле тяжести и подтверждается обоснованность использования длинноволнового приближения: длинноволновые возмущения оказываются самыми опасными. Параметрическое возбуждение системы за счет модуляции силы тяжести исследуется аналитически при малых амплитудах и численно при больших. Оба анализа проводятся как для случая дискретного спектра волновых чисел (протяженный конечный слой), так и для непрерывного (бесконечный слой).

Во второй главе исследуются нелинейные режимы стационарной термоконцентрационной конвекция бинарной смеси в тонком слое пористой среды при наличии локализованного внутреннего источника примеси или тепла: находятся течения от этих источников.

В третьей главе рассматривается двухмерная тепловая конвекция жидкости в тонком горизонтальном слое со случайной стационарной неоднородностью нагрева, обеспечиваемой фиксированным потоком тепла поперек слоя. Выводятся уравнения нелинейной длинноволновой конвекции. При неоднородности, моделируемой белым гауссовским шумом, рассматривается возможность возникновения локализованных течений, а также свойства локализации и влияние на них горизонтально прокачивания жидкости.

В четвертой главе исследуется возможность синхронизации одинаковых (или слабонеидентичных) автоколебательных динамических систем посредством воздействия общим внешним шумом. Для чего вычисляется показатель Ляпунова, количественно характеризующий, в данном случае, способность систем синхронизоваться. При малых интенсивностях шума задача решается аналитически в фазовом приближении. При умеренных интенсивностях шум обнаруживает способность десинхронизовать некоторые системы.

В заключении представлены основные выводы и результаты данной диссертации.

**Актуальность работы.** Тепловая и термоконцентрационная конвекция в пористых средах представляет интерес как в связи с прикладными задачами, связанными с технологическими (например, охлаждение реакторов, фильтрация) и природными процессами (например, течения грунтовых вод), так и с точки зрения математической физики (например, явление косимметрии). При этом, учитываемый в работе при рассмотрении термоконцентрационной конвекции эффект термодиффузии (эффект Соре) существенен в газах и некоторых жидкостях, с одной стороны, и приводит к влиянию примесей на конвекцию при реалистичных граничных условиях, с другой.

Вместе с тем, вибрационное воздействие на термоконцентрационную конвекцию, имея самые разнообразные приложения (например, вибрационный контроль процессов тепло- и массопереноса в теплообменниках, смесителях, сепараторах минеральных веществ и системах для выращивания кристаллов из расплавов), остается на сегодняшний день мало исследованным.

В силу того, что вынужденные течения в слое очень медленно затухают по мере удаления от источника тепла или примеси даже при интенсивности подогрева, не достигающей критического (для возбуждения конвекции) значения, эти источники могут очень сильно влиять на течения жидкости и процессы тепло- и массопереноса в слое.

В реальных ситуациях, как правило, сложно реализовать системы с идеально однородными свойствами. Хотя в квантовой механике явление локализации решений при случайной пространственной модуляции параметров широко известно (локализация Андерсона) и довольно хорошо изучено, упоминаний об аналогичных исследованиях для гидродинамических систем в литературе не встречается.

Другим эффектом, связанным со свойствами стохастических систем, и требующим формального математического аппарата, очень похожего на тот, что используется для исследования одномерной локализации Андерсона, является синхронизация идентичных или похожих систем общим шумом. Явление синхронизации нелинейных динамических систем настолько широко распространено в природе и фундаментально, что не имеет смысла перечислять отдельные приложения, связанные с ним напрямую. Но, в дополнение, можно упомянуть приложения, не связанные с синхронизацией напрямую, а связанные с устойчивостью отклика системы на сложный сигнал (надежность (reliability) нейронов и устойчивость (consistency) лазеров). Дело в том, что если отклик устойчив, то сравнение двух реализаций отклика позволяет судить о степени соответствия реализаций соответствующих входных сигналов. С точки зрения восприятия и передачи информации, в случае с нейронами это предоставляет (или запрещает) принципиальную возможность кодирования сложных входных сигналов последовательностью синаптических импульсов.

## Цели работы:

1. Изучение вызванной эффектом Соре термоконцентрационной конвекции бинарной смеси, насыщающей тонкий горизонтальный слой пористой среды, при заданном тепловом потоке через границы;

2. Рассмотрение резонансного и нерезонансного влияния вибраций на термоконцентрационную конвекцию в упомянутой системе;

3. Исследование поведения системы при наличии локализованного внутреннего источника тепла или примеси (в отсутствие вибраций);

4. Изучение локализации конвективных течений (однокомпонентной) жидкости в тонком горизонтальном слое пористой среды при стационарной случайной пространственной неоднородности нагрева;

5. Исследование влияния горизонтального прокачивания жидкости на свойства локализации конвективных течений при случайной неоднородности нагрева;

6. Исследование устойчивости отклика автоколебательных систем на шумовое воздействие;

7. Изучение синхронизации идентичных или похожих систем общим внешним шумом.

## Научная новизна результатов

1. Выведены уравнения длинноволновой термоконцентрационной (с учетом эффекта Соре) конвекции в тонком горизонтальном слое пористой среды при заданном потоке тепла через границы слоя, справедливые при конечных надкритичностях [97]. Обосновано использование этого приближения: доказано, что критическими являются длинноволновые возмущения.

2. Впервые исследовано резонансное параметрическое возбуждение термоконцентрационной конвекции, обусловленной эффектом Соре, в пористой среде [96].

3. Получены решения, отвечающие течениям бинарной смеси от локализованного источника тепла или примеси в слое пористой среды [97].

4. Впервые исследована локализация Андерсона в гидродинамической системе (слой пористой среды при случайно неоднородном подогреве) [98]. Обнаружено существенное нетривиальное влияние прокачивания на свойства локализации конвективных течений.

5. Обнаружено и доказано, что слабый шум всегда синхронизует системы с гладким предельным циклом [99,100] (иными словами, отклик на шумовое воздействие устойчив).

6. Обнаружено, что умеренный шум может десинхронизовать автоколебательные системы с достаточно большой неизохронностью колебаний в окрестности предельного цикла [101,102,103] (неустойчивый отклик).

7. Впервые обнаружена и исследована возможность неустойчивости отклика для нейроноподобных систем [103].

8. Исследовано влияние неидентичностей систем, подверженных общему шумовому воздействию, на степень синхронности их колебаний [101,103].

**Автором представляются к защите:**

1. Результаты исследования термоконцентрационной конвекции бинарной смеси, для которой существенен эффект Соре, в тонком горизонтальном слое пористой среды при заданном тепловом потоке через границы.

2. Описание влияния вибрационного воздействия на конвективную устойчивость бинарной смеси в слое пористой среды [96].

3. Решения, описывающие вынужденную конвекцию бинарной смеси от локализованного источника тепла или примеси в подогреваемом горизонтальном слое пористой среды [97].

4. Свойства локализации конвективных течений однокомпонентной жидкости в тонком горизонтальном слое пористой среды при стационарной случайной пространственной неоднородности нагрева и их зависимость от горизонтального прокачивания жидкости через слой [98].

5. Свойства устойчивости отклика автоколебательных систем при шумовом воздействии [99,100,101,102,103].

6. Эффекты синхронизации и десинхронизации идентичных или похожих систем общим внешним шумом [101,102,103].

7. Свойства синхронности колебаний неидентичных систем [101,103].

**Достоверность результатов.** Достоверность полученных результатов подтверждается сравнением как с известными ранее работами, так и с появляющимися работами авторов, работающих над смежными проблемами, адекватностью методов исследования и согласием результатов, полученных разными методами и с использованием разных подходов.

В частности, в работах [96,98,101,102,103] все численные результаты сходятся по мере уменьшения погрешности процедур интегрирования уравнений, не изменяются при использовании других процедур и в соответствующих пределах согласуются с результатами аналитических теорий, построенных, как в самих этих работах и работах [99,100], так и в работах других авторов.

В пользу достоверности результатов также свидетельствует то, что основные результаты работы [99] были, по-видимому, независимо получены J. Teramae и D. Tanaka и представлены в работе Phys. Rev. Lett. **93**, 204103 (2004).

**Публикации.** Материалы диссертации изложены в 6 статьях в журналах[1] [97,98,99,100,101,103], 2 трудах конференций [96,102] и тезисах 9 перечисленных ниже докладов на конференциях.

**Апробация работы.** Результаты работы были представлены в виде:

Лекционный доклад

1. D.S. Goldobin, A. Pikovsky "Synchronization and Desynchronization of Oscillators by Common Noisy Driving" on the Int. Workshop "Constructive Role of Noise in Complex Systems", Dresden, Germany, July 16–21, 2006;

Выступления на семинаре:

2. Голдобин Д.С. и Любимов Д.В. "Конвекция в слое пористой среды, индуцированная локальным источником тепла или примеси" на Пермском гидродинамическом семинар им. Г.З.Гершуни и Е.М.Жуховицкого, Пермь (Россия), 17 марта 2006 г.;

3. Голдобин Д.С. "Локализация течений в горизонтальном слое при случайно неоднородном нагреве" на Пермском гидродинамическом семинар им. Г.З.Гершуни и Е.М.Жуховицкого, Пермь (Россия), 10 января 2007 г.;

Устные доклады:

4. Goldobin D.S., Lyubimov D.V., and Mojtabi A. "Parametric instability of a conductive state of binary mixture in porous medium" on the Int. Conf. "Advanced Problems in Thermal Convection" Perm (Russia), Nov. 24–27, 2003;

---

[1] входящих в перечни ведущих рецензируемых и зарубежных научных журналов и изданий, в которых должны быть опубликованы основные научные результаты диссертации на соискание ученой степени доктора и кандидата наук

5. D.S. Goldobin, A. Pikovsky "Antireliability of Noisy-Driven Neuron-Like Oscillators" on the 6-th Crimean School and Workshops "Nonlinear Dynamics, Chaos, and Applications", Yalta, Crimea, Ukraine, May 15–26, 2006;

6. D.S. Goldobin, A. Pikovsky "Synchrony of Oscillators Subject to Common Noise" on the Int. Seminar and Workshop "Constructive Role of Noise in Complex Systems", Dresden, Germany, June 26–July 21, 2006;

7. D.S. Goldobin, A. Pikovsky "Intermittent Synchrony of Noise-Induced Bursts" on the Int. Conf. "Dynamics Days Europe 2006", Crete, Greece, September 25–29, 2006;

Стендовые доклады:

8. Голдобин Д.С. и Любимов Д.В. "Термоконцентрационная конвекция бинарной смеси в горизонтальном слое пористой среды при подогреве снизу" на 14-ой Зимней школе по механике сплошных сред, Пермь (Россия), 28 февраля – 3 марта 2005 г.;

9. D.S. Goldobin, A. Pikovsky "Noise-Induced Synchronization and Desynchronization of Self-Sustained Oscillators" on the 4-th Int. Conf. on "Unsolved Problems of Noise", Gallipoli (Lecce), Italy, June 6–10, 2005;

10. D.S. Goldobin, A. Pikovsky "Noise-Induced Synchronization and Desynchronization of Neural Oscillators" on the 13-th Int. IEEE Workshop on "Nonlinear Dynamics of Electronic Systems", Potsdam, Germany, September 18–22, 2005.

**Практическая ценность.** Полученные результаты могут быть использованы для создания методов изучения физических параметров и свойств жидкости. Вибрационное управление конвективной устойчивостью бинарной смеси в пористой среде может быть использовано при фильтрации и, например, разделении изотопов. Точечные источники тепла/примеси в слое пористой среды, насыщенном бинарной смесью, как и случайная неоднородность нагрева слоя, насыщенного однокомпонентной жидкостью, вызывают существенный конвективный перенос тепла даже в подкритической области, что может радикально сказываться на интегральных теплопроводящих(-изолирующих) свойствах таких слоев. Контроль синхронности работы автоколебательных систем находит самое широкое применение в технике и проявление в природе; устойчивость отклика нейронов на шумовое воздействие позволяет допустить принципиальную возможность кодирования сложных входных сигналов последовательностью синаптических импульсов.

# Глава 1: Параметрическое возбуждение термоконцентрационной конвекции в слое пористой среды

Данная глава посвящена исследованию термоконцентрационной конвекции бинарной смеси в подогреваемом снизу горизонтальном слое пористой среды при наличии модуляции поля тяжести. Известно, что при фиксированном потоке тепла через границы в системе возможна монотонная длинноволновая тепловая конвективная неустойчивость, а при фиксированном диффузионном потоке примеси – монотонная длинноволновая концентрационная конвективная неустойчивость. Конкуренция этих механизмов вполне ожидаемо приводит к возникновению в системе длинноволновой колебательной неустойчивости. В связи с этой неустойчивостью вызывает интерес возможность параметрического резонанса в данной системе. Однако, фиксированный ненулевой градиент концентрации примеси около границы при реальных граничных условиях может быть достигнут лишь тогда, когда существенен эффект Соре и поток вещества через границу равен нулю. Соответственно, приняты граничные условия, требующие фиксированный тепловой поток и отсутствие потока примеси через границы; движения жидкости описываются в приближении Дарси-Буссинеска с учетом эффекта Соре.

Для описанной конфигурации системы выводятся уравнения, описывающие нелинейную (нелинейные режимы рассматриваются в следующей главе) динамику длинноволновых структур при конечной надкритичности.

Далее аналитически исследуется линейная устойчивость состояния механического равновесия по отношению к длинноволновым возмущениям в отсутствии модуляции. Для проверки обоснованности использования длинноволнового приближения исследуется линейная устойчивость системы по отношению к возмущениям с конечной длиной волны. Это исследование показывает, что длинноволновые возмущения, действительно, являются самыми опасными.

Задачи о конечной области, когда спектр волновых векторов дискретен, и бесконечной, когда спектр непрерывен, оказываются существенно различными. Влияние модуляции исследуется аналитически для малых амплитуд модуляции: рассматриваются изменение монотонного уровня и первые три резонанса – и численно для конечных амплитуд. То и другое выполняется как для дискретного, так и для непрерывного спектров волновых векторов.

Оказывается, что параметрическое воздействие не оказывает влияния на модули комплексных мультипликаторов возмущения с фиксированной длиной волны, а влияет лишь на границы области, где эти мультипликаторы остаются комплексными. В результате устойчивость системы по отношению к квазипериодическим колебаниям возмущения с данной длиной волны не зависит от амплитуды модуляции – дестабилизация может быть связана исключительно с резонансами и монотонной неустойчивостью немодулированной системы. При дискретном спектре граница устойчивости системы может определяться для сколь угодно малых амплитуд модуляции резонансами как первого порядка, так и старших, а при непрерывном спектре и малой модуляции параметрическая дестабилизация колебательного уровня определяется резонансом исключительно первого порядка. Для некоторых значений параметров системы монотонный уровень неустойчивости может дестабилизироваться модуляцией. При этой дестабилизации наиболее опасными оказываются однородные возмущения при непрерывном спектре, а при дискретном – возмущения с наибольшей возможной длиной волны. Стабилизация монотонного уровня невозможна ни при дискретном, ни при непрерывном спектрах волновых чисел. Изложение базируется в основном на работе [96].

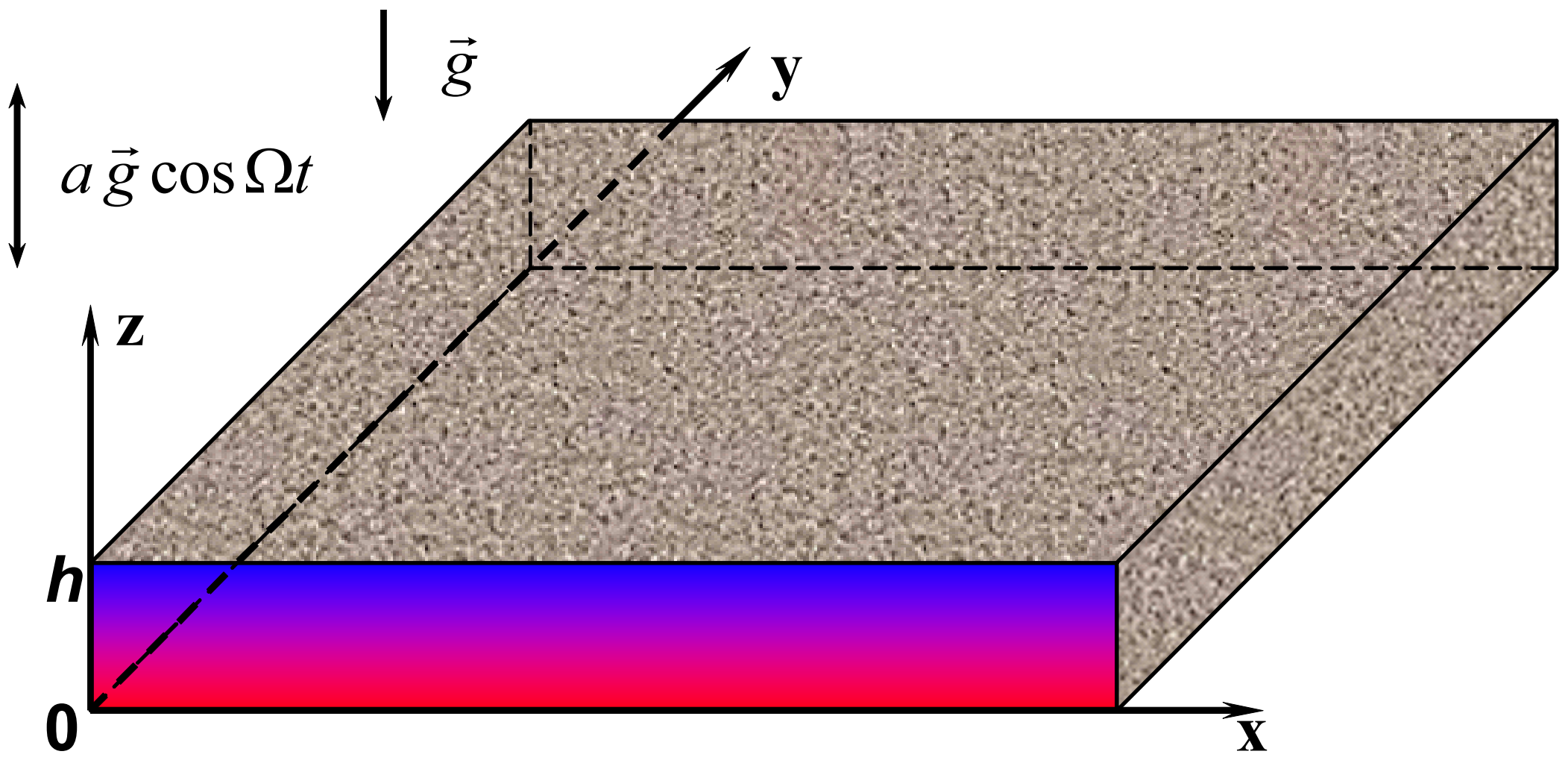

Рис. 1.1: Геометрия задачи и система координат

## 1.1. Уравнения термоконцентрационной конвекции с учетом эффекта Соре в слое пористой среды

Рассматривается термоконцентрационная конвекция бинарной смеси в подогреваемом снизу горизонтальном слое пористой среды. Границы слоя полагаются непроницаемыми (в том числе для примеси), тепловой поток – фиксированным. Принимается во внимание эффект термодиффузии (эффект Соре): поток концентрации примеси

$$\vec{j} = -D\,\nabla C + \alpha\, T^{-1}\nabla T\,,$$

где $D$ и $\alpha$ – коэффициенты диффузии и термодиффузии, соответственно. Предполагается, что выравнивание температуры между жидкостью и твердым скелетом происходит достаточно быстро, и отдельных температур для них не вводится. При малых перепадах температуры и концентрации можно полагать, что плотность жидкости зависит от них линейно:

$$\rho = \rho_0\Big(1 - \beta\big(T - T_0\big) + \beta_C\big(C - C_0\big)\Big),$$

где $C$ – концентрация тяжелой компоненты, $\rho_0$ – плотность смеси при температуре $T_0$ и концентрации $C_0$, $\beta$ – коэффициент теплового расширения, а $\beta_C = \rho_0^{-1}\left(\partial\rho / \partial C\right)_{T,p}$ определяет зависимость плотности от концентрации. Для потока концентрации изменения концентрации и температуры учитываются лишь в градиентах. Система координат выбирается так, что плоскость $(x, y)$ горизонтальна, $z = 0$ и $z = h$ – нижняя и верхняя границы слоя, соответственно (Рис. 1.1).

Для описания поведения системы используется модель Дарси-Буссинеска [4]:

$$0 = -\frac{1}{\rho_0}\nabla P - \frac{\nu m}{K}\vec{v} + g\left(\beta T - \beta_C C\right)\vec{e}_z,$$

$$\frac{\partial C}{\partial t} + \vec{v}\cdot\nabla C = D\,\Delta C - \frac{\alpha}{T_0}\Delta T,$$

$$\frac{\partial T}{\partial t} + b^{-1}\,\vec{v}\cdot\nabla T = \chi\,\Delta T,$$

$$\nabla\cdot\vec{v} = 0,$$

$$z = 0{,}1: \ \frac{\partial T}{\partial z} = A,\ \ v_z = 0,\ \ j_z = 0,$$

где $\vec{v}$ – средняя (осредненная на масштабах пор) скорость жидкости в порах, $\nu$ – кинематическая вязкость смеси, $m$ – пористость среды (отношение объема пор в элементе пористой среды к объему этого элемента), $K$ – коэффициент проницаемости, $\vec{g} = -g\,\vec{e}_z$ – поле тяжести, $b$ – отношение теплоемкости пористой среды, насыщенной жидкостью, к части этой теплоемкости, приходящейся на жидкость в порах, $b > 1$, $\chi$ – температуропроводность пористой среды, насыщенной жидкостью.

В данной задаче удобной единицей измерения длины является толщина слоя $h$, времени – $h^2\chi^{-1}$, скорости – $h^{-1}b\chi$, температуры и концентрации – разности температур и концентраций на границах слоя в основном состоянии $Ah$ и $\alpha Ah/(DT_0)$, соответственно, и давления $b\rho_0\nu\chi m/K$.

Характер поведения системы будет определяться четырьмя безразмерными параметрами:

$$N \equiv \frac{\beta_C\,\alpha}{\beta\,DT_0}, \quad Ra \equiv \frac{\beta Ah^2 gK}{mb\,\nu\chi}, \quad S \equiv \frac{D}{\chi} \quad \text{и} \quad b,$$

где $N$ – параметр плавучести, $Ra$ – число Релея-Дарси (концентрационное число Релея-Дарси $Rc \equiv Ra\,N$), и $S \equiv Le^{-1}$ – обратное число Льюиса.

Для учета модуляции силы тяжести достаточно вынести из чисел Релея-Дарси (обычного и концентрационного) множитель $1 + a\cos\Omega t$, где $a$ определено так, что $a\cos\Omega t$ есть мгновенная вибрационная перегрузка, а $\Omega$ – безразмерная частота вибраций. При выводе последующих уравнений негласно уделяется внимание тому, что числа Релея-Дарси могут зависеть от времени, и для всех уравнений, приведенных ниже в этой главе, указанный способ формального учета модуляции остается справедливым.

Система безразмерных уравнений, описывающая поведение конечных возмущений состояния механического равновесия:

$$T = -z\,, \quad C = -z\,,$$

имеет вид:

$$-\nabla p-\vec{v}+Ra\Big(\big(1-N\,b\big)\theta+N\phi\Big)\vec{e}_z=0,\qquad(1.1)$$

$$\frac{\partial\phi}{\partial t}+b\,\vec{v}\cdot\nabla\big(\phi-(b-1)\theta\big)=b\Delta\theta+S\,\Delta\big(\phi-(b-1)\theta\big),\qquad(1.2)$$

$$\frac{\partial\theta}{\partial t}+\vec{v}\cdot\nabla\theta=\Delta\theta+w,\qquad(1.3)$$

$$\nabla\cdot\vec{v}=0\qquad(1.4)$$

с граничными условиями:

$$z=0,1:\quad w=\frac{\partial\theta}{\partial z}=\frac{\partial\phi}{\partial z}=0,\qquad(1.5)$$

где $\theta$ – возмущение поля температуры, $\phi$ – возмущение поля $\varPhi\equiv bT-C$ (на которое можно смотреть как на некоторый эффективный химический потенциал), $w$ – $z$-компонента поля скорости.

Замена переменных и параметров:

$$\phi^*=\big(b-S(b-1)\big)^{-1}\phi,\ \ Ra^*=\big(1-NS(b-1)\big)Ra,$$
$$N^*=\big(1-NS(b-1)\big)^{-1}\big(Nb-NS(b-1)\big)\qquad(1.6)$$

– позволяет увидеть, что в линеаризованной задаче фигурируют только три параметра. Знак "*" далее опускается. Исключая горизонтальную компоненту скорости

$$\vec{v}_2=-\nabla_2 p\qquad(1.7)$$

(индекс "2" у векторов и операторов означает горизонтальную компоненту и дифференцирование лишь по горизонтальным координатам, соответственно; выражение для горизонтальной компоненты поля скорости получается при проектировании уравнения (1.1) на горизонтальную плоскость), можно получить

$$\Delta w = Ra\ \Delta_2\big((1-N)\theta + N\phi\big), \tag{1.8}$$

$$\frac{\partial\phi}{\partial t} + b\boldsymbol{\Gamma}\left(\phi - \frac{b-1}{b-S(b-1)}\theta\right) = S\Delta\phi + \Delta\theta, \tag{1.9}$$

$$\frac{\partial\theta}{\partial t} + \boldsymbol{\Gamma}\theta = \Delta\theta + w, \tag{1.10}$$

$$\frac{\partial w}{\partial z} = \Delta_2 p, \tag{1.11}$$

где $\boldsymbol{\Gamma} \equiv w\dfrac{\partial}{\partial z} - \nabla_2 p\cdot\nabla_2$ – оператор конвективной производной. Первое уравнение последней системы – это $z$-компонента результата двукратного применения операции ротора к закону сохранения импульса (1.1) с учетом того, что $\mathrm{rot}\ \mathrm{rot}\big(f\,\vec\gamma\big) = \nabla\big(\vec\gamma\cdot\nabla f\big) - \vec\gamma\,\Delta f$. Уравнение (1.11) – это уравнение непрерывности (1.4), переписанное с учетом (1.7).

## 1.2. Длинноволновое приближение при конечных надкритичностях – нелинейные уравнения

Хотя в данной главе мы интересуемся только линейными уравнениями, имеет смысл выводить сразу нелинейные, поскольку они будут фигурировать в следующей. Благодаря принятым граничным условиям в системе не исключено существование длинноволновой неустойчивости [34]. В связи с этим уравнения поведения длинноволновых возмущений представляют особый интерес. Суть приближения заключается в предположении о малости горизонтальных производных поля скорости по сравнению с вертикальной. Для учета нелинейности уравнений, связанной с конвективной производной, вертикальную компоненту поля скорости нужно полагать малой: $w \sim L^{-1}$, где $L$ – характерный горизонтальный масштаб течений, – а скалярные поля большими: $\big(p,\ \phi,\ \theta\big) \sim L$. Тогда из системы (1.8)–(1.11) с граничными условиями (1.5) можно последовательно получать:

В порядке $L^{1}$:

$$\phi^{(-1)} = \varphi\left(x,\, y\right), \qquad \theta^{(-1)} = \vartheta\left(x,\, y\right),$$

$$w^{(1)} = -\frac{Ra}{12}\Delta_2\left((1-N)\vartheta + N\varphi\right)\left(6z - 6z^2\right),$$

$$p^{(-1)} = -\frac{Ra}{12}\left((1-N)\vartheta + N\varphi\right)\left(6 - 12\,z\right) \equiv -P\left(x,\, y\right)\left(6 - 12\,z\right).$$

В порядке $L^{0}$:

$$\phi^{(0)} = \mathit{\Phi}\left(x,\, y\right)\left(3z^2 - 2z^3\right),$$

$$\mathit{\Phi}\left(x,\, y\right) \equiv S^{-1}\,\nabla_2 P\cdot\nabla_2\left(b\,\varphi - \left(b\,\sigma + 1\right)\vartheta\right),$$

$$\theta^{(0)} = \mathit{\Theta}\left(x,\, y\right)\left(3z^2 - 2z^3\right),$$

$$\mathit{\Theta}\left(x,\, y\right) \equiv \nabla_2 P\cdot\nabla_2\,\vartheta\,.$$

В порядке $L^{-1}$:

С учетом того, что

$$\begin{aligned}
&\mathbf{\Gamma}^{(2)}f^{(-1)} + \mathbf{\Gamma}^{(1)}f^{(0)} = \\
&\quad = w^{(2)}f_z^{(-1)} - \nabla_2 p^{(0)}\cdot\nabla_2 f^{(-1)} + w^{(1)}f_z^{(0)} - \nabla_2 p^{(-1)}\cdot\nabla_2 f^{(0)} = \\
&\qquad\qquad \partial_z f^{(-1)} = 0, \qquad p^{(0)} = \Delta_2^{-1}w_z^{(2)} \\
&\quad = \partial_z\left(\ldots\right) - \left(w_z^{(1)} - \mathit{\Delta}_2 p^{(-1)}\right)f^{(0)} - \nabla_2\cdot\left(f^{(0)}\,\nabla_2 p^{(-1)}\right),
\end{aligned}$$

а, следовательно,

$$\int_0^1 \left(\mathbf{\Gamma}^{(2)} f^{(-1)} + \mathbf{\Gamma}^{(1)} f^{(0)}\right) dz =$$

$$= \nabla_2 \cdot \left(F\, \nabla_2 P\right) \int_0^1 \left(6 - 12\, z\right)\left(3z^2 - 2z^3\right) dz = -\frac{6}{5} \nabla_2 \cdot \left(F\, \nabla_2 P\right)$$

(здесь вместо функций $f$ и $F$ нужно подставлять соответствующие функции $\phi$, $\theta$ и $\Phi$, $\Theta$), окончательно получается

$$\frac{\partial \varphi}{\partial t} - \frac{6b^2}{5S} \nabla_i \left( \nabla_j \left[ \varphi - \frac{b\, \vartheta}{b - S(b-1)} \right] \nabla_i P\, \nabla_j P \right) = S\Delta\varphi + \Delta\vartheta, \quad (1.12)$$

$$\frac{\partial \vartheta}{\partial t} - \frac{6}{5} \nabla_i \left( \nabla_j \vartheta\, \nabla_i P\, \nabla_j P \right) = \Delta\vartheta - \Delta P, \quad (1.13)$$

где $P(x,\, y) \equiv Ra \,/\, 12\, \left((1-N)\vartheta + N\varphi\right)$ и индексы $i$ и $j$ пробегают значения $1,\, 2$; все поля в последней системе зависят лишь от горизонтальных координат и времени.

## 1.3. Устойчивость состояния механического равновесия в статическом поле тяжести

### 1.3.1. Статическое поле тяжести, длинноволновое приближение

После исключения нелинейных слагаемых система (1.12),(1.13) принимает вид:

$$\frac{\partial \varphi}{\partial t} = S\Delta\varphi + \Delta\vartheta, \quad (1.14)$$

$$\frac{\partial \vartheta}{\partial t} = \Delta\vartheta - \frac{Ra}{12} \Delta\left((1-N)\vartheta + N\varphi\right). \quad (1.15)$$

Рассматривая нормальные возмущения ~ $\exp\left(i\vec{k}\cdot\vec{r}\right)$ и изменяя масштаб времени $k^2 t \to t$, последнюю систему можно переписать в виде

$$\frac{d\varphi}{dt} = -S\varphi - \vartheta\,, \tag{1.16}$$

$$\frac{d\vartheta}{dt} = -\vartheta + \frac{Ra}{12}\big((1-N)\vartheta + N\varphi\big). \tag{1.17}$$

Возмущения температуры можно исключить из второго уравнения, выразив их из первого:

$$\varphi_{tt} + \left(S + 1 + \frac{Rc - Ra}{12}\right)\varphi_t + \left(S + \frac{Rc + S\,Rc - S\,Ra}{12}\right)\varphi = 0. \tag{1.18}$$

Важно, что при последней подстановке $Ra$ не выносилось из-под оператора дифференцирования по времени, а значит, последнее уравнение остается справедливым и при переменном поле тяжести, что позволяет использовать его при рассмотрении параметрического резонанса.

Для постоянных $Ra$ и $Rc$ положение равновесия является седлом при

$$Ra > \left(1 + S^{-1}\right) Rc + 12, \tag{1.19}$$

центром с частотой $\omega^2 = \dfrac{Rc}{12} - S^2$ при

$$Ra = Rc + 12 + 12S \tag{1.20}$$

и неустойчивым фокусом при

$$Rc + 12 + 12S < Ra < \left(\sqrt{Rc} + \sqrt{12}\right)^2 - 12S\,.$$

Границы колебательной и монотонной неустойчивостей пересекаются при

$$Rc_0 = 12S^2, \;\; Ra_0 = 12\left(S^2 + S + 1\right);$$

система неравенств, определяющая область существования колебательной неустойчивости, имеет решения при $Rc > Rc_0$. Результаты можно видеть на Рис. 1.2.

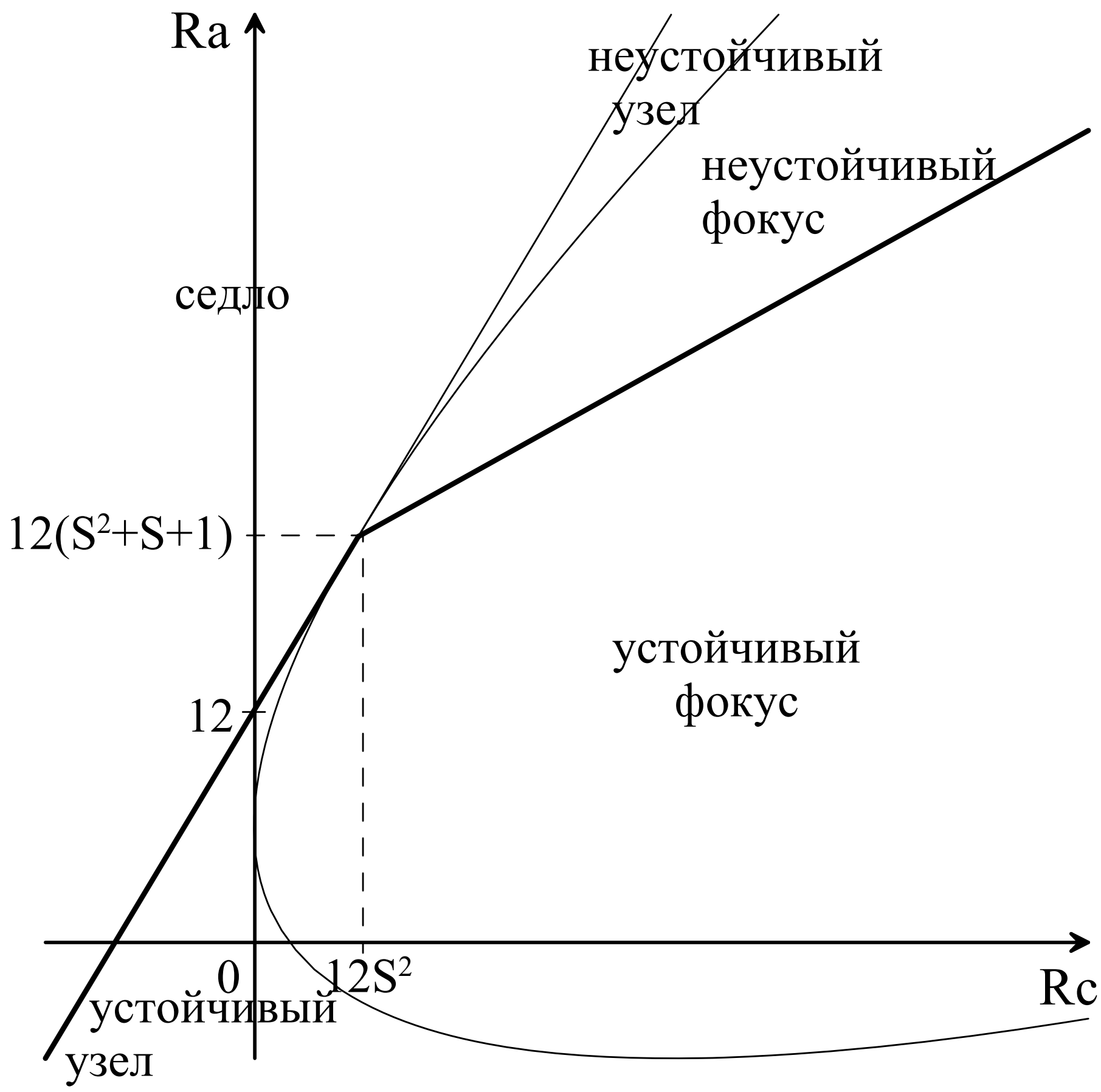

Рис. 1.2: *Статическое поле тяжести.* Тип стационарной точки фазового пространства, отвечающей состоянию механического равновесия системы, в подпространстве длинноволновых возмущений.

Следует обратить внимание на то, что физически содержательными являются значения обратного числа Льюиса $S$ в диапазоне от ~ $0.01$ (жидкости) до $\sim 1$ (газы). В первом случае угол между границами монотонной и колебательной неустойчивостей на плоскости $\left(Rc,\, Ra\right)$ очень мал. Поэтому естественная для резонансных явлений координатная сетка – сетка нормальных отклонений от границ устойчивости:

$$D \equiv S + \frac{Rc + S\ Rc - S\ Ra}{12}, \tag{1.21}$$

$$r \equiv \frac{Ra - Rc}{12} - 1 - S \tag{1.22}$$

– сильно "сплюснута". Вследствие чего, многие результаты существенно удобнее представлять на плоскости $\left(D,\, r\right)$, а не $\left(Rc,\, Ra\right)$. Тогда (1.19) и (1.20) примут вид

$$D < 0 \quad \text{и} \quad r = 0,$$

соответственно.

### 1.3.2. Статическое поле тяжести, возмущения конечной длины волны

Перед тем, как продолжать исследование задачи в коротковолновом приближении, необходимо проверить ценность такого приближения, рассмотрев устойчивость системы по отношению к возмущениям конечной длины в постоянном поле тяжести. Для малых нормальных возмущений с конечным горизонтальным волновым вектором $\vec{k}$ и инкрементом $\lambda$ из системы уравнений (1.8)–(1.11) с граничными условиями (1.5) можно получить следующую линейную задачу:

$$\left(\partial_z^2 - k^2\right)w + k^2 Rc\ \phi + k^2\left(Ra - Rc\right)\theta = 0, \tag{1.23}$$

$$\lambda\phi = S\left(\partial_z^2 - k^2\right)\phi + \left(\partial_z^2 - k^2\right)\theta, \tag{1.24}$$

$$\lambda\theta = \left(\partial_z^2 - k^2\right)\theta + w\,, \tag{1.25}$$

$$z = 0, 1: \quad w = \partial_z\theta = \partial_z\phi = 0\,. \tag{1.26}$$

**Монотонные возмущения.** При $\lambda = 0$ из второго уравнения последней системы с учетом граничных условий сразу же следует

$$\phi = -S^{-1}\theta\,, \tag{1.27}$$

и получающееся для системы

$$\partial_z^2 w = k^2 w - k^2\left(Ra - \left(1 + S^{-1}\right)Rc\right)\theta\,, \tag{1.28}$$

$$\partial_z^2\theta = -w + k^2\theta\,, \tag{1.29}$$

биквадратное характеристическое уравнение может быть решено аналитически. Линейные комбинации собственных решений системы (1.28),(1.29), удовлетворяющие граничным условиям, дают симметричные:

$$w_{s,n}\left(z, k\right) = \frac{\cos\left[\sqrt{k\left(q_{s,n} - k\right)}\left(z - \frac{1}{2}\right)\right]}{\cos\frac{\sqrt{k\left(q_{s,n} - k\right)}}{2}} - \frac{\operatorname{ch}\left[\sqrt{k\left(q_{s,n} + k\right)}\left(z - \frac{1}{2}\right)\right]}{\operatorname{ch}\frac{\sqrt{k\left(q_{s,n} + k\right)}}{2}}\,, \tag{1.30}$$

$$\theta_{s,n}\left(z,k\right)=\frac{\cos\left[\sqrt{k\left(q_{s,n}-k\right)}\left(z-\frac{1}{2}\right)\right]}{kq_{s,n}\cos\frac{\sqrt{k\left(q_{s,n}-k\right)}}{2}}+ \\ +\frac{\operatorname{ch}\left[\sqrt{k\left(q_{s,n}+k\right)}\left(z-\frac{1}{2}\right)\right]}{kq_{s,n}\operatorname{ch}\frac{\sqrt{k\left(q_{s,n}+k\right)}}{2}} \qquad (1.31)$$

и антисимметричные решения линейной задачи:

$$w_{a,n}\left(z,k\right)=\frac{\sin\left[\sqrt{k\left(q_{a,n}-k\right)}\left(z-\frac{1}{2}\right)\right]}{\sin\frac{\sqrt{k\left(q_{a,n}-k\right)}}{2}}- \\ -\frac{\operatorname{sh}\left[\sqrt{k\left(q_{a,n}+k\right)}\left(z-\frac{1}{2}\right)\right]}{\operatorname{sh}\frac{\sqrt{k\left(q_{a,n}+k\right)}}{2}}, \qquad (1.32)$$

$$\theta_{a,n}\left(z,k\right)=\frac{\sin\left[\sqrt{k\left(q_{a,n}-k\right)}\left(z-\frac{1}{2}\right)\right]}{kq_{a,n}\sin\frac{\sqrt{k\left(q_{a,n}-k\right)}}{2}}+ \\ +\frac{\operatorname{sh}\left[\sqrt{k\left(q_{a,n}+k\right)}\left(z-\frac{1}{2}\right)\right]}{kq_{a,n}\operatorname{sh}\frac{\sqrt{k\left(q_{a,n}+k\right)}}{2}}, \qquad (1.33)$$

где $n \in N$ и $q^2_{\dots,n} \equiv Ra - Rc\left(1 + S^{-1}\right) \equiv 12 - 12\,S^{-1}D$, при котором нейтрально устойчивы возмущения $n$-го порядка и соответствующей симметрии с волновым числом $k$, определяется из трансцендентного уравнения

$$\sqrt{q_s - k}\,\mathrm{tg}\frac{\sqrt{k\left(q_s - k\right)}}{2} - \sqrt{q_s + k}\,\mathrm{th}\frac{\sqrt{k\left(q_s + k\right)}}{2} = 0 \qquad (1.34)$$

для симметричных возмущений и

$$\sqrt{q_a - k}\,\mathrm{ctg}\frac{\sqrt{k\left(q_a - k\right)}}{2} + \sqrt{q_a + k}\,\mathrm{cth}\frac{\sqrt{k\left(q_a + k\right)}}{2} = 0 \qquad (1.35)$$

– для антисимметричных. В область $q < k$ (как и в область чисто мнимых $q$) можно строить аналитическое продолжение этих уравнений, но там нет нетривиальных решений.

В окрестности границы монотонной неустойчивости системы по отношению к длинноволновым возмущениям из трансцендентного уравнения (1.34), связывающего $q$ и $k$ для симметричных возмущений, можно аналитически получить:

$$k^2 \approx \frac{7}{8}\left(q^2 - 12\right) = \frac{7}{8}\left(Ra - Rc\left(1 + S^{-1}\right) - 12\right) = -\frac{21}{2}\frac{D}{S}. \qquad (1.36)$$

Т.е., как минимум, из возмущений с малыми волновыми числами наиболее опасны длинноволновые. Кроме того, аналитически можно получить, что

$$q_{s,n} \in \left(k + \frac{4\pi^2}{k}(n-1)^2,\, k + \frac{4\pi^2}{k}\left(n - \frac{1}{2}\right)^2\right], \qquad (1.37)$$

$$q_{a,n} \in \left(k + \frac{4\pi^2}{k}\left(n - \frac{1}{2}\right)^2,\, k + \frac{4\pi^2}{k}n^2\right], \qquad (1.38)$$

где $n \in N$. Для произвольных волновых чисел трансцендентное уравнение может быть точно решено лишь численно и соответствующие результаты, представленные на Рис. 1.3, демонстрируют, что самыми опасными стационарными возмущениями являются длинноволновые.

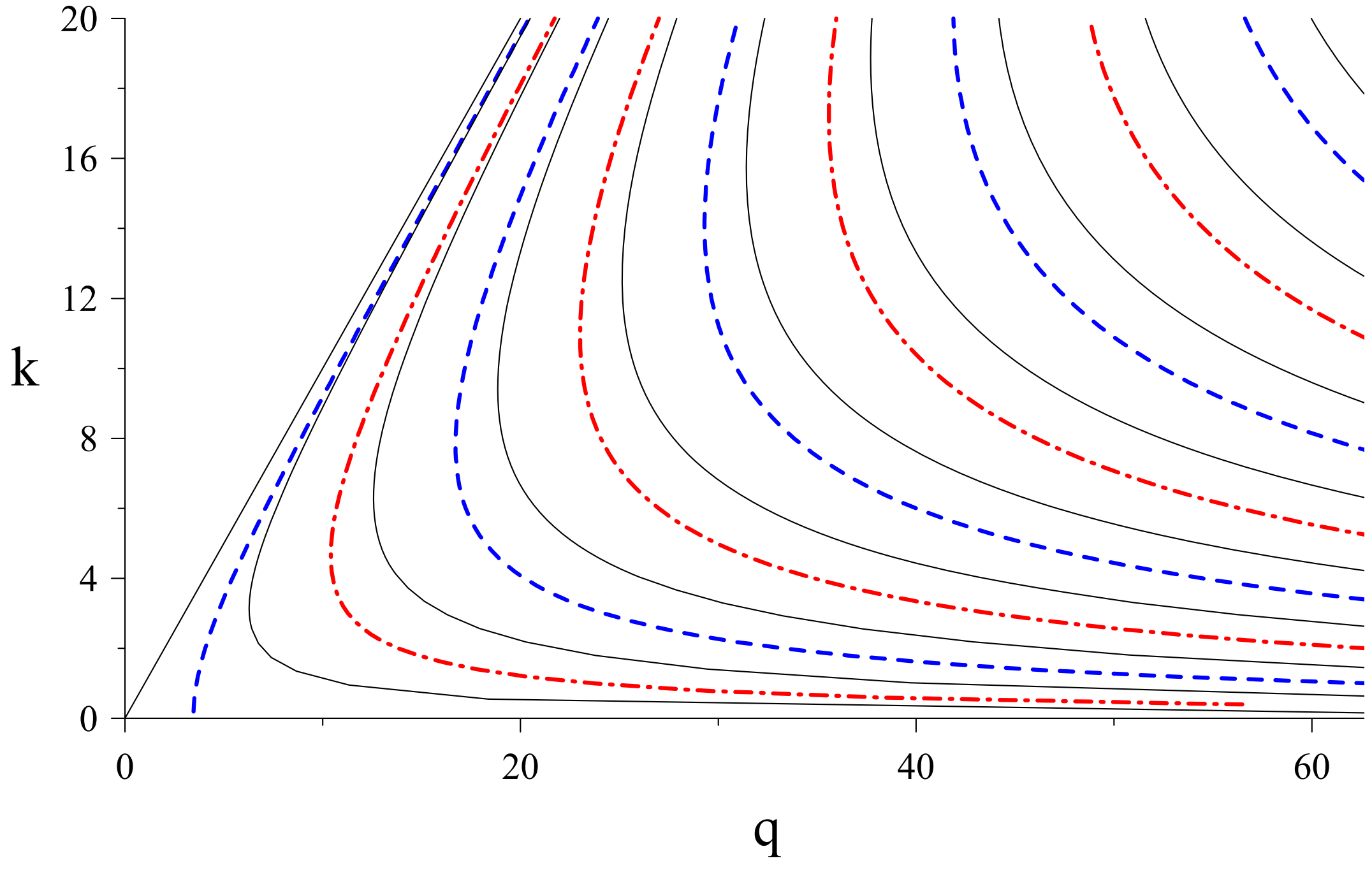

Рис. 1.3: *Статическое поле тяжести.* Уровни монотонной неустойчивости системы по отношению к возмущениям с конечной длиной волны $2\pi k^{-1}$. Сплошные линии – границы интервалов, определяемых (1.37),(1.38), штриховые – границы устойчивости по отношению к симметричным возмущениям, штрихпунктирные – по отношению к антисимметричным.

**Колебательные возмущения.** При $\lambda = i\omega$ система

$$\partial_z^2 w = k^2 w - k^2 Rc\,\phi - k^2 (Ra - Rc)\theta, \tag{1.39}$$

$$\partial_z^2 \phi = S^{-1} w + \left(k^2 + i\omega S^{-1}\right)\phi - i\omega S^{-1}\theta, \tag{1.40}$$

$$\partial_z^2 \theta = -w + \left(k^2 + i\omega\right)\theta, \tag{1.41}$$

аналогичная (1.28),(1.29), имеет бикубическое характеристическое уравнение общего вида, аналитическое решение которого является довольно сложной задачей. Поэтому для нахождения границы колебательной неустойчивости используется следующая стандартная процедура. Численно строится соответствующее системе (1.39)–(1.41) линейное отображение $\mathbf{A}$:

$$\begin{bmatrix} w \\ \partial_z w \\ \phi \\ \partial_z \phi \\ \theta \\ \partial_z \theta \end{bmatrix} (z=1) = \mathbf{A} \begin{bmatrix} w \\ \partial_z w \\ \phi \\ \partial_z \phi \\ \theta \\ \partial_z \theta \end{bmatrix} (z=0),$$

и находятся значения параметров, при которых могут быть удовлетворены принятые в задаче граничные условия, что математически выражается в требовании

$$\begin{Vmatrix} A_{12} & A_{13} & A_{15} \\ A_{42} & A_{42} & A_{42} \\ A_{62} & A_{62} & A_{62} \end{Vmatrix} = 0.$$

Для проверки гипотезы о том, что самыми опасными колебательными возмущениями являются длинноволновые, достаточно найти границу устойчивости для самых опасных при фиксированном значении волнового числа возмущений. Очевидно, что для колебательной неустойчивости (как и для монотонной) существует бесконечное множество уровней, но старшие уровни могут представлять лишь академический интерес. Результаты численного счета подтверждают актуальность длинноволнового приближения. Примеры полученных границ устойчивости представлены на Рис. 1.4.

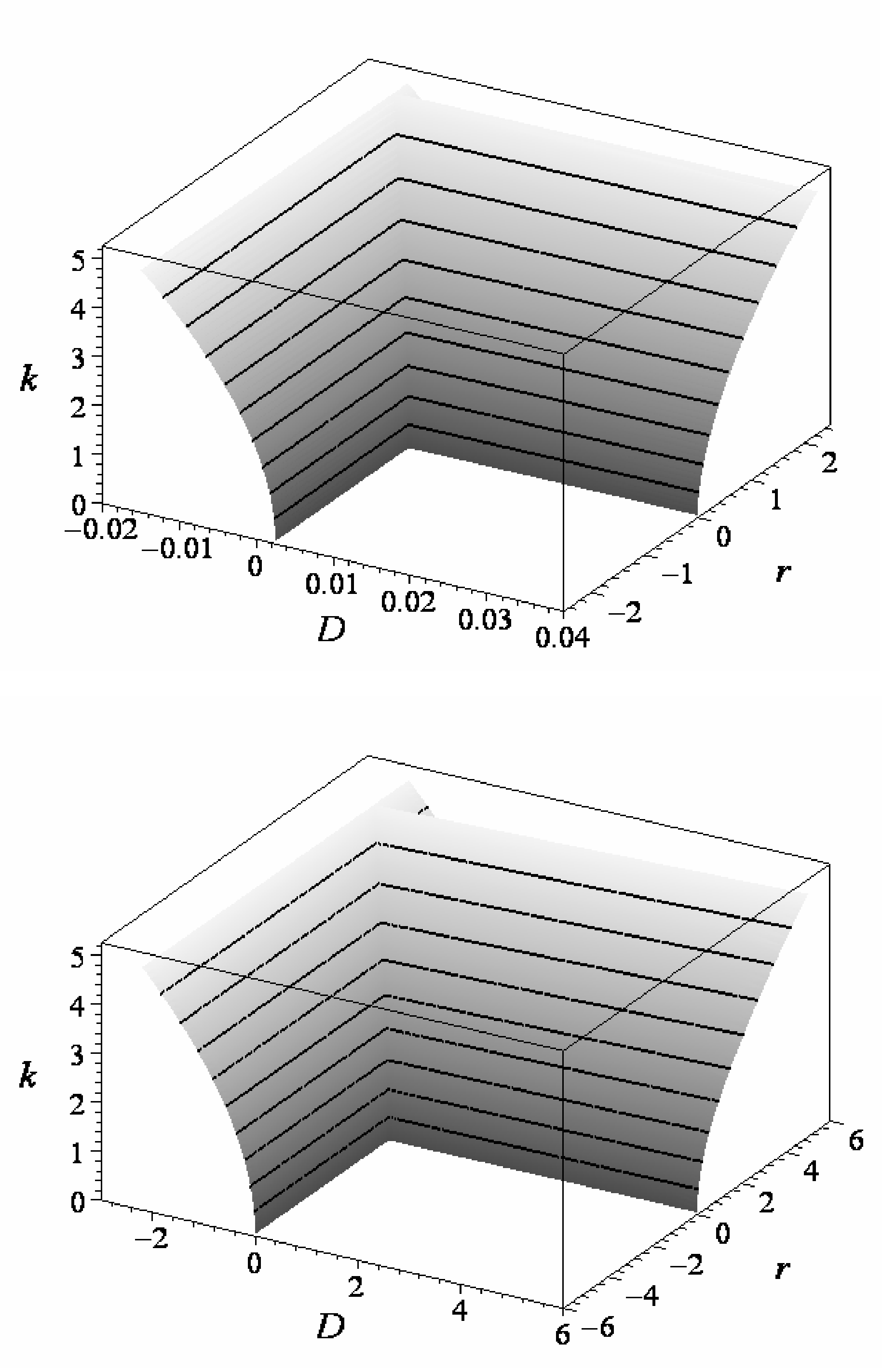

$S = 0.01$

$S = 1$

Рис. 1.4: *Статическое поле тяжести.* Уровни неустойчивости системы по отношению к возмущениям с конечной длиной волны $2\pi k^{-1}$. Правая поверхность – граница колебательной неустойчивости, левая – монотонной. Видно, что самыми опасными являются длинноволновые возмущения.

## 1.4. Параметрическое возбуждение конвекции при низкочастотной модуляции поля тяжести

В данном разделе рассматривается резонансное параметрическое возбуждение длинноволновых течений. Поскольку в исходных единицах измерения времени эволюция длинноволновых возмущений происходит медленно, резонансные для этих возмущений частоты малы (в исходных единицах измерения). Таким образом, здесь низкочастотность параметрического возбуждения означает длинноволновость возбуждаемых течений.

### 1.4.1. Дискретный спектр волновых чисел (протяженная ограниченная область)

Уточнение относительно дискретности спектра волновых чисел необходимо в связи с тем, что в длинноволновом приближении все возмущения возбуждаются одновременно и от длины волны зависит лишь характерный временной масштаб их эволюции. Вследствие чего, отдельные возмущения с фиксированной длинной волны можно рассматривать лишь при дискретном спектре. При упоминании ограничения горизонтальных размеров области будет иметься в виду, что слой принимает форму тонкого прямоугольного параллелепипеда с теплоизолированными и непроницаемыми боковыми границами. Когда будет учитываться лишь одно горизонтальное направление, будет неявно подразумеваться, что размеры области во втором горизонтальном направлении много меньше, чем в учитываемом.

Как уже было отмечено выше, при выводе уравнения (1.18) для эволюции малых возмущений числа Релея не выносились из-под оператора дифференцирования по времени, и потому это уравнение остается справедливым при переменном поле тяжести. Выполнив подстановку $Ra \to Ra(1 + a\cos\Omega t)$, $Rc \to Rc(1 + a\cos\Omega t)$ и изменив масштаб времени в $\Omega/2$ раз, можно получить

$$\ddot{\varphi}+\left(-\frac{2r}{\Omega}-\varepsilon\cos 2t\right)\dot{\varphi}+\left(\frac{4D}{\Omega^2}-\varepsilon\beta\cos 2t\right)\varphi=0, \qquad (1.42)$$

где $\varepsilon\equiv\dfrac{a(Ra-Rc)}{6\Omega}=\dfrac{2a}{\Omega}(r+1+S)$ и

$$\varepsilon\beta\equiv\frac{a\left(S\,Ra-S\,Rc-Rc\right)}{3\Omega^2}=\frac{4a}{\Omega^2}(S-D).$$

Рассмотрим систему (1.42). Ее фазовый поток на плоскости $(\varphi,\dot{\varphi})$ имеет дивергенцию

$$\nabla\cdot\vec{F}=\frac{2r}{\Omega}+\varepsilon\cos 2t. \qquad (1.43)$$

Произведение мультипликаторов, отвечающих функциям Флоке, определяет изменение элемента фазового объема, занимаемого некоторым множеством близких состояний системы, за период модуляции:

$$\Lambda_1\Lambda_2=\exp\left(\int_0^{\pi}\nabla\cdot\vec{F}\,dt\right)=e^{\frac{2\pi r}{\Omega}}. \qquad (1.44)$$

Таким образом, если мультипликаторы комплексные, то $\Lambda_{1,2}=R\,e^{\pm i\kappa}$ и

$$\Lambda_1\Lambda_2=R^2=e^{\frac{2\pi r}{\Omega}}, \qquad (1.45)$$

а значит, модуляция влияет лишь на $\kappa$. Т.е. в областях, где мультипликаторы при наложении модуляции остаются комплексными, модуль этих мультипликаторов

$$|\Lambda_{1,2}|=\exp\left(\frac{\pi r}{\Omega}\right)=\exp\left(\frac{\pi}{\Omega}\left(\frac{Ra-Rc}{12}-1-S\right)\right) \qquad (1.46)$$

не зависит от амплитуды и частоты модуляции. Но от последних зависят границы этих областей. Кроме того, из (1.45) следует, что при $r > 0$ равновесие всегда неустойчиво.

Дальнейшее исследование уравнения (1.42) требует привлечения методов теории возмущений (для малых амплитуд модуляции) и численного счета (для конечных амплитуд модуляции).

**Малые амплитуды вибраций.** Рассмотрим сначала поправки к границе неустойчивости, связанной с колебательной неустойчивостью немодулированной системы (в актуальной здесь области параметров $|r| \ll 1$). Комплексным мультипликаторам на границе устойчивости соответствуют решения со спектром $F(\omega) = \sum_{n,m} a_{n\,m} \delta\left(m + n\omega_0\right)$, где $n = \pm 1$. Нерезонансные поправки дают для этих решений такой сдвиг частот, что данное значение $\omega_0$ наблюдается (на границе устойчивости, т.е. при $r = 0$) при

$$D \equiv \frac{\Omega^2 \omega_0^2}{4} + \delta = \frac{\Omega^2 \omega_0^2}{4} + \frac{\Omega^2 \varepsilon^2}{32} \frac{\omega_0^2 + \beta^2}{\omega_0^2 - 1} + O\left(\varepsilon^4\right). \tag{1.47}$$

Благодаря этому же сдвигу резонансные мешки смещены вдоль границы колебательной неустойчивости исходной системы. Для первого резонанса это выражение, однако, несправедливо в силу своей расходимости. Но нерезонансные поправки получаются из более высоких порядков разложения уравнения (1.42), чем первый резонанс и, потому, для него не существенны.

Нижеприведенные уравнения задают в пространстве параметров множества, на которых один из мультипликаторов в связи с резонансом указанного порядка равен $\pm 1$.

$\omega_0 = 1$: $\Lambda = -1$,

$$\left(\delta + O_\delta\left(\varepsilon^2\right)\right)^2 + \left(\frac{2r}{\Omega} + O_r\left(\varepsilon^2\right)\right)^2 = \frac{\varepsilon^2\left(1+\beta^2\right)}{4};$$

$\omega_0 = 2$: $\Lambda = +1$,

$$\left(\delta - \frac{\varepsilon^2\left(4+\beta^2\right)}{24} + O_\delta\left(\varepsilon^3\right)\right)^2 + \left(\frac{4r}{\Omega} + O_r\left(\varepsilon^3\right)\right)^2 = \frac{\varepsilon^4\beta^2\left(4+\beta^2\right)}{256};$$

$\omega_0 = 3$: $\Lambda = -1$,

$$\left(\delta - \frac{\varepsilon^2\left(9+\beta^2\right)}{64} + O_\delta\left(\varepsilon^4\right)\right)^2 + \left(\frac{9r}{\Omega} + O_r\left(\varepsilon^4\right)\right)^2 = \frac{\varepsilon^6\left(1+\beta^2\right)^2\left(9+\beta^2\right)}{262144}.$$

На части множества, лежащей вне области неустойчивости немодулированной системы, через $1$ проходит модуль большего по абсолютному значению мультипликатора.

Теперь интерес представляют поправки к границе монотонной неустойчивости немодулированной системы. В окрестности этой границы $r < 0$ и немало, а $|D| \ll 1$. С учетом этого можно получить, что граница неустойчивости системы по отношению к периодическим движениям с ненулевым средним при данном $r$ смещается на

$$D = \frac{\Omega^4}{32\left(r^2+\Omega^2\right)}\varepsilon^2\beta\left(-\frac{2r}{\Omega} - \beta\right) + O\left(\varepsilon^4\right) = \\ = -\frac{a^2 S}{2}\frac{Rc\left(Rc - 12S^2 + 12S\right)}{\left(Rc - 12S^2\right)^2 + 144S^2\Omega^2} + O\left(a^4\right) \qquad (1.48)$$

(напомним, что без модуляции границе отвечает $D = 0$). Таким образом, вибрации могут не только стабилизировать монотонный уровень, но и дестабилизируют его при

$$Rc \in \begin{cases} \left(-12S(1-S),\, 0\right), & \text{ï ðè} \quad S < 1; \\ \left(0,\, -12S(1-S)\right), & \text{ï ðè} \quad S > 1. \end{cases} \tag{1.49}$$

Точка пересечения границ устойчивости по отношению к периодическим движениям с ненулевым средним и квазипериодическим движениям специального рассмотрения не требует, поскольку модули комплексных мультипликаторов модуляции "не чувствуют", а для поправок к "бывшему монотонному" уровню справедлив предельный переход $r \to 0$. Т.е. границы устойчивости вблизи описанной точки – это кривые:

$$D = -\frac{\left(\varepsilon\beta\Omega\right)^2}{2} + O\left(\varepsilon^4, r\varepsilon^2\right), \ \ r < 0 \quad \text{для} \quad \Lambda = 1; \tag{1.50}$$

$$r = 0, D > -\frac{\left(\varepsilon\beta\Omega\right)^2}{2} + O\left(\varepsilon^4\right) \ \ \text{для} \ \ \left|\Lambda\right| = 1, \arg\Lambda \neq 0. \tag{1.51}$$

**Конечные амплитуды вибраций.** Как и было сказано выше, рассмотрение случая конечных амплитуд вибраций требует численного счета. На Рис. 1.5 представлены примеры результатов такого счета при значениях $S$, характерных для жидкостей ($S$ ~ $0.01$) и газов ($S$ ~ $1$).

Поскольку волновое число на самом деле не фиксировано, а есть некоторый набор волновых чисел (для ограниченного слоя – дискретный), нам следует вспомнить, что для нормального возмущения с волновым числом $k$ вводилась замена времени $k^2 t \to t$. Следовательно

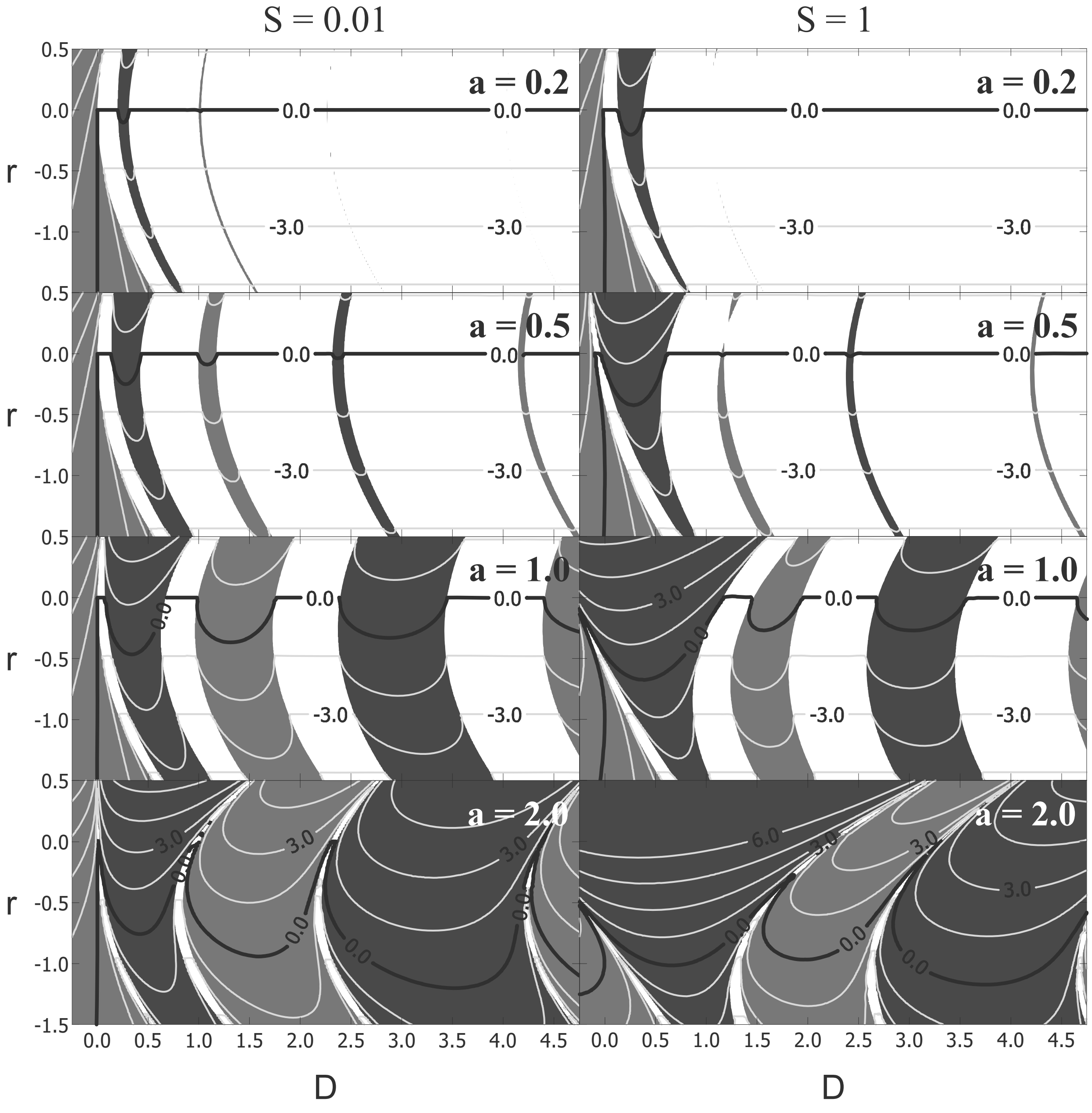

Рис. 1.5: *Модуляция поля тяжести.* На графике отложен модуль большего по абсолютной величине мультипликатора для нормального возмущения с большой длиной волны при $\Omega = 1$. Жирная черная линия – граница устойчивости по отношению к данному возмущению; в темно-серых областях мультипликаторы отрицательные, в светло-серых – положительные, в белых – комплексные.

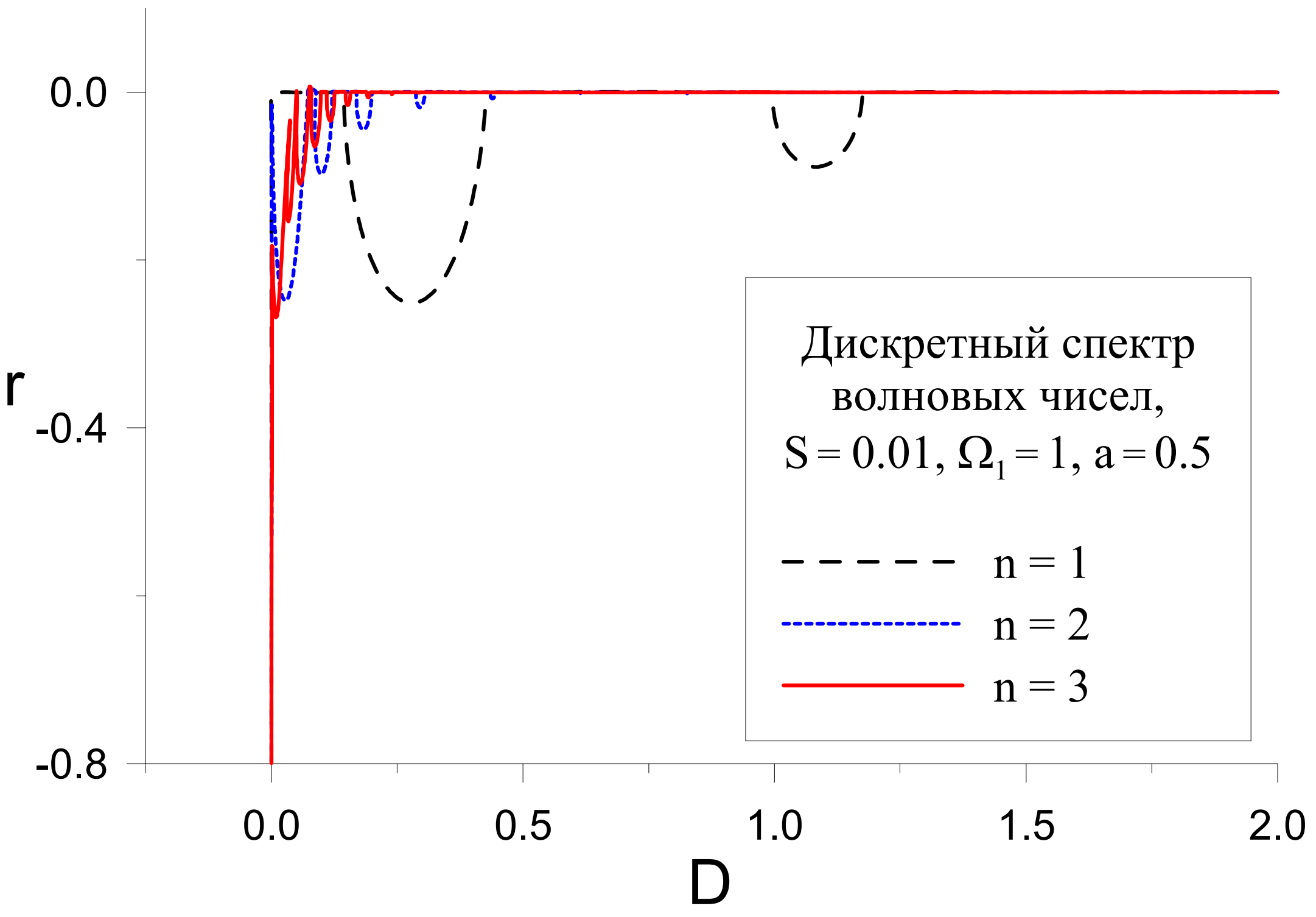

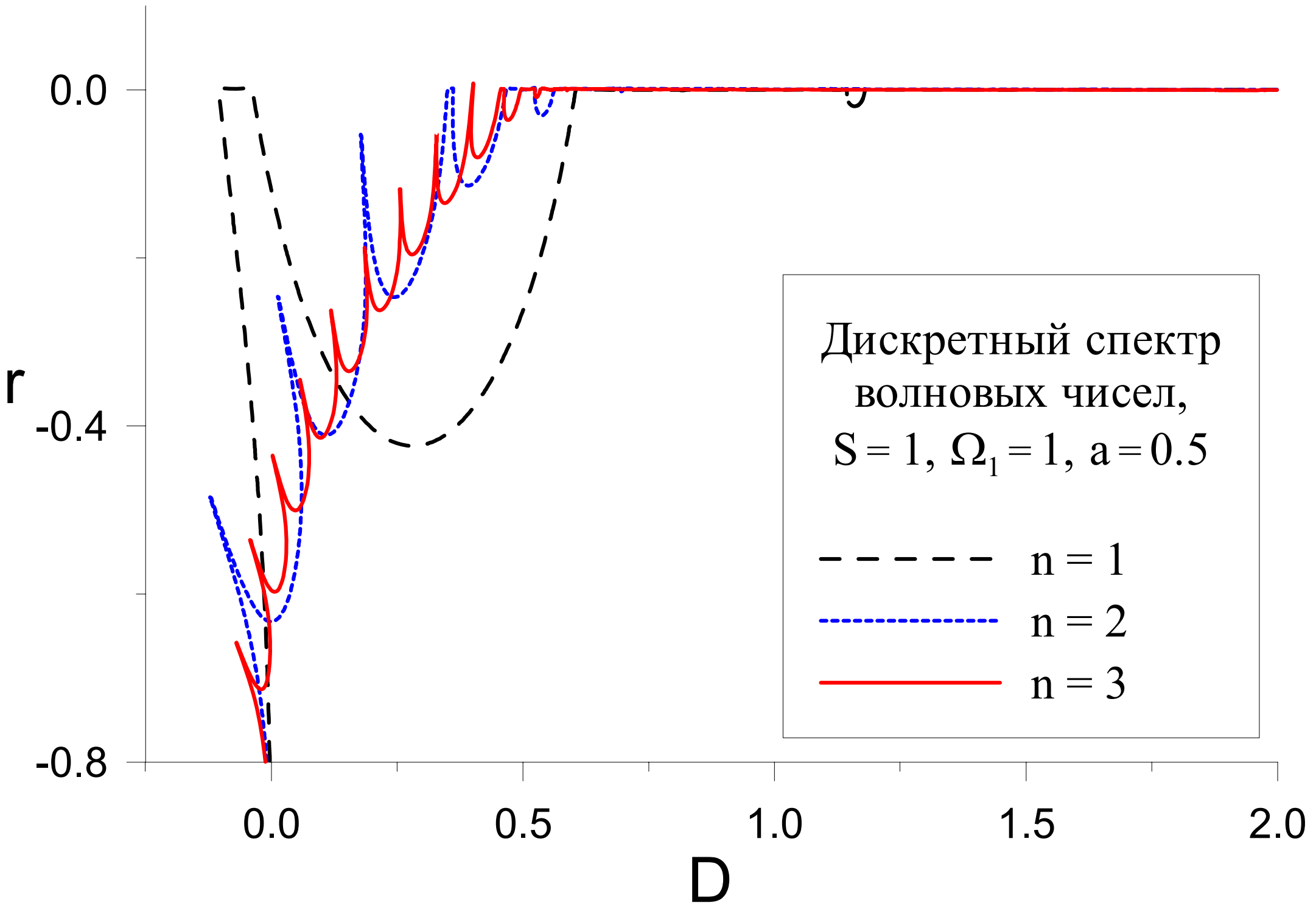

Рис. 1.6: *Модуляция поля тяжести.* Границы устойчивости системы по отношению к нескольким первым гармоникам в случае дискретного спектра волновых чисел. Соответствующие значения параметров указаны на графиках. В нижнем правом углу стратифицированное состояние устойчиво.

$$\Omega = k^{-2}\,\Omega_0, \tag{1.52}$$

где $\Omega_0$ – частота модуляции поля тяжести в тепловых единицах, а $\Omega$ – частота, входящая в уравнение (1.42) для эволюции возмущения с длиной волны $k$. Таким образом, область неустойчивости исходной гидродинамической системы будет представлять собой объединение областей неустойчивости динамической системы (1.42), соответствующих дискретному набору значений $\Omega$ из последовательности

$$\left\{\Omega_n = \frac{\Omega_0}{k_{x\min}^2 n_x^2 + k_{y\min}^2 n_y^2};\, n_x,\, n_y \in Z\right\}. \tag{1.53}$$

Одновременное обращение $n_x$ и $n_y$ в ноль означает однородное состояние. На Рис. 1.6 представлены примеры такого объединения областей неустойчивости для конечной амплитуды вибраций при $n_y = 0$ (горизонтальные размеры вдоль $y$ малы по сравнению с горизонтальными размерами вдоль $x$).

### 1.4.2. Непрерывный спектр волновых чисел (бесконечный слой)

Для бесконечной области последовательность (1.53) переходит в непрерывное множество, а граница устойчивости гидродинамической системы будет представлять собой огибающую к однопараметрическому множеству границ устойчивости динамической системы (1.42), отвечающих разным значениям $\Omega$. В точке касания огибающей и границы устойчивости, отвечающей заданному значению $\Omega$, система нейтрально устойчива по отношению к возмущению с волновым числом $k = \sqrt{\Omega_0 / \Omega}$.

**Поправки к колебательному уровню при малых амплитудах вибраций.** При малых амплитудах модуляции искомая огибающая может быть получена аналитически. В этом случае граница области дестабилизации колебатель-ного уровня будет определяться лишь первым резонансом[1]. При малой модуляции

$$\overline{\left(\frac{\partial r}{\partial \delta}\right)}_{a,S,\Omega} \gg \left(\frac{\partial \varepsilon}{\partial \Omega}\right)_{a,S,D=\frac{\Omega^2}{4},r=0},$$

где черта сверху обозначает характерное значение данной величины. Это позволяет получить верное в ведущем порядке выражение для огибающей, лишь максимизируя $|r|$ для фиксированного $\Omega$. Последнее достигается при $\delta = 0$. Тогда граница устойчивости задается в параметрическом виде системой:

$$D = \frac{\Omega^2}{4}, \quad r = -\frac{\Omega}{4}\sqrt{\varepsilon^2 + \varepsilon^2\beta^2}. \tag{1.54}$$

Параметром здесь выступает $\Omega \in \left(0, \infty\right)$. Исключая из этой системы частоту и помня о малости $r$, можно получить явное выражение для огибающей вблизи границы колебательной неустойчивости немодулированной системы:

$$r_{cr} = -\frac{a}{2}\sqrt{1 + S^2 + D + \frac{S^2}{D}} + O\left(a^2\right). \tag{1.55}$$

Ввиду важности последнего выражения, имеет смысл привести его и в терминах чисел Релея:

$$Ra_{cr} = Rc + 12 + 12S - 6a\sqrt{1 + \frac{Rc}{12} + \frac{12S^2}{Rc - 12S^2}} + O\left(a^2\right). \tag{1.56}$$

---

[1] В принципе, возможна ситуация, когда резонанс первого порядка может пропадать при некоторых значениях параметров, и тогда наше утверждение будет неверно, но, как показано ниже, поправки к границе устойчивости, связанные с первым резонансом, всегда $\sim a$, и для рассматриваемой системы верхнее утверждение справедливо всегда.

Нейтрально устойчивые на этой огибающей возмущения имеют мультипликатор $\Lambda = -1$ и волновой вектор

$$k_{cr} = \sqrt{\frac{\Omega_0}{2}} D^{-\frac{1}{4}} \left(1 + O(a)\right) = \sqrt{\frac{\Omega_0}{2}} \left(\frac{Rc}{12} - S^2\right)^{-\frac{1}{4}} \left(1 + O(a)\right). \qquad (1.57)$$

Следует отметить, что выражение (1.55) (а, следовательно, и (1.56)) расходятся вблизи границы монотонной неустойчивости исходной системы. Впрочем, вблизи точки этой расходимости волновое число самых опасных возмущений становится большим и предположение о длинноволновости нарушается.

**Поправки к монотонному уровню при малых амплитудах вибраций.** Для нахождения искомой огибающей вблизи монотонного уровня нужно максимизировать отклонение (1.48) границы неустойчивости системы по отношению к периодическим движениям с ненулевым средним по времени вглубь области устойчивости немодулированной системы, варьируя $\Omega$. Это отклонение экстремально при $\Omega = 0$, когда $|D|$ максимален, и при $\Omega = \infty$, когда $|D| = 0$. В результате чего формально получается, что

$$D_{cr} = \frac{a^2 S}{2} \frac{Rc\left(-Rc + 12S^2 - 12S\right)}{\left(Rc - 12S^2\right)^2} + O\left(a^4\right),$$
$$k_{cr} = \infty, \quad \text{ï ðè} \quad Rc\left(Rc - 12S^2 + 12S\right) < 0, \qquad (1.58)$$

$$D_{cr} = 0, \;\; k_{cr} = 0, \quad \text{ï ðè} \quad Rc\left(Rc - 12S^2 + 12S\right) > 0. \qquad (1.59)$$

В первом случае наиболее опасны коротковолновые возмущения, для которых наше длинноволновое приближение имеет ограниченную ценность: выражение для границы, скорее всего, при малых амплитудах вибраций остается справедливым, но критическое волновое число – конечно. Во втором случае наиболее опасны однородные возмущения. В терминах чисел Релея выражения (1.58)–(1.59) принимают вид

$$
\begin{aligned}
Ra_{cr} &= 12 + \left(1 + S^{-1}\right) Rc \\
&+ \begin{cases} 6a^2 \dfrac{Rc\left(Rc - 12S^2 + 12S\right)}{\left(Rc - 12S^2\right)^2}, & Rc\left(Rc - 12S(S-1)\right) < 0; \\ 0, & Rc\left(Rc - 12S(S-1)\right) > 0. \end{cases}
\end{aligned}
\qquad (1.60)
$$

На этой границе критические возмущения имеют мультипликатор $\Lambda = +1$.

При **конечных амплитудах вибраций** в окрестности границы устойчивости системы по отношению к колебаниям с ненулевым средним (по времени) расходимостей не возникает: граница неустойчивости по отношению к колебаниям с нулевым средним (по времени) ведет себя гладко, а критические волновые числа малы. Характерные примеры огибающих для жидкостей и газов представлены на Рис. 1.7.

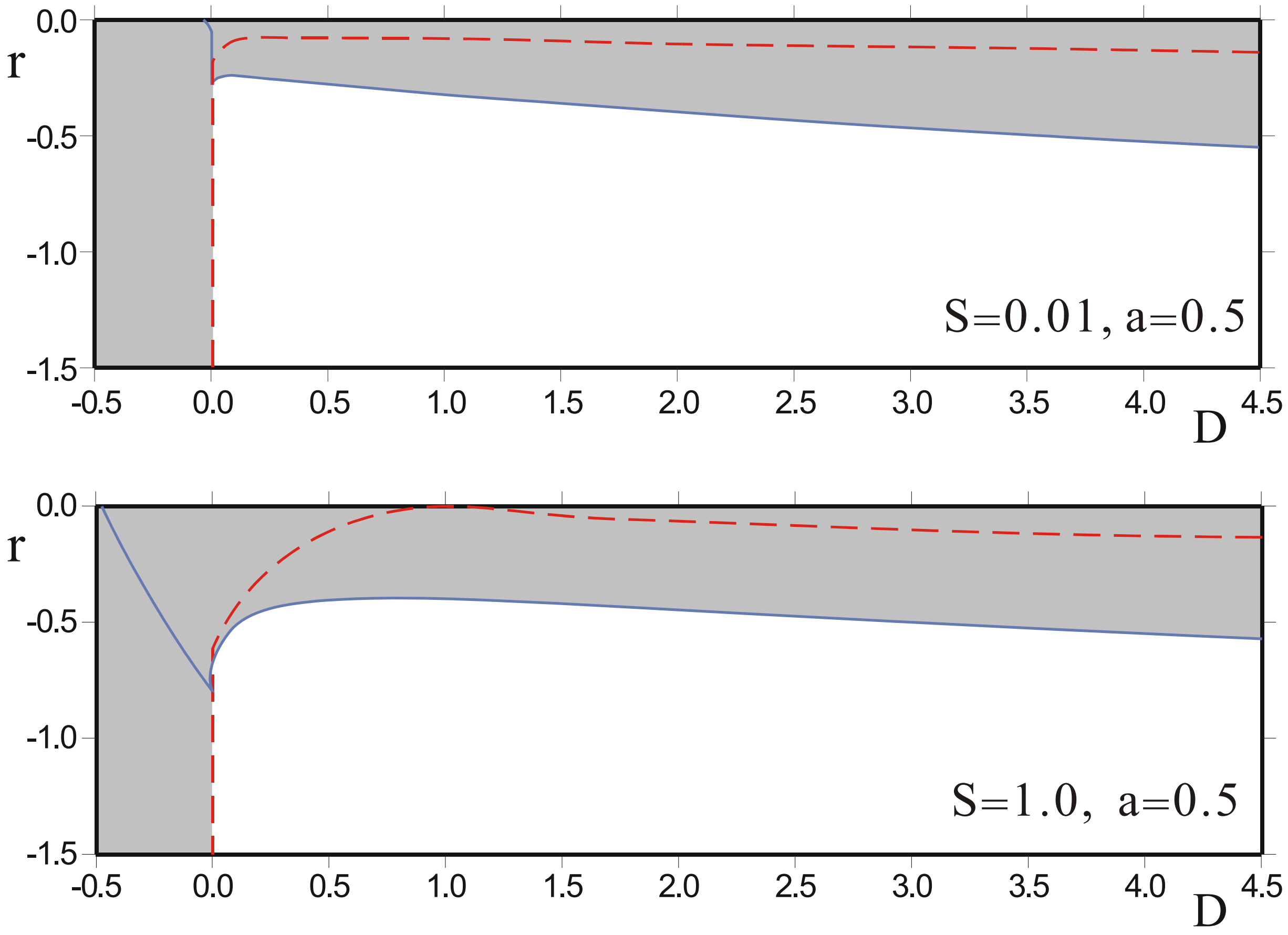

Рис. 1.7: *Модуляция поля тяжести.* Границы устойчивости системы в случае непрерывного спектра волновых чисел. В пределах серой области система неустойчива; штриховая линия ограничивает область параметров, в которой могут возбуждаться возмущения с положительным мультипликатором; сплошная – с отрицательным.

# Глава 2: Термоконцентрационная конвекция в слое пористой среды от источников тепла или примеси

В данной главе, как и в предыдущей, исследуется термоконцентрационная конвекция бинарной смеси в тонком слое пористой среды, но теперь основной интерес уделяется поведению системы при наличии внутреннего источника примеси или тепла.

Конкретно, рассматриваются режимы длинноволновой стационарной конвекции от локализованного источника/стока тепла или примеси. За исключением окрестности источника, для описания системы используются те же граничные условии и уравнения, что и в предыдущей главе, а, следовательно, и те же уравнения эволюции длинноволновых возмущений. Поскольку оказывается, что решения определяются локальными характеристиками потоков (тепла и примеси), то структура источника и то, как он организован, не влияют на течение в конкретной точке пространства: это течение полностью определяется интенсивностью источника. При этом нужно иметь в виду, что получаемые решения справедливы лишь начиная с некоторого конечного удаления от источника, которое при слабом источнике может быть малым – порядка толщины слоя.

Показывается, что в случаях, когда стационарное течение устойчиво, области длинноволнового течения могут быть разделены одним или несколькими кольцами переходного течения (в некотором смысле это аналогично гидравлическому скачку [55]). Кроме того, оказывается, что в случае локализованного источника примеси, ее вынос из непосредственной окрестности источника осуществляется преимущественно конвективным образом. В то же время, для локализованного источника тепла возможны два типа режимов: с преимущественно конвективным и преимущественно диффузионным механизмами выноса тепла из окрестности источника. В некоторых случаях между этими режимами с разными механизмами выноса тепла возможна мультистабильность. Результаты данной части диссертации опубликованы в работе [97].

## 2.1. Стационарные конвективные течения в тонком слое

Длинноволновая термоконцентрационная конвекция всюду, кроме окрестности источника, описывается нелинейными уравнениями (1.12),(1.13), полученными в начале главы 1. Однако при исследовании нелинейных режимов конвекции удобнее переписать эти уравнения в следующем виде:

$$\dot{\vartheta} - \frac{6}{5}\operatorname{div}\big(\nabla P\left(\nabla\vartheta \cdot \nabla P\right)\big) = \Delta\vartheta - \Delta P, \tag{2.1}$$

$$\sigma\dot{P} + \sigma F\,\dot{\vartheta} - \frac{6}{5}\operatorname{div}\Big(\ \nabla P\left(\nabla P\right)^2\Big) = A_1\Delta\vartheta + A_2\Delta P, \tag{2.2}$$

где $\sigma = b^{-2}S$, $F = \dfrac{Ra}{12}\left[\left(\dfrac{b^2}{S} - 1\right)(1-N) + \dfrac{Nb^3}{Sb - S^2(b-1)}\right]$,

$A_1 = \left[\dfrac{Ra}{12}\big(1 + S(N-1)\big) + F\right]\dfrac{S}{b^2}$ и $A_2 = \left[S + \dfrac{Ra}{12}(N-1) - F\right]\dfrac{S}{b^2}$. Следует иметь в виду, что длинноволновые уравнения рассматриваются на плоскости, т.е. все операции дифференцирования производятся лишь по горизонтальным координатам.

Также будут полезны те результаты предыдущей главы, что при

$$Ra > \left(1 + S^{-1}\right) Rc + 12 \tag{2.3}$$

или, что то же самое, $A_1 + A_2 < 0$ в системе возникает монотонная линейная неустойчивость состояния механического равновесия (здесь и далее, когда речь будет идти об устойчивости состояния механического равновесия, будет подразумеваться отсутствие источников тепла и примеси, в противном случае, это состояние просто не существует), а при

$$Rc + 12 + 12S < Ra < \left(1 + S^{-1}\right) Rc + 12 \qquad (2.4)$$

– колебательная.

Система уравнений (2.1)–(2.2) имеет дивергентный вид и в стационарном случае может быть однократно проинтегрирована:

$$\nabla\vartheta - \nabla P + \frac{6}{5}\nabla P(\nabla\vartheta \cdot \nabla P) = -\vec{k}_1, \qquad (2.5)$$

$$A_1\nabla\vartheta + A_2\nabla P + \frac{6}{5}\nabla P(\nabla P)^2 = -\vec{k}_2, \qquad (2.6)$$

где $\vec{k}_1 = \vec{q}_T$ – тепловой поток,

$$\vec{k}_2 = \gamma\vec{k}_1 - \frac{RaNS\,\vec{q}_C}{12b^2\left[b - S(b-1)\right]},$$

$$\gamma = \frac{S}{b^2}\left[F + \frac{Ra}{12}\left(1 - N + \frac{Nb}{b - S(b-1)}\right)\right],$$

$\vec{q}_C$ – поток примеси; причем $\operatorname{div}\vec{k}_i = 0$. Приведенные выше соотношения между $\vec{k}_i$ и реальными потоками тепла $\vec{q}_T$ и примеси $\vec{q}_C$ в исходной системе восстановлены на основании тех фактов, что в нестационарном случае $\dot{\vartheta} = -\operatorname{div}\vec{k}_1$ (ср. (2.1)), $\sigma\dot{P} + \sigma F\dot{\vartheta} = -\operatorname{div}\vec{k}_2$ (ср. (2.2)), и было выполнено преобразование возмущений химического потенциала (1.6).

Стационарные решения, соответствующие локализованному источнику тепла или примеси, рассматриваются, начиная с некоторого конечного, но малого удаления от источника, где справедливо приближение тонкого слоя. Чем меньше интенсивность источника, тем меньше область больших градиентов, где длинноволновое приближение несправедливо.

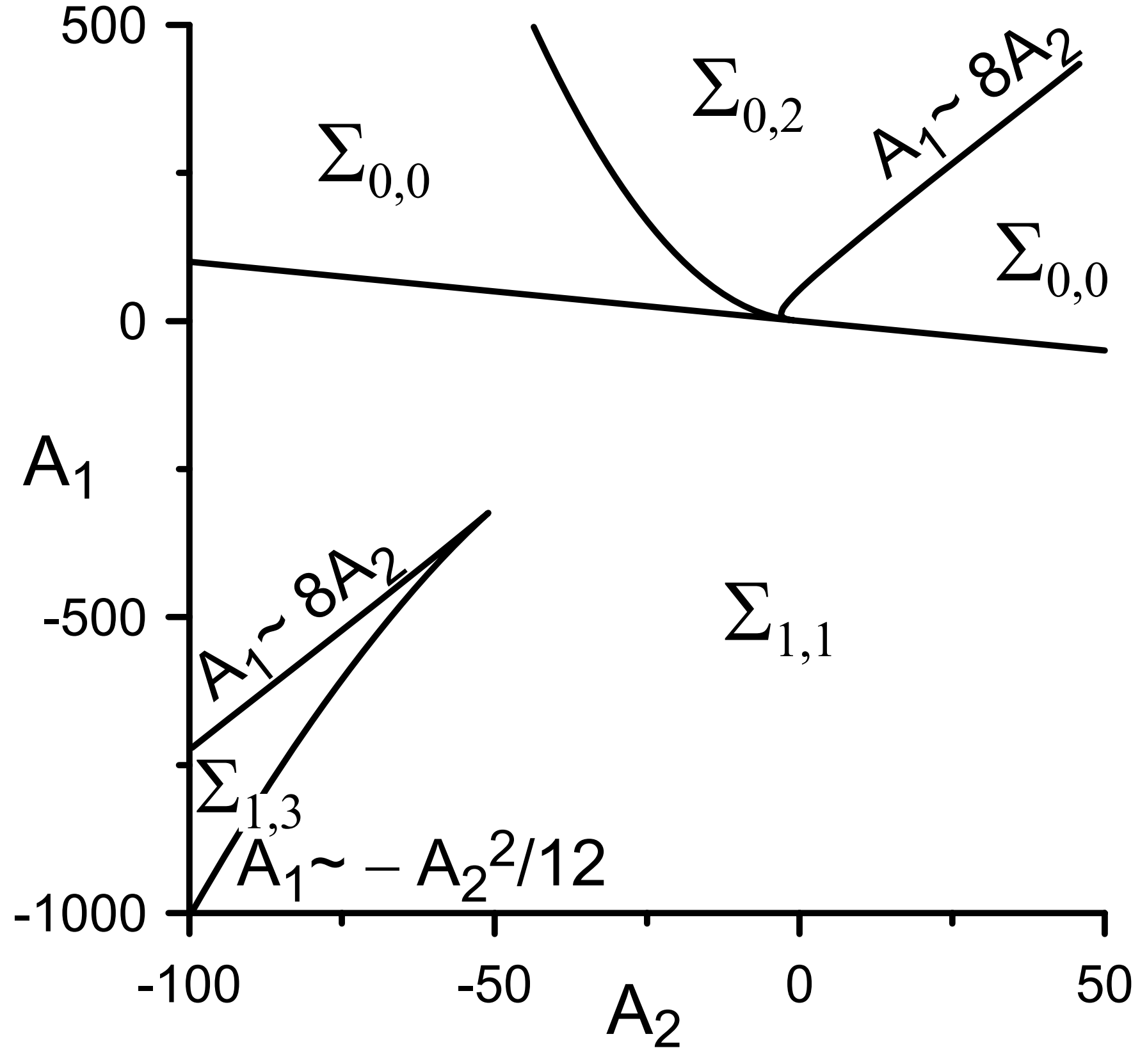

Рис. 2.1: Области различных решений при наличии источника примеси. Решения, возникающие в областях $\Sigma_{i,j}$ (при маркировке областей первый индекс определяет количество положительных решений системы (2.8) относительно $P'$ при $r \to \infty$, а второй – количество особенностей $P'' = \infty$ у формальных решений во всем диапазоне изменения $r$), представлены на Рис. 2.2.

## 2.2. Локализованный источник примеси

В случае источника примеси для осесимметричных решений

$$\vec{k}_1 = 0, \quad \vec{k}_2 = -\frac{RaNSQ_C}{24\pi b^2\left(b - S(b-1)\right)}\frac{\vec{e}_r}{r} \equiv -\frac{k}{r}\vec{e}_r, \tag{2.7}$$

где $Q_C$ – приток примеси в систему. Тогда система (2.5) принимает алгебраический вид относительно производных полей:

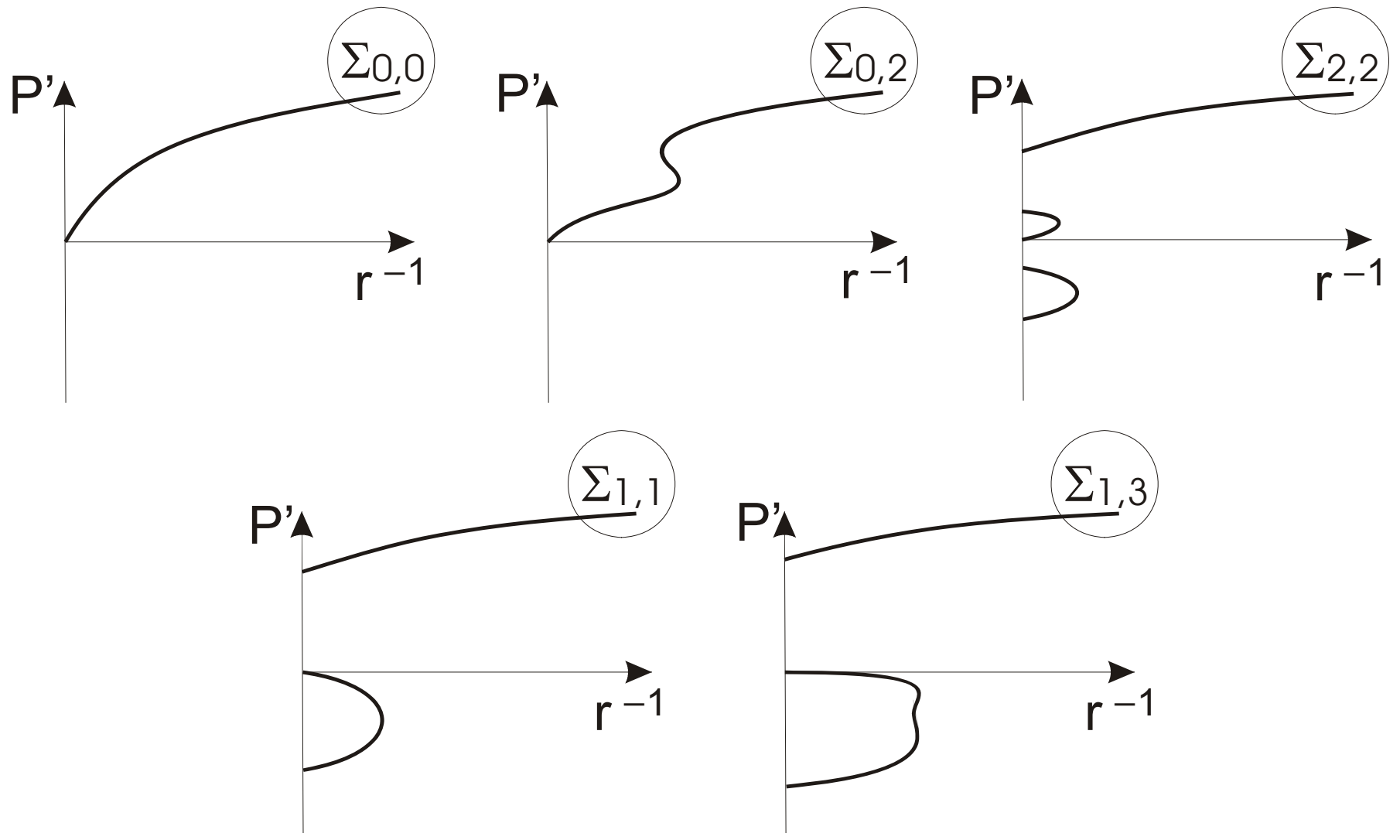

Рис. 2.2: Формальные решения системы (2.8), соответствующей источнику примеси.

$$\vartheta' = \frac{P'}{1+6/5(P')^2}, \quad A_1\vartheta' + A_2P' + \frac{6}{5}(P')^3 = \frac{k}{r}. \tag{2.8}$$

Здесь штрих означает производную по $r$.

На Рис. 2.1 представлена диаграмма формальных решений системы (2.8), сами решения для $k > 0$ представлены на Рис. 2.2. Хотя отрицательные значения $r$ физического смысла не имеют, для лучшего понимания того, как решения системы (2.8) ведут себя по мере изменения параметров, следует отметить, что эта система инвариантна относительно преобразования $(r, \vartheta', P') \to (-r, -\vartheta', -P')$. По этой же причине для перехода к случаю отрицательных $k$ достаточно поменять знак всех скалярных полей.

Верхняя граница области $\Sigma_{1,1}$ задается уравнением

$$A_1 + A_2 = 0,$$

граница $\Sigma_{0,2}$ и $\Sigma_{2,2}$ –

$$A_1 = \left(A_2 - 1\right)^2 \big/ 4, \quad A_2 < -1,$$

а параметрически задаваемая кривая

$$A_2 = -6a^2 - 9 - 24 \big/ \left(a^2 - 3\right),$$
$$A_1 = 6a^4\left(a^2 + 1\right) + \left(a^4 - 1\right)A_2$$

при $a < \sqrt{3}$ дает границу областей $\Sigma_{0,0}$ и $\Sigma_{0,2}$, а при $a > \sqrt{3}$ – $\Sigma_{1,1}$ и $\Sigma_{1,3}$. Асимптотическое поведение этой кривой при больших $A_i$ представлено на Рис. 2.1.

Примечательно также, что вблизи источника (здесь и далее удаление от источника остается достаточно большим, чтобы не была существенна геометрия источника, и было справедливо длинноволновое приближение)

$$\vartheta \approx \vartheta_0 + \frac{3}{4}\left(\frac{5}{6}\right)^{\frac{2}{3}} k^{-\frac{1}{3}} r^{\frac{4}{3}}, \quad P \approx P_0 + \frac{3}{2}\left(\frac{5}{6}\right)^{\frac{1}{3}} k^{\frac{1}{3}} r^{\frac{2}{3}},$$

что соответствует конвективному выносу примеси из окрестности источника (в уравнении переноса (2.2) нелинейное слагаемое, связанное с конвективной производной, доминирует над диффузионным линейным слагаемым). На бесконечности, в случае, если состояние механического равновесия устойчиво, производные полей должны быть равны нулю. Конкретно,

$$P \approx \vartheta \approx \frac{k \ln r}{A_1 + A_2}.$$

Примечательно, что в случае неоднозначной зависимости $P'(r)$ корректные асимптоты полей температуры и давления при $r \to 0$ и $r \to \infty$ при-надлежат разным веткам формальных решений системы (2.8), т.е. между этими ветками должен быть скачкообразный переход (скачок, с точки зрения длинноволновой теории, испытывают производные полей, но не сами поля), для которого длинноволновое приближение не справедливо. Если стационарное источниковое течение при этом устойчиво, имеет смысл ожидать, что переходные течения, связанные с разрывом, затухают по мере удаления от него и оказываются локализованы в узкой, с точки зрения длинноволнового приближения, области.

Получившийся разрыв связан с переходом между различными режимами переноса: при малых градиентах полей в уравнениях (2.5), (2.6) доминируют линейные слагаемые, описывающие диффузионный перенос примеси и тепла, а при больших – нелинейные, описывающие конвективный перенос. Заслуживает упоминания тот факт, что этот разрыв не является уникальным в своем роде: в гидродинамике широко известен другой пример – гидравлический скачок в радиально растекающемся потоке жидкости от падающей на горизонтальную поверхность струи (см., например, [55]).

Вопрос о точном положении "колец" переходного течения (очевидно, что они лежат где-то в области, где формальное решение неоднозначно) требует выхода за пределы длинноволнового приближения, довольно сложного численного анализа и выходит за рамки настоящей работы. Можно, однако, отметить, что внутренний и внешний радиусы границы области, где это кольцо может располагаться, соответствуют конкретным значениям $k/r$ (см. (2.8)), т.е. прямо пропорциональны интенсивности источника.

Строгий анализ устойчивости источниковых решений для бинарной смеси также остается за рамками настоящей работы, однако можно с уверенностью говорить о том, что в области параметров, где состояние механического равновесия неустойчиво, неустойчивы будут и источниковые решения (вдали от источника будут возбуждаться конвективные течения, которые будут проникать в область, близкую к источнику). Обратное же утверждение, в общем случае, не обязательно справедливо, поскольку при устойчивом состоянии механического равновесия могут нарастать возмущения, локализованные вблизи источника. Таким образом, граница устойчивости состояния механического равновесия дает лишь внешнее ограничение для области устойчивости источниковых решений.

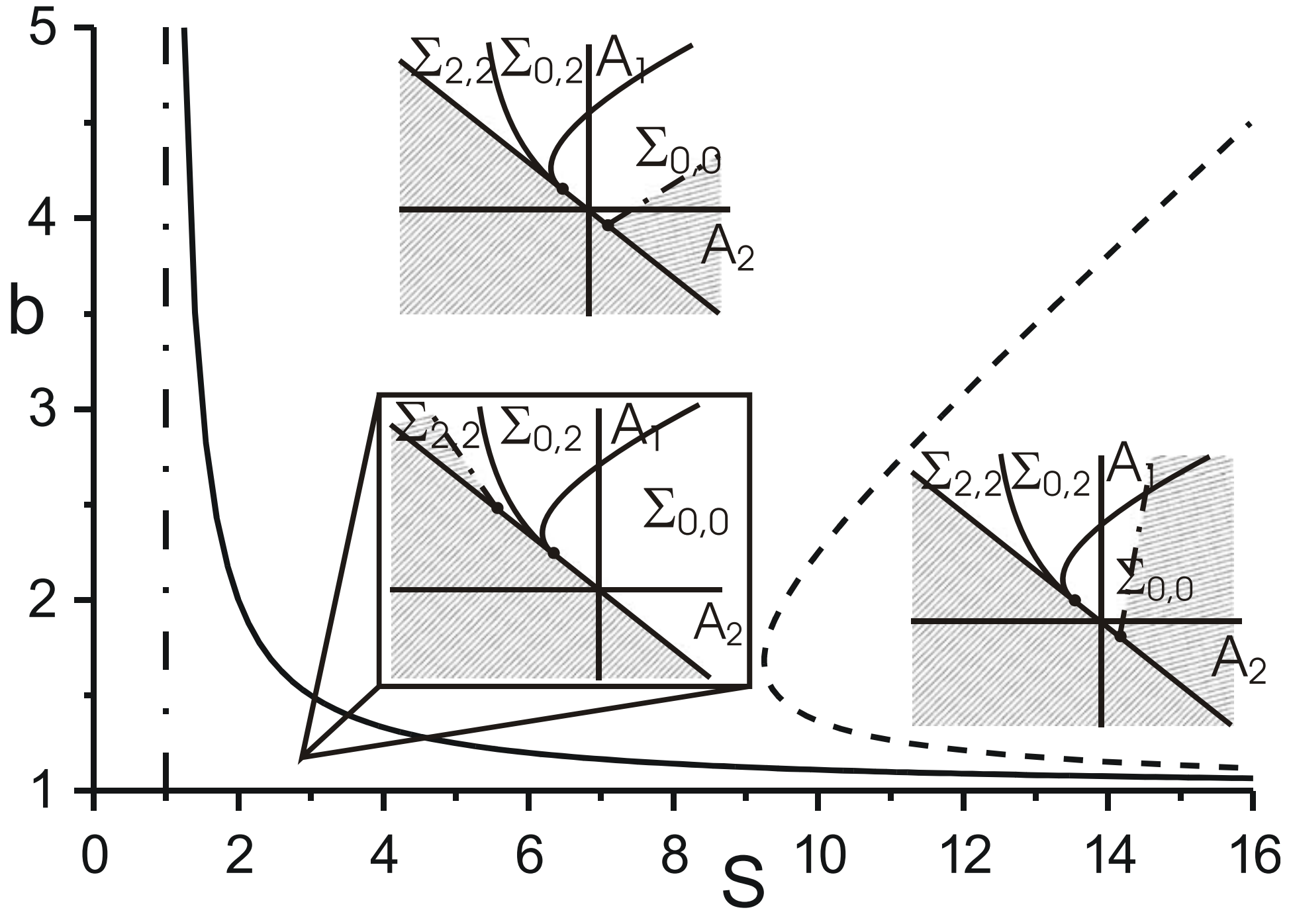

Рис. 2.3: Зависимость свойств устойчивости состояния механического равновесия от параметров $b$ и $S$.

Из сопоставления выражений для $A_i$ с выражениями для границ устойчивости состояния механического равновесия (2.3) и (2.4), следует, что совокупность областей $\Sigma_{1,i}$ соответствует области линейной монотонной неустойчивости состояния механического равновесия. Колебательная неустойчивость наблюдается в области, ограниченной двумя лучами:

$$-A_2 < A_1 < -b^{-3}(S+b)\big(b^2 - S(b-1)\big)A_2 + \\ +b^{-5}(S+b)\Big(b^2\big(S^2 - S - 1\big) - S^3(b-1)\Big). \tag{2.9}$$

При $b=1$ (теплоемкость скелета пористой среды пренебрежимо мала по сравнению с таковой для жидкости, ее насыщающей) формула (2.9) принимает вид $-A_2 < A_1 < -(S+1)(A_2+S)$; и дает область, которая принадлежит $\Sigma_{2,2}$ и касается области $\Sigma_{0,2}$ в точке $A_2 = -2S-1$, $A_1 = (S+1)^2$ (см. Рис. 2.3). По мере увеличения теплоемкости пористой среды относительно теплоемкости жидкости вплоть до $b = S/(S-1)$ (сплошная линия на Рис. 2.3) верхняя граница области колебательной неустойчивости состояния механического равновесия сдвигается вглубь области $\Sigma_{2,2}$. При этом устойчивыми могут быть режимы, в области $\Sigma_{0,0}$, $\Sigma_{0,2}$ и части области $\Sigma_{2,2}$, лежащей вне (2.9); в последних двух областях параметров не исключены устойчивые источниковые течения с кольцами переходного течения.

При $b = S/(S-1)$ область (2.9) исчезает и состояние механического равновесия в $\Sigma_{0,0}$, $\Sigma_{0,2}$, и $\Sigma_{2,2}$ устойчиво. При дальнейшем увеличении $b$ ($b > S/(S-1)$) область (2.9) снова возникает, но уже как часть области $\Sigma_{0,0}$ и растет по мере увеличения $b$. Теперь область колебательной неустойчивости состояния механического равновесия либо целиком лежит в области $\Sigma_{0,0}$, либо пересекается с $\Sigma_{0,2}$ при достаточно больших значениях $A_2$. Условие этого пересечения выражается неравенством

$$S > \frac{b}{b-1}\left(\frac{1}{2} + \sqrt{\frac{1}{4} + 9b(b-1)}\right)$$

(штриховая линия на Рис. 2.3) и является условием того, что в части области $\Sigma_{0,2}$ состояние механического равновесия становиться неустойчивым.

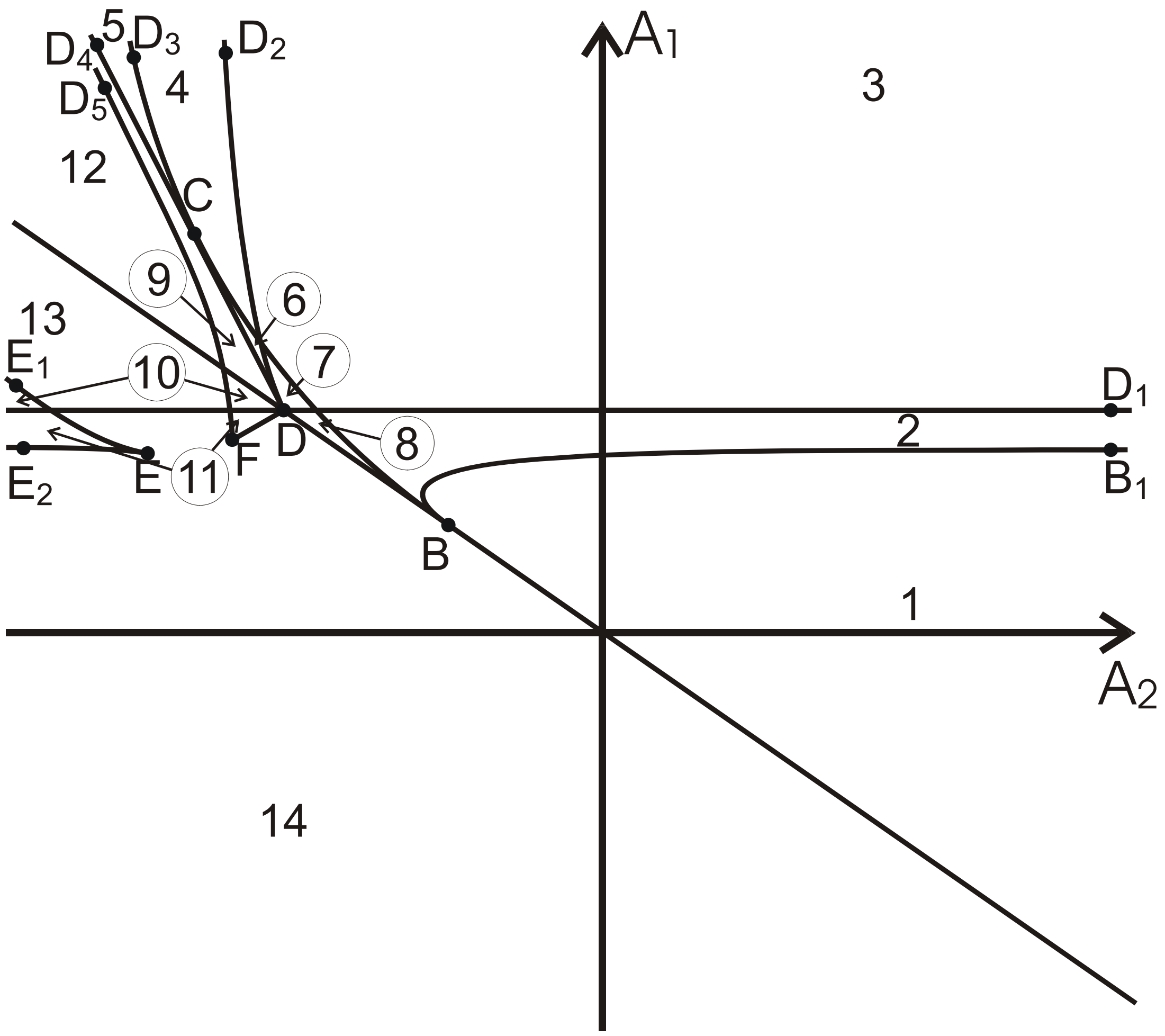

Рис. 2.4: Диаграмма формальных стационарных решений при наличии источника тепла и $\gamma > 1$. Сами решения, помеченные цифрами, представлены на Рис. 2.7.

## 2.3. Локализованный источник тепла

В случае источника тепла для осесимметричных решений

$$\vec{k}_1 = \frac{Q_T}{2\pi r}\vec{e}_r, \quad \vec{k}_2 = \frac{\gamma Q_T}{2\pi r}\vec{e}_r. \tag{2.10}$$

Тогда

$$\vartheta' = \frac{5}{5+6(P')^2}\left(P' - \frac{Q_T}{2\pi r}\right), \; A_1\vartheta' + A_2 P' + \frac{6}{5}(P')^3 = -\frac{\gamma Q_T}{2\pi r}. \tag{2.11}$$

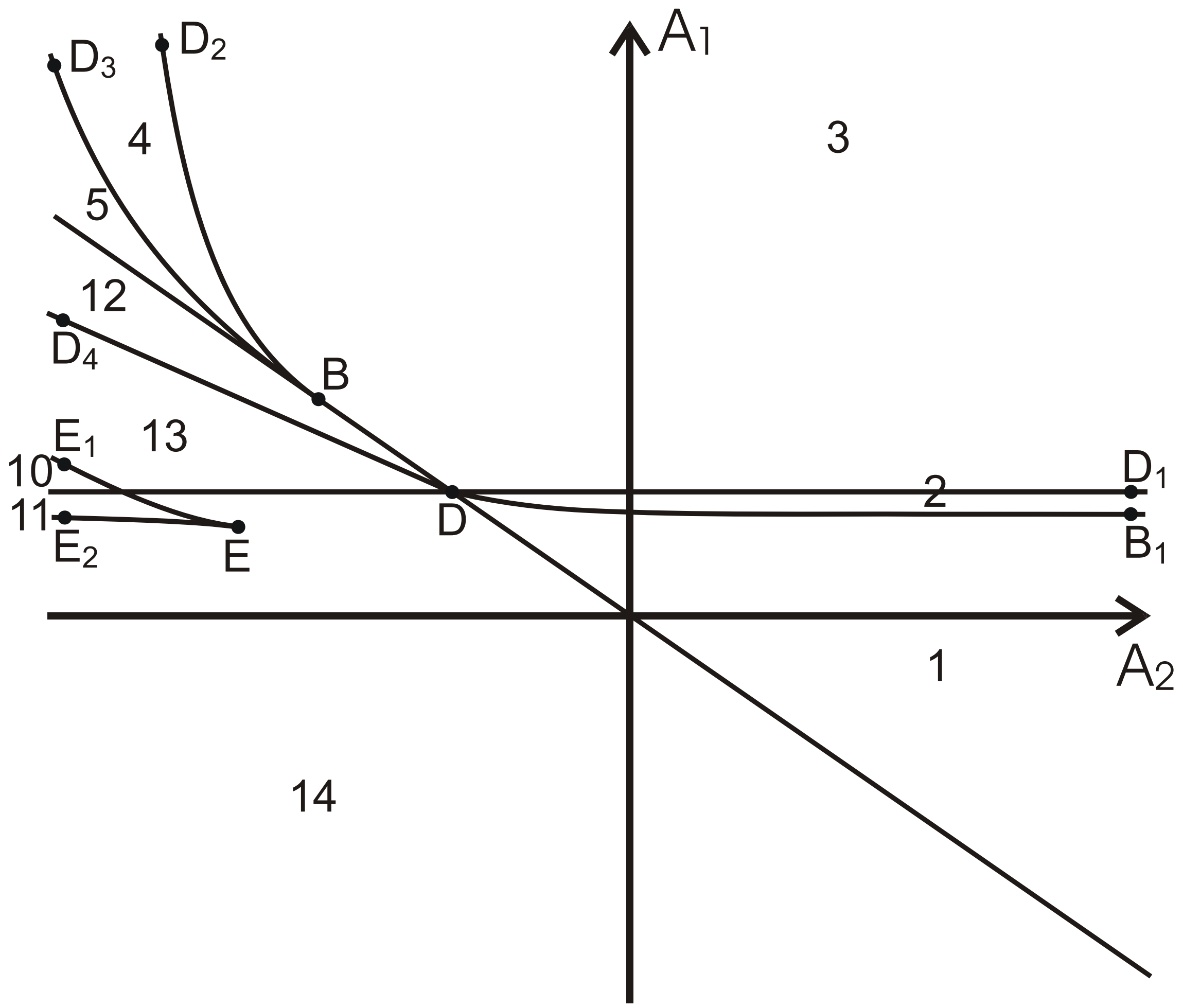

Рис. 2.5: Диаграмма формальных стационарных решений при наличии источника тепла и $0<\gamma<1$. Сами решения представлены на Рис. 2.7.

Вновь поведение системы описывается системой алгебраических уравнений. Диаграммы формальных решений этих уравнений приведены на Рис. 2.4–2.6. Решения на Рис. 2.7 представлены для положительных $Q_T$. Как и в случае источника примеси, в системе имеется симметрия $(r,\vartheta',P') \rightarrow (-r,-\vartheta',-P')$, и для перехода к случаю отрицательных $Q_T$ (т.е. случая оттока тепла) достаточно поменять знак всех скалярных полей.

Границы областей на диаграммах, представленных на Рис. 2.4–2.6., описываются следующими уравнениями:

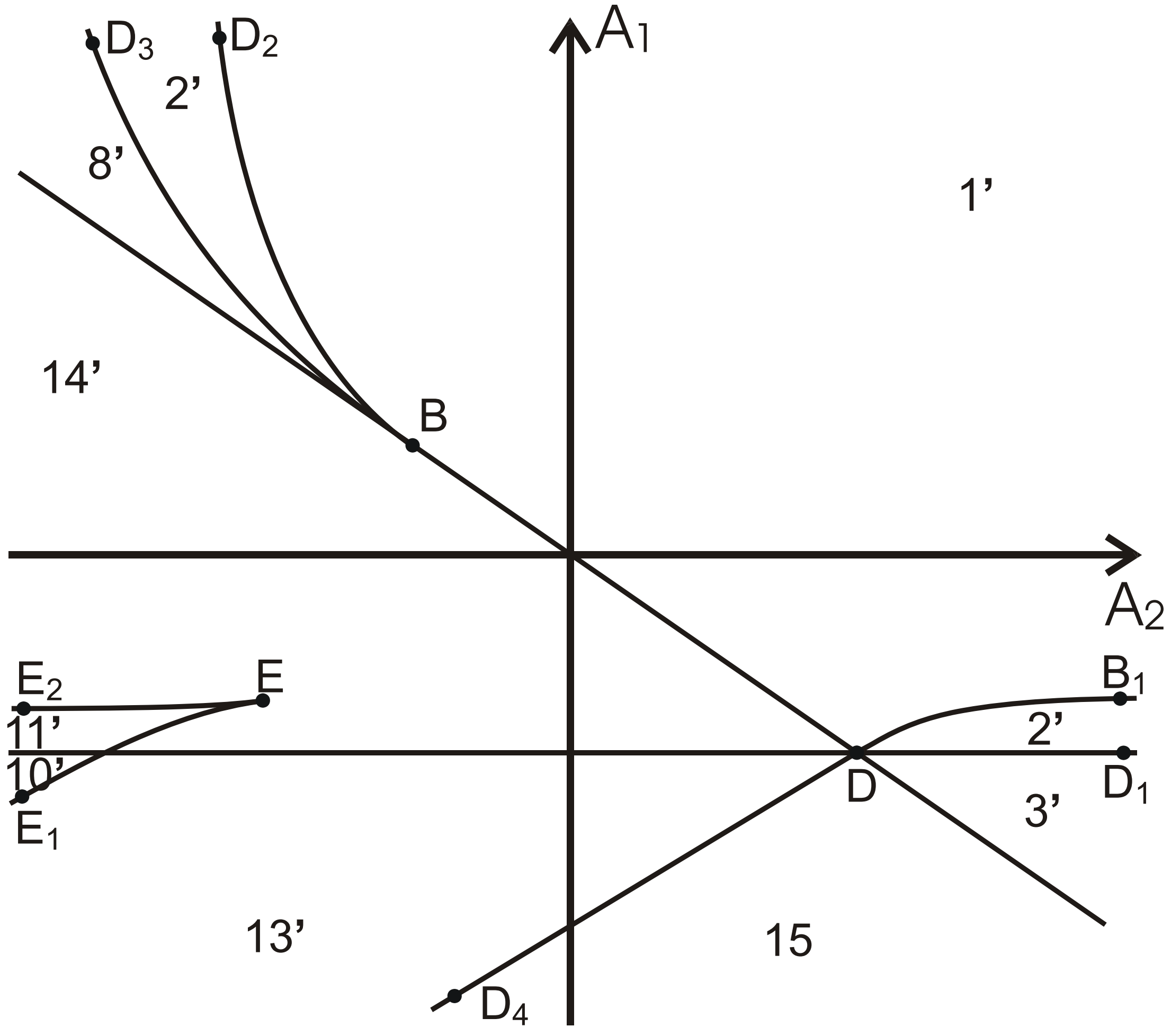

Рис. 2.6: Диаграмма формальных стационарных решений при наличии источника тепла и $\gamma < 0$. Сами решения представлены на Рис. 2.7. Режимы, маркированные цифрой со штрихом, отличаются от режимов, маркированных той же цифрой, но без штриха, знаком решений.

– прямая $\mathrm{BD}$: $A_1 + A_2 = 0$;

– прямая $\mathrm{DD}_1$: $A_1 = \gamma$;

– кривая $\mathrm{BD}_3$: $A_1 = (A_2 - 1)^2 / 4,\ A_2 < -1$;

– луч $\mathrm{DD}_4$: $A_1 = \gamma(1 - \gamma - A_2),\ A_2 < -\gamma$;

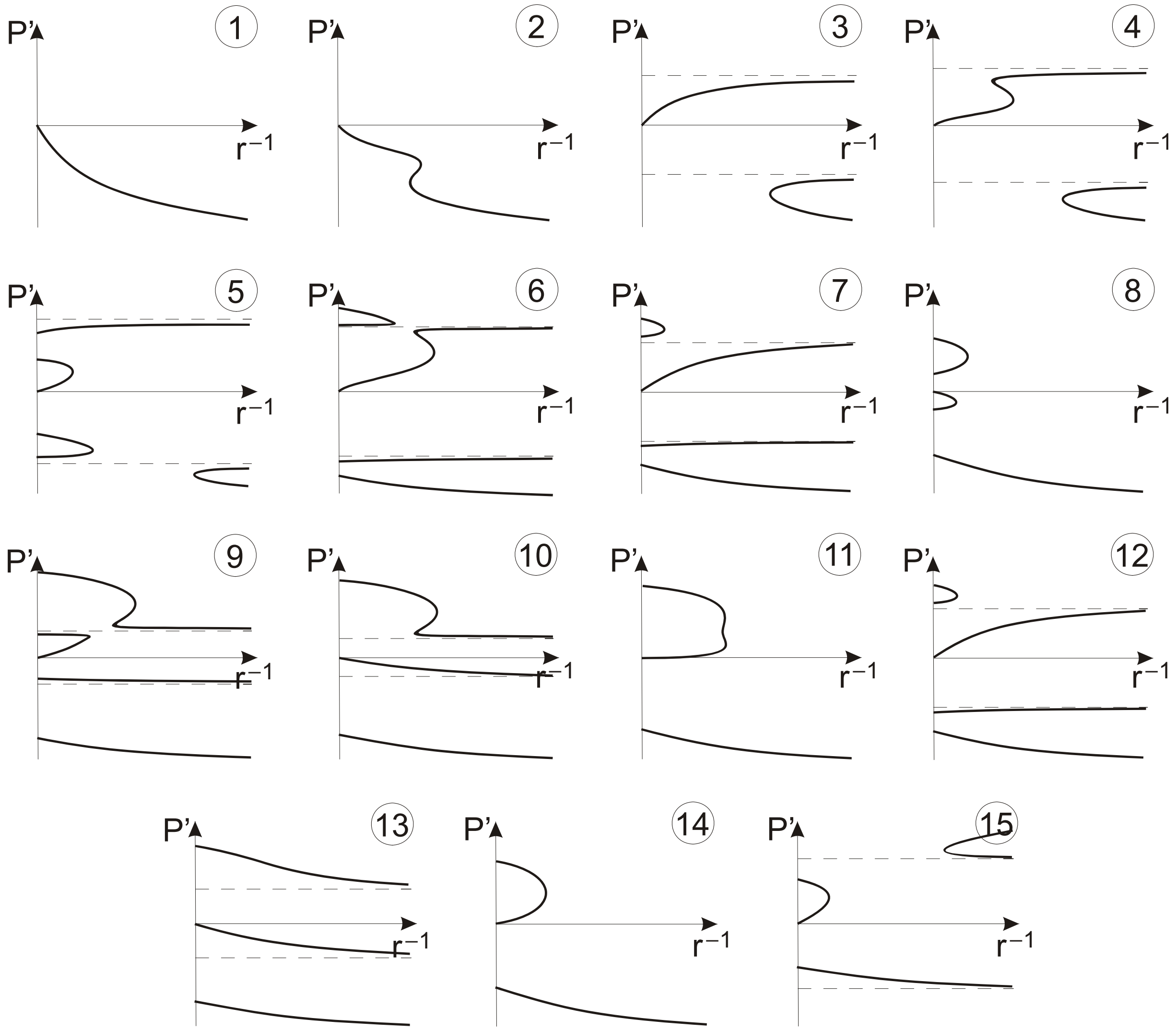

Рис. 2.7: Формальные стационарные решения системы (2.11), соответствующей источнику тепла.

– пара кривых $BB_1$ и $E_1EE_2$ и кривые $DFD_5$ и $DD_2$ на Рис. 2.4 как и пара кривых $DB_1$ и $E_1EE_2$ и кривая $BD_2$ на Рис. 2.5,2.6 являются различными ветками решений системы уравнений

$$3+2A_2-A_1-3(A_2+1)\gamma^{-1}A_1+2a^2\left(6+A_2-5\gamma^{-1}A_1\right)+9a^4=0,$$

$$\left(A_1+A_2\right)\left(1-\gamma^{-1}A_1\right)-a^4\left(6+A_2-5\gamma^{-1}A_1\right)-6a^6=0,$$

где $a$ параметризует эти решения. Следует отметить, что полупрямая $A_1 = \gamma(1-\gamma-A_2)$, $A_2 < -\gamma$ так же является решением последней системы. Аналитически можно получить асимптотическое поведение для этих кривых при больших $A_i$:

$$\begin{aligned} \mathrm{BB_1,\ DB_1\ и\ EE_2}: &\quad A_1 \approx \frac{8\gamma}{9}; \\ \mathrm{EE_1}: &\quad A_1 \approx -\frac{\gamma A_2}{25}; \\ \mathrm{DD_2\ и\ BD_2}: &\quad A_1 \approx \frac{9A_2^2}{20}; \\ \mathrm{FD_5}: &\quad A_1 \approx -\gamma A_2 - \gamma(\gamma-1). \end{aligned}$$

Можно также отметить, что точки $\mathrm{B}$ и $\mathrm{D}$ имеют координаты $\left(-1;1\right)$ и $\left(-\gamma;\gamma\right)$, соответственно.

В отличие от предыдущих случаев, тепло из окрестности источника может выноситься как конвективным образом:

$$\vartheta \approx \vartheta_0 - \frac{3}{2}\left(\frac{5Q_T}{12\pi\gamma^2}\right)^{\frac{1}{3}} r^{\frac{2}{3}}, \quad P \approx P_0 - \frac{3}{2}\left(\frac{5\gamma Q_T}{12\pi}\right)^{\frac{1}{3}} r^{\frac{2}{3}}$$

– так и диффузионно (решения с $\lim\limits_{r\to 0} P' = \pm\sqrt{5\left(\gamma^{-1}A_1 - 1\right)/6}$):

$$\vartheta \approx -\frac{\gamma Q_T}{2\pi A_1}\ln r, \quad P \approx P_0 \pm r\sqrt{\frac{5}{6}\left(\frac{A_1}{\gamma} - 1\right)}.$$

Причем, как видно из диаграмм, возможна мультистабильность между режимами с этими механизмами выноса тепла. На бесконечности, в случае, ес-ли состояние механического равновесия устойчиво, производные полей должны быть равны нулю:

$$\vartheta \approx -\frac{Q_T\left(A_2+\gamma\right)}{2\pi\left(A_1+A_2\right)}\ln r,\quad P \approx \frac{Q_T\left(A_1-\gamma\right)}{2\pi\left(A_1+A_2\right)}\ln r.$$

Аналогично предыдущему случаю, при $A_1 < -A_2$ состояние механического равновесия монотонно неустойчиво. В общих чертах можно также отметить, что при $b = S/(S-1)$ состояние механического равновесия устойчиво во всей области $A_1 + A_2 > 0$, при $b < S/(S-1)$ – в правой ее части, а при $b > S/(S-1)$ - в верхней. Более детальный анализ взаимного расположения областей устойчивости состояния механического равновесия и границ областей, где наблюдаются различные режимы стационарного длинноволнового течения, для источника тепла чрезвычайно громоздок и остается за рамками настоящего исследования. В области параметров, где состояние механического равновесия без источников устойчиво и асимптоты для $r \to 0$ и $r \to \infty$ принадлежат разным веткам решений, область длинноволнового течения вблизи источника отделена от области длинноволнового течения на бесконечности кольцом (или несколькими) переходного течения.

# Глава 3: Локализация течений в горизонтальном слое при случайно неоднородном нагреве

В данной главе рассматривается двухмерная тепловая конвекция жидкости в тонком горизонтальном слое со случайной стационарной неоднородностью нагрева, обеспечиваемой фиксированным потоком тепла поперек слоя. Уравнения длинноволнового приближения, соответствующего тонкому слою, выводятся для слоя пористой среды, однако для двухмерных течений они совпадают с уравнениями конвекции однородной жидкости [36].

Впервые локализация решений при случайной модуляции параметров рассматривалась в квантово-механических системах, где это явление фигурирует под названием "локализация Андерсона" [56]. Случай рассматриваемой ниже гидродинамической системы примечателен, помимо прочего, принципиальным отличием в физической интерпретации и наблюдаемости эффектов, связанных с формальными свойствами уравнений, в данной нелинейной гидродинамической задаче и линейном уравнении Шредингера. В уравнении Шредингера различные локализованные решения линейной задачи принципиально не взаимодействуют между собой и соответствуют связанным состояниям частиц в случайном потенциале, тогда как в данной задаче все такие моды взаимодействуют между собой через нелинейность и вместе формируют некоторое стационарное течение, которое при большой пространственной плотности локализованных мод может иметь примерно постоянную в пространстве интенсивность.

Таким образом, для того, чтобы наблюдать локализованные течения в данной гидродинамической системе (если они имеют место в линеаризованной стационарной задаче), пространственная плотность возбуждаемых мод должна быть невелика. Такая ситуация реализуется при достаточно большом отрицательном среднем отклонении теплового потока от критического значения для случая однородного нагрева. Действительно, при неоднородности, моделируе-мой белым гауссовским шумом, в системе возможны локальные превышения критического значения теплового потока, приводящие, как показывается, к возбуждению локализованных течений, изучению которых и посвящена данная глава.

Конкретно, исследуются свойства локализации нелинейных течений и влияние на них прокачивания жидкости в горизонтальном направлении. Обнаружено, что прокачивание приводит к локализации не только течений, но и возмущений поля температуры (в отсутствии прокачивания локализовано течение, но не возмущения температуры). Вычисляются показатели локализации $\gamma$ в направлении по потоку прокачивания и против. Аналитически находится показатель экспоненциального роста $\mu$ среднеквадратичных по реализациям шума значений полей, дающий оценку показателей локализации, определяемых численно. В соответствии с предсказаниями теории, численное интегрирование полной нелинейной системы выявляет радикальное влияние прокачивания на свойства локализации в направлении против потока прокачивания: длина локализации при малых конечных скоростях прокачивания может увеличиваться на порядок. Описанные результаты представлены в работе [98].

В связи с показателями локализации следует сделать существенное замечание. Показателями локализации $\gamma$ в данной задаче являются показатели Ляпунова, описывающие здесь асимптотическую эволюцию возмущений состояния системы на больших расстояниях при конкретной реализации шума и, в этом плане, соответствующие осреднению по состояниям системы. Именно такие характеристики описывают локализацию течений в конкретной гидродинамической системе с заданным случайно неоднородным нагревом. В тоже время, показатели роста $\mu$ (используемые для аналитической оценки $\gamma$) описывают поведение средних по реализациям шума значений полей, что не в точности соответствует наблюдаемым в конкретной задаче явлениям.

Хотя в теории стохастических процессов существенно шире распространены ситуации, когда требуется производить осреднение по реализациям шума, есть два классических круга задач, отвечающих осреднению по состояниям системы (или вдоль траектории системы) при заданной реализации шума. Так или иначе, эти задачи связаны с вычислением показателей Ляпунова, определяемых для стохастических систем в таком смысле, что они описывают эволюцию малых возмущений траектории системы при заданной реализации шума, т.е. описывают устойчивость отклика не к возмущениям шума, а к возмущениям состояния системы. Первый круг задач связан с явлениями локализации в распределенных системах со случайной пространственной модуляцией параметров – задача такого типа рассматривается в данной главе. Второй круг задач упомянутого типа связан с синхронизацией нелинейных систем общим шумом, и некоторым фундаментальным результатам относительно этого явления посвящена следующая глава диссертации.

## 3.1. Уравнения тепловой конвекции в тонком горизонтальном слое пористой среды при неоднородном нагреве

Рассматривается тепловая конвекция в подогреваемом снизу горизонтальном слое пористой среды. Границы слоя полагаются непроницаемыми для жидкости, тепловой поток – фиксированным, но не постоянным вдоль слоя. Предполагается, что выравнивание температуры между жидкостью и твердым скелетом происходит достаточно быстро, и отдельных температур для них не вводится. При малых перепадах температуры можно полагать, что плотность жидкости зависит от них линейно:

$$\rho = \rho_0(1 - \beta(T - T_0)),$$

где $\rho_0$ – плотность жидкости при температуре $T_0$, $\beta$ – коэффициент теплового расширения. Система координат выбирается так, что плоскость $(x, y)$ горизон-тальна, $z=0$ и $z=h$ – нижняя и верхняя границы слоя, соответственно. Для описания поведения системы используется модель Дарси-Буссинеска:

$$0=-\frac{1}{\rho_0}\nabla p-\frac{\nu m}{K}\vec{v}+g\beta T\vec{e}_z,$$

$$\frac{\partial T}{\partial t}+b^{-1}\operatorname{div}\left(\vec{v}T\right)=\chi\,\Delta T,$$

$$\operatorname{div}\vec{v}=0,$$

$$z=0,h:\ \ v_z=0,\ \ \frac{\partial T}{\partial z}=-A(1+q(x,y)),$$

где $\vec{v}$ – средняя (осредненная на масштабах пор) скорость жидкости в порах, $\nu$ – кинематическая вязкость смеси, $m$ – пористость среды (отношение объема пор в элементе пористой среды к объему этого элемента), $K$ – коэффициент проницаемости, $\vec{g}=-g\,\vec{e}_z$ – ускорение свободного падения, $b$ – отношение теплоемкости пористой среды, насыщенной жидкостью, к части этой теплоемкости, приходящейся на жидкость в порах, $b>1$, $\chi$ – температуропроводность пористой среды, насыщенной жидкостью, $q(x,y)$ описывает неоднородность нагрева.

В данной задаче удобной единицей измерения длины является толщина слоя $h$, времени – $h^2\chi^{-1}$, скорости – $h^{-1}b\chi$, температуры – $Ah$, и давления – $b\rho_0\nu\chi m/K$. Характер поведения системы определяется числом Релея-Дарси:

$$Ra\equiv\frac{\beta A h^2 gK}{mb\,\nu\chi}.$$

При таком выборе единиц уравнения принимают вид:

$$-\nabla p-\vec{v}+Ra\,T\vec{e}_z=0, \qquad (3.1)$$

$$\frac{\partial T}{\partial t} + \operatorname{div}\left(\vec{v}T\right) = \Delta T, \tag{3.2}$$

$$\nabla \cdot \vec{v} = 0 \tag{3.3}$$

с граничными условиями:

$$z = 0, 1: \quad v_z = 0, \; \frac{\partial T}{\partial z} = -1 - q(x, y). \tag{3.4}$$

Здесь не стоит переходить к уравнениям для возмущений некоторого основного состояния, как это делается при однородном нагреве, поскольку из-за неоднородности нагрева состояние механического равновесия в системе невозможно.

### 3.1.1. Длинноволновое приближение (тонкий слой)

При однородном вдоль слоя фиксированном тепловом потоке критическими в данной системе являются длинноволновые возмущения [34]. В связи с этим имеет смысл рассматривать задачу в длинноволновом приближении, полагая $q(x,y)$ медленно меняющимся вдоль слоя: $|\,\partial q \,/\, \partial x\,| \,/\, |\,q\,| \sim \varepsilon \ll 1$. Во избежание возникновения больших градиентов температур, которые в длинноволновом приближении соответствуют разрывам производных полей, следует ограничиться случаем малых надкритичностей (в процессе решения будет явно найдено критическое значение $Ra_{\mathrm{C}}$ для случая однородного нагрева, и будет положено $Ra = Ra_{\mathrm{C}}$), и, соответственно, малых $q(x,y)$, которые будут описывать относительное отклонение локального значения числа Релея-Дарси от критического.

Из уравнения непрерывности (3.3) следует, что горизонтальная компонента скорости велика по сравнению с вертикальной (поскольку изменение полей в горизонтальном направлении происходит медленнее, чем в вертикальном). Учитывая это явно, можно записать поле скорости в виде:

$$\vec{v} = w\vec{e}_z + \varepsilon^{-1}\vec{u},$$

где $\vec{u}$ – горизонтальная компонента поля скорости. Изменяя масштаб в горизонтальных направлениях: $x \to \varepsilon^{-1}x$, $y \to \varepsilon^{-1}y$ – полагая $q(x,y) = \varepsilon^2 q_2(x,y)$ и проектируя закон сохранения импульса (3.1) на вертикальное и горизонтальное направления, систему (3.1)–(3.3) с граничным условием (3.4), можно переписать в виде, удобном для последующего анализа:

$$-\partial_z p - w + Ra\ T = 0, \tag{3.5}$$

$$\vec{u} = -\varepsilon^2 \nabla_2 p, \tag{3.6}$$

$$\partial_t T + \partial_z (wT) + \varepsilon^2 \nabla_2 \cdot (\vec{u}T) = \partial_z^2 T + \varepsilon^2 \Delta_2 T, \tag{3.7}$$

$$\partial_z w + \nabla_2 \cdot \vec{u} = 0, \tag{3.8}$$

$$z = 0, 1: \quad w = 0, \quad \partial_z T = -1 - \varepsilon^2 q_2(x,y), \tag{3.9}$$

где индекс 2 у операторов дифференцирования по пространственным координатам означает, что дифференцирование производится лишь по горизонтальным координатам.

Очевидно, что разложение идет лишь по четным степеням $\varepsilon$ (во всех уравнениях задачи (3.5)–(3.9) $\varepsilon$ появляется в квадрате). Также, можно сразу учесть, что длинноволновое приближение предполагает медленное изменение полей с координатами, а слабая неоднородность в пространстве не может вызвать быстрой эволюции системы во времени. Это значит, что эволюция во времени будет медленной. Поскольку разложение будет производится по четным степеням $\varepsilon$, имеет смысл ожидать, что характерные времена эволюции длинноволновых возмущений будут не быстрее, чем $\propto \varepsilon^{-2}$.

$\underline{\varepsilon^0}$: В нулевом порядке задача (3.5)–(3.9) дает

$$-\partial_z p_0 - w_0 + Ra\ T_0 = 0, \tag{3.10}$$

$$\vec{u}_0 = 0, \tag{3.11}$$

$$\partial_z(w_0 T_0) + \nabla_2 \cdot (\vec{u}_0 T_0) = \partial_z^2 T_0, \tag{3.12}$$

$$\partial_z w_0 + \nabla_2 \cdot \vec{u}_0 = 0, \tag{3.13}$$

$$z = 0, 1: \quad w_0 = 0, \quad \partial_z T_0 = -1. \tag{3.14}$$

Из (3.13), (3.11): $\partial_z w_0 = 0$, $w_0 = C_1(x, y) = 0$ ($C_1 = 0$ в силу граничных условий (3.14)).

Из (3.12): $\partial_z^2 T_0 = 0$, т.е. $T_0 = C_2(x, y) z + \theta(x, y)$, с учетом граничных условий (3.14):

$$T_0 = -z + \theta(x, y), \tag{3.15}$$

где $\theta(x, y)$ остается на данный момент произвольной функцией горизонтальных координат.

Из (3.10): $\partial_z p_0 = Ra\ T_0 = -Ra\ z + Ra\ \theta(x, y)$,

$$p_0 = -\frac{Ra}{2} z^2 + Ra\ \theta(x, y) z + \Pi_0(x, y), \tag{3.16}$$

где $\Pi_0(x, y)$ в данном порядке разложения не определено.

$\underline{\varepsilon^2}$:

$$-\partial_z p_2 - w_2 + Ra\ T_2 = 0, \tag{3.17}$$

$$\vec{u}_2 = -\nabla_2 p_0, \tag{3.18}$$

$$\partial_{t_2} T_0 + \partial_z(w_2 T_0) + \nabla_2 \cdot (\vec{u}_2 T_0) = \partial_z^2 T_2 + \Delta_2 T_0, \tag{3.19}$$

$$\partial_z w_2 + \nabla_2 \cdot \vec{u}_2 = 0, \tag{3.20}$$

$$z = 0, 1: \quad w_2 = 0, \quad \partial_z T_2 = -q_2(x, y). \tag{3.21}$$

Из (3.18): $\vec{u}_2 = -\nabla_2 p_0 = -Ra\ z \nabla_2 \theta(x, y) - \nabla_2 \Pi_0(x, y)$.

Из (3.20): $\partial_z w_2 = -\nabla_2 \cdot \vec{u}_2 = Ra\ z\ \Delta_2\theta(x,y) + \Delta_2 \Pi_0(x,y)$,

$$w_2 = \frac{1}{2} Ra\ z^2 \Delta_2\theta(x,y) + \Delta_2\Pi_0(x,y) z + C_3(x,y).$$

Граничные условия (3.21) для скорости дают:

$$z = 0:\ C_3 = 0, \qquad z = 1:\ \Delta_2 \Pi_0(x,y) = -\frac{1}{2} Ra\, \Delta_2\theta(x,y).$$

На основании последнего, можно написать

$$\Pi_0(x,y) = -\frac{1}{2} Ra\ \theta(x,y) + \pi_0(x,y), \ \ \Delta_2 \pi_0(x,y) = 0.$$

Можно также видеть, что $\langle \vec{u}_2 \rangle = -\nabla_2 \pi_0$ (здесь и далее $\langle f \rangle \equiv \int_0^1 f\, dz$). Положим, имеется ограниченный (в горизонтальных направлениях) фрагмент слоя, на границе $\Gamma$ (пусть, для простоты, она будет вертикальна, что, вообще говоря, не обязательно) которого навязан приток жидкости в этот фрагмент (или, например, его отсутствие; достаточно отсутствия притока – поток через границу допускается, но он должен иметь нулевое среднее по $z$). Для этого слоя нормальная к границе $\Gamma$ компонента $\langle \vec{u}_2 \rangle$ задана: $\langle u_2 \rangle_n |_\Gamma = Q$ ($Q$ может зависеть от координат вдоль границы слоя). Таким образом, получается граничная задача для $\pi_0(x,y)$:

$$\Delta_2 \pi_0 = 0, \quad \left. \frac{\partial \pi_0}{\partial n} \right|_\Gamma = -Q \tag{3.22}$$

($\partial / \partial n$ – производная по нормали к границе), которая имеет единственное (с точностью до несущественной аддитивной константы) решение, однозначно определяемое граничными условиями. Т.о. $\pi_0(x,y)$ описывает принудительное прокачивание жидкости через слой за счет воздействия на его боковых грани-цах, причем однозначно. Поскольку рассматривается задача вблизи порога устойчивости, то имеет смысл не допускать прокачивания в ведущем порядке, иначе в первую очередь оно будет определять поведение системы, а не неоднородность нагрева (которая мала). Т.о. имеет смысл положить

$$\pi_0(x,y) = 0 \tag{3.23}$$

и попытаться учесть возможное прокачивание в следующих порядках малости.

Тогда,

$$\vec{u}_2 = -Ra\left(z - \frac{1}{2}\right)\nabla_2\theta(x,y), \tag{3.24}$$

$$w_2 = \frac{1}{2}Ra(z^2 - z)\Delta_2\theta(x,y). \tag{3.25}$$

Проинтегрируем уравнение (3.19) по $z$:

$$\partial_{t_2}\langle T_0\rangle + \langle\partial_z(w_2T_0)\rangle + \nabla_2\cdot\langle\vec{u}_2T_0\rangle = \langle\partial_z^2T_2\rangle + \Delta_2\langle T_0\rangle. \tag{3.26}$$

С учетом граничных условий (3.21),

$$\int_0^1 \partial_z(w_2T_0)\,dz = w_2T_0\Big|_{z=0}^{1} = 0,$$

$$\int_0^1 \partial_z^2T_2\,dz = \partial_zT_2\Big|_{z=0}^{1} = -q_2(x,y)\Big|_{z=0}^{1} = 0.$$

В свою очередь,

$$\nabla_2\cdot\langle\vec{u}_2T_0\rangle = -\nabla_2\cdot\langle\vec{u}_2z\rangle + \nabla_2\cdot(\langle\vec{u}_2\rangle\theta)$$

Подставляя в первое слагаемое выражение (3.24), и учитывая, что во втором, заведомо, $\langle\vec{u}_2\rangle = -\nabla_2\pi_0 = 0$, можно получить:

$$\nabla_2 \cdot \left\langle \vec{u}_2 T_0 \right\rangle = Ra\, \Delta_2 \theta \left\langle z^2 - z/2 \right\rangle = (Ra/12)\, \Delta_2 \theta .$$

В то же время, $\partial_{t_2} T_0 = \partial_{t_2} \theta$, $\Delta_2 T_0 = \Delta_2 \theta$, и уравнение (3.26) принимает вид:

$$\partial_{t_2} \theta = \left(1 - Ra/12\right) \Delta_2 \theta .$$

Это уравнение при $Ra < 12$ является обычным уравнением теплопроводности и при тривиальных граничных условиях все возмущения температуры затухают, при $Ra > 12$ получается уравнение с отрицательной теплопроводностью, т.е. все возмущения нарастают; $Ra = 12$ – граница линейной устойчивости системы. При этом, в данном порядке не проявляет себя нелинейность. Т.о., для учета нелинейности и зависимости свойств устойчивости от длины волны (сейчас либо все возмущения затухают, либо все нарастают), необходимо ограничиться малой окрестностью границы устойчивости, полагая (всюду далее)

$$Ra = Ra_{\mathrm{C}} = 12$$

и учитывая надкритичность через $q$. Заслуживает упоминания, что локальное число Релея-Дарси ("локальное" значит, в некоторой области слоя) $Ra_{\mathrm{local}} = Ra(1 + q)$, т.е. положительные $q$ соответствуют надкритическим режимам, а отрицательные – подкритическим. Кроме того, при $Ra = 12$, получается $\partial_{t_2} \theta = 0$ и, следовательно, стоит рассматривать более медленную эволюцию $\theta$, а именно полагать $\partial_t \theta = \varepsilon^4 \partial_{t_4} \theta$.

Теперь обратимся к уравнению (3.19) и найдем из него $T_2$:

$$\begin{aligned}
\partial_z^2 T_2 &= \partial_z (w_2 T_0) + \nabla_2 \cdot (\vec{u}_2 T_0) - \Delta_2 T_0 = \\
&= -\partial_z (w_2 z) + \theta \partial_z w_2 - \nabla_2 \cdot (\vec{u}_2 z) + \nabla_2 \cdot (\vec{u}_2 \theta) - \Delta_2 \theta = \\
&= -\partial_z (6z^3 - 6z^2) \Delta_2 \theta + \partial_z (6z^2 - 6z) \theta \Delta_2 \theta + \\
&\quad +(12z^2 - 6z) \Delta_2 \theta - (12z - 6) \nabla_2 \cdot (\theta \nabla_2 \theta) - \Delta_2 \theta,
\end{aligned}$$

$$T_2 = -\left(\frac{3}{2}z^4 - 2z^3\right)\Delta_2\theta + (2z^3 - 3z^2)\theta\Delta_2\theta +$$
$$+(z^4 - z^3)\Delta_2\theta - (2z^3 - 3z^2)\nabla_2 \cdot (\theta\nabla_2\theta) - \frac{1}{2}z^2\Delta_2\theta +$$
$$+C_4(x,y)z + \theta_2(x,y).$$

Используя соотношение $\nabla_2 \cdot (\theta\nabla_2\theta) = \theta\Delta_2\theta + (\nabla_2\theta)^2$, можно видеть, что слагаемое $\theta\Delta_2\theta$ исчезает из выражения для $T_2$:

$$T_2 = \left(-\frac{1}{2}z^4 + z^3 - \frac{1}{2}z^2\right)\Delta_2\theta + + (-2z^3 + 3z^2)(\nabla_2\theta)^2 + C_4(x,y)z + \theta_2(x,y), \quad (3.27)$$

Учет граничных условий (3.21) дает, что $C_4(x,y) = -q_2(x,y)$, в $\theta_2(x,y)$ остается произвол.

Можно видеть, что $\theta(x,y)$ и $\theta_2(x,y)$ имеют одинаковый характер зависимости от $z$ (т.е. не зависят) и являются произвольными функциями. Поэтому $\theta_2$ можно полагать любой функцией и это автоматически отразится в $\theta$. Это обстоятельство удобно использовать. Дело в том, что

$$\langle T\rangle = -\frac{1}{2} + \theta(x,y) + \varepsilon^2\langle T_2\rangle + \mathrm{O}(\varepsilon^4),$$

и, если использовать $\theta_2$ для обращения $\langle T_2\rangle$ в ноль, то $\theta$ можно будет интерпретировать как среднюю по толщине слоя температуру с большей точностью (конкретнее, с той точностью, до которой рассматривается здесь разложение уравнений, т.е. $\varepsilon^4$). Итак,

$$\langle T_2\rangle = -\frac{1}{60}\Delta_2\theta + \frac{1}{2}(\nabla_2\theta)^2 - \frac{1}{2}q_2(x,y) + \theta_2(x,y) = 0,$$

откуда

$$\theta_2 = \frac{1}{60}\Delta_2\theta - \frac{1}{2}(\nabla_2\theta)^2 + \frac{q_2}{2}. \tag{3.28}$$

Таким образом,

$$\begin{gathered} T_2 = \left(-\frac{1}{2}z^4 + z^3 - \frac{1}{2}z^2 + \frac{1}{60}\right)\Delta_2\theta + \\ + \left(-2z^3 + 3z^2 - \frac{1}{2}\right)(\nabla_2\theta)^2 + \left(-z + \frac{1}{2}\right)q. \end{gathered} \tag{3.29}$$

Далее, из уравнения (3.17) следует $\partial_z p_2 = -w_2 + Ra\ T_2$ и, интегрируя по $z$, можно получить:

$$\begin{gathered} p_2 = \left(-\frac{6}{5}z^5 + 3z^4 - 4z^3 + 3z^2 + \frac{1}{5}z\right)\Delta_2\theta + \\ + \left(-6z^4 + 12z^3 - 6z\right)(\nabla_2\theta)^2 + \left(-6z^2 + 6z\right)q + \Pi_2(x,y). \end{gathered} \tag{3.30}$$

$\underline{\varepsilon^4}$: Если ограничиваться данным порядком разложения, то не требуются все уравнения – достаточно

$$\vec{u}_4 = -\nabla_2 p_2, \tag{3.31}$$

$$\begin{gathered} \partial_{t_4}T_0 + \partial_z(w_4T_0) + \nabla_2\cdot(\vec{u}_4T_0) + \\ + \partial_z(w_2T_2) + \nabla_2\cdot(\vec{u}_2T_2) = \partial_z^2T_4 + \Delta_2T_2, \end{gathered} \tag{3.32}$$

$$\partial_z w_4 + \nabla_2\cdot\vec{u}_4 = 0, \tag{3.33}$$

$$z = 0, 1: \quad w_4 = 0, \quad \partial_z T_4 = 0. \tag{3.34}$$

Нет необходимости вычислять выражение для $\vec{u}_4$, а достаточно сразу подставить (3.31) в (3.33):

$$\partial_z w_4 = \Delta_2 p_2,$$

тогда

$$w_4 = \left(-\frac{1}{5}z^6 + \frac{3}{5}z^5 - z^4 + z^3 + \frac{1}{10}z^2\right)\Delta_2^2\theta + \\ + \left(-\frac{6}{5}z^5 + 3z^4 - 3z^2\right)\Delta_2(\nabla_2\theta)^2 + \\ + \left(-2z^3 + 3z^2\right)\Delta_2 q_2 + z\Delta_2 \Pi_2. \tag{3.35}$$

Поскольку

$$w_4\big|_{z=1} = 0 = \frac{1}{2}\Delta_2^2\theta - \frac{6}{5}\Delta_2(\nabla_2\theta)^2 + \Delta_2 q_2 + \Delta_2 \Pi_2,$$

то

$$\Pi_2 = -\frac{1}{2}\Delta_2\theta + \frac{6}{5}(\nabla_2\theta)^2 - q_2 + \pi_2, \text{ где } \Delta_2\pi_2 = 0.$$

Следует принять во внимание, что

$$\left\langle \vec{u}_4 \right\rangle = -\nabla_2 \left\langle p_2 \right\rangle = \\ = -\nabla_2 \left( \left\langle -\frac{6}{5}z^5 + 3z^4 - 4z^3 + 3z^2 + \frac{1}{5}z \right\rangle \Delta_2\theta + \right. \\ + \left\langle -6z^4 + 12z^3 - 6z \right\rangle (\nabla_2\theta)^2 + \\ \left. + \left\langle -6z^2 + 6z \right\rangle q_2 + \Pi_2(x,y) \right) = \\ = -\nabla_2 \pi_2(x,y).$$

Выше была установлена связь между $\left\langle \vec{u}_2 \right\rangle$ и $\pi_0$ и было показано, что $\pi_0$ описывает прокачивание. В том порядке прокачивание не вводилось, однако теперь имеет смысл это сделать: оставить $\pi_2$ и иметь в виду, что оно описывает принудительное прокачивание жидкости через слой, навязанное граничными условиями на боковых границах слоя, а именно заданным средним по $z$ потоком жидкости через боковую границу. Подставляя $\Pi_2$ в (3.35), можно выписать окончательное выражение

$$\begin{aligned} w_4 = &\left(-\frac{1}{5}z^6 + \frac{3}{5}z^5 - z^4 + z^3 + \frac{1}{10}z^2 - \frac{1}{2}z\right)\Delta_2^2\theta + \\ &+\left(-\frac{6}{5}z^5 + 3z^4 - 3z^2 + \frac{6}{5}z\right)\Delta_2(\nabla_2\theta)^2 + \\ &+\left(-2z^3 + 3z^2 - z\right)\Delta_2 q_2. \end{aligned} \tag{3.36}$$

Теперь проинтегрируем уравнение (3.32) по $z$ (это даст искомое уравнение эволюции $\theta$):

$$\begin{aligned} &\partial_{t_4}\theta + \left\langle\partial_z(w_4T_0)\right\rangle + \nabla_2\cdot\left\langle\vec{u}_4T_0\right\rangle + \\ &+\left\langle\partial_z(w_2T_2)\right\rangle + \nabla_2\cdot\left\langle\vec{u}_2T_2\right\rangle = \left\langle\partial_z^2T_4\right\rangle + \Delta_2\left\langle T_2\right\rangle, \end{aligned}$$

Средние от всех производных по $z$ в силу граничных условий равны нулю, также $\left\langle T_2\right\rangle = 0$. Остается:

$$\partial_{t_4}\theta + \nabla_2\cdot\left\langle\vec{u}_4T_0 + \vec{u}_2T_2\right\rangle = 0, \tag{3.37}$$

где

$$\begin{aligned} \nabla_2\cdot\left\langle\vec{u}_4T_0\right\rangle &= -\nabla_2\cdot\left\langle T_0\nabla_2 p_2\right\rangle = -\nabla_2\cdot\left(-\left\langle z\nabla_2 p_2\right\rangle + \left\langle\theta\nabla_2 p_2\right\rangle\right) = \\ &= \left\langle z\Delta_2 p_2\right\rangle - \nabla_2\cdot\left(\theta\,\nabla_2\left\langle p_2\right\rangle\right) = \\ &= \left\langle z\partial_z w_4\right\rangle - \nabla_2\cdot\left(\theta\,\nabla_2\pi_2\right) = -\left\langle w_4\right\rangle - \nabla_2\cdot\left(\theta\,\nabla_2\pi_2\right), \end{aligned}$$

$$\left\langle w_4\right\rangle = -\frac{2}{21}\Delta_2^2\theta,$$

$$\nabla_2\cdot\left\langle\vec{u}_4T_0\right\rangle = \frac{2}{21}\Delta_2^2\theta - \nabla_2\cdot\left(\theta\,\nabla_2\pi_2\right); \tag{3.38}$$

$$\nabla_2 \cdot \left\langle \vec{u}_2 T_2 \right\rangle = -\nabla_2 \cdot \left\langle T_2 \nabla_2 p_0 \right\rangle = -Ra\nabla_2 \cdot \left\langle \left(z - \tfrac{1}{2}\right) T_2 \nabla_2 \theta \right\rangle$$
$$= -Ra\nabla_2 \cdot \left( \left\langle \left(z - \tfrac{1}{2}\right) T_2 \right\rangle \nabla_2 \theta \right) = -Ra\nabla_2 \cdot \left( \left\langle z T_2 \right\rangle \nabla_2 \theta \right),$$
$$\left\langle z T_2 \right\rangle = \frac{1}{10} (\nabla_2 \theta)^2 - \frac{q_2}{12},$$
$$\nabla_2 \cdot \left\langle \vec{u}_2 T_2 \right\rangle = -\nabla_2 \cdot \left[ \frac{6}{5} \nabla_2 \theta (\nabla_2 \theta)^2 - q_2 \nabla_2 \theta \right]. \qquad (3.39)$$

Подставляя (3.38) и (3.39) в (3.37), можно получить искомое уравнение медленной эволюции:

$$\dot{\theta} + \frac{2}{21} \Delta^2 \theta - \operatorname{div}\left( \theta \, \nabla \pi_2 \right) - \operatorname{div}\left[ \frac{6}{5} \nabla \theta (\nabla \theta)^2 - q_2 \nabla \theta \right] = 0. \qquad (3.40)$$

Индексы 2 у операторов градиента опущены, поскольку все поля двухмерны.

Следует отметить, что неоднородность нагрева делает невозможным состояние механического равновесия, а из (3.29) и (3.36) видно, что в подкритической области параметров, когда возмущения $\theta$ затухают, в системе устанавливается режим с

$$w_{4,\mathrm{BG}} = \left( -2z^3 + 3z^2 - z \right) \Delta_2 q_2 \quad \text{и} \quad T_{2,\mathrm{BG}} = \left( -z + \frac{1}{2} \right) q.$$

Однако, эти поля скорости и температуры малы по сравнению с характерными значениями полей возбуждаемых в надкритической области конвективных течений.

Изменением масштабов

$$t \to \frac{2}{21} t, \; x \to \sqrt{\frac{2}{21}} x \;\; \text{и} \;\; \theta \to \sqrt{\frac{5}{63}} \theta$$

в уравнении (3.40) можно обратить все коэффициенты в 1. Тогда, можно окончательно получить

$$\dot{\theta} - \nabla\pi_2 \cdot \nabla\theta + \Delta^2\theta - \mathrm{div}\Big(\nabla\theta(\nabla\theta)^2 - q_2\nabla\theta\Big) = 0, \tag{3.41}$$

$$\Delta\pi_2 = 0. \tag{3.42}$$

где $-\nabla\pi_2$, как было отмечено выше, описывает принудительное прокачивание и можно видеть, что второе слагаемое уравнения (3.41) фактически является конвективной производной. Подобные уравнения, отличающиеся лишь возможным нелинейным слагаемым в $\pi_2$ (которое исключено в случае двухмерной конвекции) описывают длинноволновую тепловую конвекцию в однородной жидкости [36] и слое турбулентной жидкости [37].

## 3.2. Локализация течений в слое пористой среды

В данной работе рассматривается случай $q_2(x,y) = q_0 + \sqrt{D}\,\xi(x)$, где $\xi(x)$ – нормированный гауссовский $\delta$-коррелированный шум, $\big\langle\xi(x)\big\rangle = 0$, $\big\langle\xi(x+x')\xi(x)\big\rangle = 2\delta(x')$. Приближение $\delta$-коррелированного шума не противоречит длинноволновому приближению – достаточно соблюсти следующую иерархию малости величин: $|\,q\,| \ll (|\,\partial q/\partial x\,|\,/\,|\,q\,|)^2 \ll 1$ – так чтобы две указанные малые величины проявляли себя в одном порядке разложения. Интенсивность шума $D$ может быть обращена в единицу подходящим выбором единиц измерения: $q \to D^{2/3}q$, $x \to D^{-1/3}x$, $t \to D^{4/3}t$, $\pi \to D^{2/3}\pi$. В связи с однородностью нагрева вдоль одного из горизонтальных направлений, имеет смысл ограничиться исследованием однородных вдоль $y$ решений, для которых скорость прокачивания $-\nabla\pi_2 = u\vec{e}_x$ ($\vec{e}_x$ – орт оси $x$), $u = const$. Тогда уравнение (3.41) принимает вид

$$\dot{\theta} = (-u\theta - \theta_{xxx} - [q_0 + \xi(x)]\theta_x + (\theta_x)^3)_x. \tag{3.43}$$

Поле скорости жидкости в главном порядке, определяемое формулами (3.24),(3.25), может быть выражено через функцию тока:

$$\vec{v} = \frac{\partial \Psi}{\partial z}\vec{e}_x - \frac{\partial \Psi}{\partial x}\vec{e}_z, \quad \Psi = f(z)\psi(x) = \Psi_0 z(1-z)\theta_x,$$

где $\psi(x) \equiv \theta_x(x)$ – амплитуда функции тока; константа $\Psi_0$ определяется выбором единиц измерения. Здесь не фигурирует вклад в скорость от прокачивания, поскольку его характерные значения малы на фоне конвективных течений. Существенное же влияние прокачивания на систему связано с тем, что его пространственная симметрия отличается от симметрии конвективных течений: у конвективных течений суммарный поток жидкости через любое сечение слоя равен нулю, а для течения, связанного с прокачиванием, это не так. Важно, что хотя в длинноволновом приближении уравнения эволюции возмущений температуры при двухмерной конвекции в однородной жидкости и при конвекции жидкости в пористой среде совпадают (с точностью до перенормировки параметров), функции $f(z)$ различны.

Численный счет выявил, что при достаточно малых $u$ в системе (3.43) реализуются устойчивые стационарные решения. В данной главе рассматриваются свойства локализации нетривиальных стационарных решений. Если некоторое решение будет локализовано в окрестности точки $x_0$, то вдали от нее $\theta_x$ ~ $\exp(-\gamma_\pm \mid x - x_0 \mid)$ (при $(x-x_0)u > 0$ выбирается показатель локализации в направлении по потоку $\gamma_-$, при $(x-x_0)u < 0$ – показатель локализации в направлении против потока $\gamma_+$; обратную величину $\lambda_\pm = \gamma_\pm^{-1}$ принято называть длиной локализации). В области этих экспоненциальных "хвостов" решения малы и показатели могут быть определены из линеаризации уравнения (3.43), которое в стационарном случае может быть однократно проинтегрировано по $x$:

$$u\theta + \theta''' + [q_0 + \xi(x)]\theta' = const \equiv S \tag{3.44}$$

(штрих обозначает производную по координате $x$), при $u \neq 0$ замена $\theta \to \theta + u^{-1}S$ позволяет обратить поток тепла $S$ в ноль.

Примечательно, что при $u = S = 0$ для $\theta'$ получается стационарное уравнение Шредингера (уравнение при $q(x,y)$, зависящем от обеих горизонтальных координат, к уравнению Шредингера уже не сводится), свойства локализации в котором при $\delta$-коррелированном потенциале хорошо исследованы (см., например, [61]): все состояния с любой энергией (соответствующей в нашем случае средней надкритичности $q_0$) локализованы. Однако данный случай примечателен не только наличием прокачивания, но и принципиальным отличием в физической интерпретации и наблюдаемости эффектов, связанных с формальными свойствами уравнений, в данной нелинейной гидродинамической задаче и линейном уравнении Шредингера: в уравнении Шредингера различные локализованные решения линейной задачи принципиально не взаимодействуют между собой и соответствуют связанным состояниям частиц в случайном потенциале, тогда как в данной задаче все такие моды взаимодействуют между собой через нелинейность и вместе формируют некоторое стационарное течение, которое при большой пространственной плотности локализованных мод может иметь примерно постоянную в пространстве интенсивность.

Таким образом, для того, чтобы наблюдать локализованные течения в данной системе (если они имеют место в линеаризованной стационарной задаче), пространственная плотность возбуждаемых мод должна быть невелика. Такая ситуация реализуется при достаточно большом отрицательном среднем отклонении теплового потока от критического значения: $q_0 < 0$. Для пояснения этого утверждения введем локальное среднее

$$q_l(x) \equiv l^{-1} \int_{x-l/2}^{x+l/2} q(x_1)\, dx_1 \,.$$

Если в окрестности некоторой точки $x$ значение $q_l(x)$ положительно для достаточно больших $l$ ("достаточно большие $l$" значит $l \sim 1$), в окрестности этой точки может развиться конвективное течение, в то время, как в областях с отрицательными $q_l$ течение в среднем демпфируется. Для гауссовского $\delta$-коррелированного шума, $q_l$ – гауссовская случайная величина со средним $\langle q_l \rangle = q_0$ и дисперсией $D(q_l) = 2/l$, т.е. $q_l$ принимает положительное значение с вероятностью

$$P(q_l > 0) = \Phi\left(q_0 / \sqrt{D(q_l)}\right) = \Phi\left(q_0 \sqrt{l/2}\right)$$

(здесь $\Phi$ – функция распределения нормальной случайной величины). При больших отрицательных $q_0$, $P(q_l > 0) \approx (-q_0)^{-1}(\pi l)^{-1/2} \exp(-q_0^2 l/4)$ является малой величиной, и области возбуждения конвективного течения располагаются в пространстве достаточно разрежено, что может позволить наблюдать локализованные течения.

## 3.3. Показатель локализации

### 3.3.1. Показатели роста поля температуры

Рассмотрим стохастическую динамическую систему

$$\theta' = \psi, \quad \psi' = \phi, \quad \phi' = -[q_0 + \xi(x)]\psi - u\theta. \tag{3.45}$$

Спектр показателей Ляпунова данной системы состоит из трех элементов: $\gamma_1 \geq \gamma_2 \geq \gamma_3$. В силу инвариантности системы (3.45) относительно преобразования $(u, x, \theta, \psi, \phi) \to (-u, -x, \theta, -\psi, \phi)$,

$$\gamma_1(q_0, u) = -\gamma_3(q_0, -u) \quad \text{и} \quad \gamma_2(q_0, u) = -\gamma_2(q_0, -u). \tag{3.46}$$

Помимо прочего, из этих соотношений следует, что $\gamma_2(q_0, u = 0) = 0$; с другой стороны, при $u = 0$ система пропускает однородное решение $\{\theta, \psi, \phi\} = \{1, 0, 0\}$, которое и соответствует $\gamma_2 = 0$. Поскольку дивергенция фазового потока системы (3.45) равна нулю, имеем

$$\gamma_1 + \gamma_2 + \gamma_3 = 0 \quad \text{и} \quad \gamma_2(q_0, u) = -\gamma_1(q_0, u) + \gamma_1(q_0, -u). \qquad (3.47)$$

Таким образом, достаточно вычислять максимальный показатель Ляпунова $\gamma_1$. На Рис. 3.1 можно видеть зависимость спектра показателей Ляпунова от $u$ при $q_0 = 0$ как пример, демонстрирующий проявление симметрий (3.46), (3.47); зависимость $\gamma_1$ и $\gamma_2$ от $u$ и $q_0$ в широком диапазоне параметров приведена на Рис. 3.2.

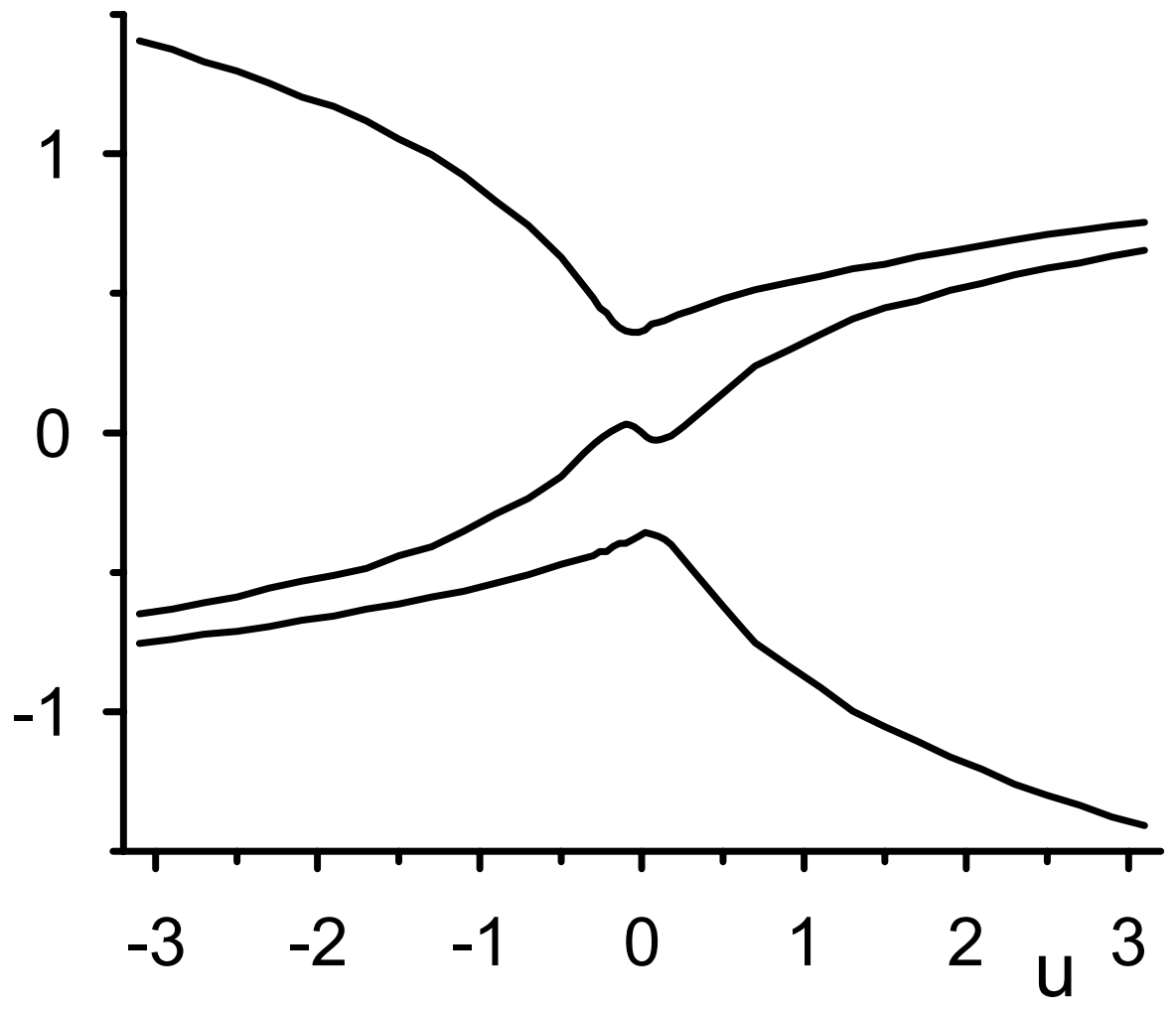

Рис. 3.1: Спектр показателей Ляпунова $\gamma_1 \geq \gamma_2 \geq \gamma_3$ при $q_0 = 0$.

При $u = 0,\ S = 0$ для течения, локализованного в окрестности $x_0$, на достаточном удалении от $x_0$

$$\theta(x) \approx \begin{cases} \Theta_{2,-} + \Theta_1(x) e^{\gamma_1(x - x_0)}, & \text{ï ðè } x < x_0; \\ \Theta_{2,+} + \Theta_3(x) e^{\gamma_3(x - x_0)}, & \text{ï ðè } x > x_0. \end{cases}$$

где $\Theta_1(x)$ и $\Theta_3(x)$ – ограниченные функции (не затухающие и не нарастающие на больших расстояниях), зависящие от конкретной реализации шума. Собирая эти асимптоты, можно выписать решение между двумя областями возбуждения конвективных течений, локализованными в $x_1$ и $x_2 > x_1$:

$$\theta(x_1 < x < x_2) \approx \Theta_3(x)e^{\gamma_3(x-x_1)} + \Theta_2 + \Theta_1(x)e^{\gamma_1(x-x_2)}, \qquad (3.48)$$

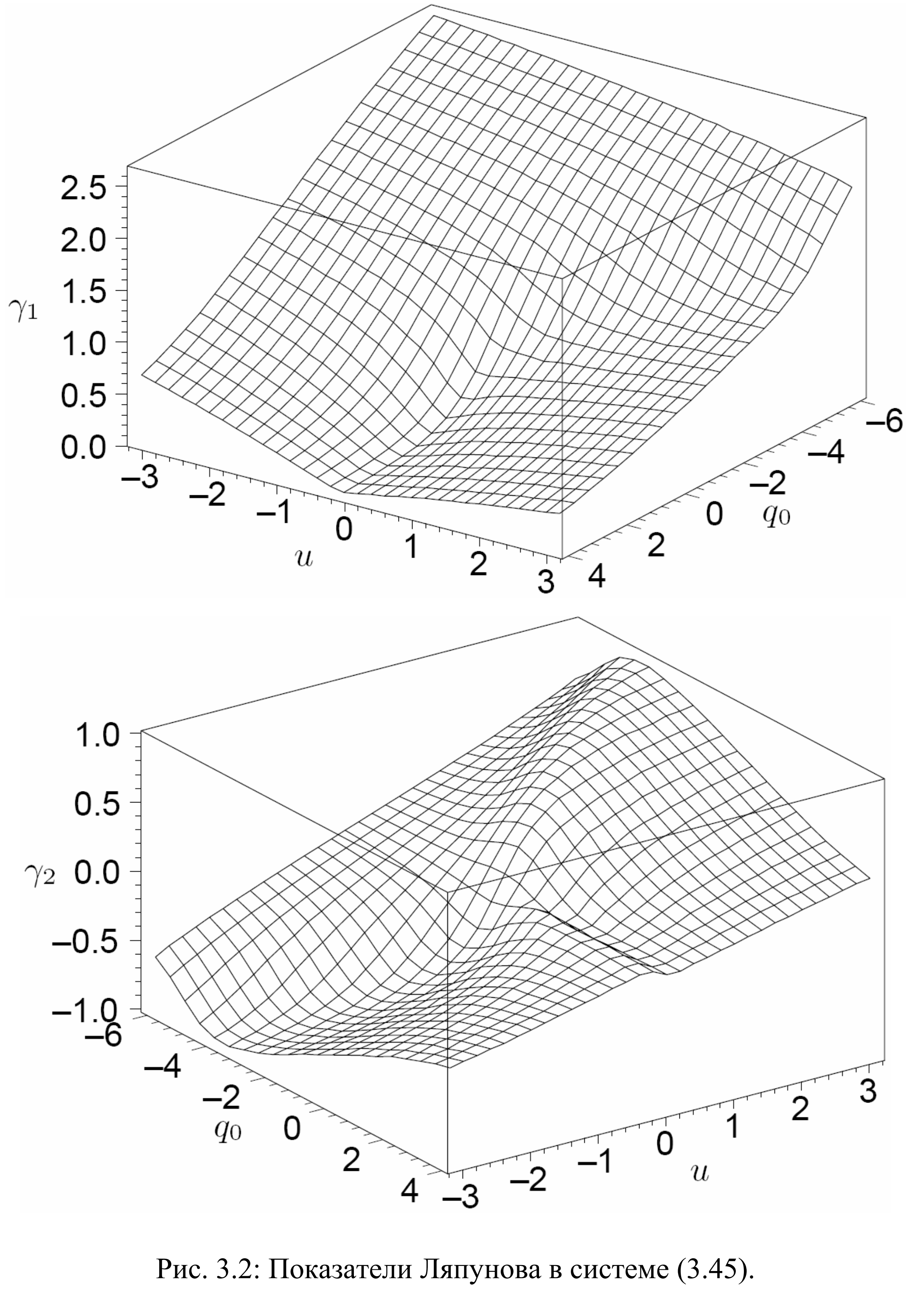

Рис. 3.2: Показатели Ляпунова в системе (3.45).

амплитуда соответствующей функции тока

$$\psi(x_1 < x < x_2) \approx \Psi_3(x)e^{\gamma_3(x-x_1)} + \Psi_1(x)e^{\gamma_1(x-x_2)} \tag{3.49}$$

($\Psi_{1,3}(x)$ ограничены подобно $\Theta_{1,3}(x)$) локализована в окрестности $x_1$ и $x_2$ с показателем $\gamma_1$ ($\gamma_3(q_0, u=0) = -\gamma_1(q_0, u=0)$), тогда как возмущения поля температуры не локализованы.

При $u > 0$ (случай $u < 0$ подобен и отдельного рассмотрения не требует) сдвигом начала отсчета температуры $\theta \to \theta + u^{-1}S$ всегда можно добиться отсутствия потока тепла: $S = 0$. Из требования $S = 0$ при $u \neq 0$ следует, что в области демпфирования течения, где $\psi \to 0$, и возмущение температуры $\theta \to 0$. В самом деле, решение (3.48) принимает вид

$$\theta(x_1 < x < x_2) \approx \Theta_3(x)e^{\gamma_3(x-x_1)} + \Theta_2(x)e^{\gamma_2(x-x_2)} + \Theta_1(x)e^{\gamma_1(x-x_2)} \tag{3.50}$$

(поскольку $\gamma_2(q_0 < q_*, u > 0) > 0$, где $q_*$ – небольшое отрицательное число, см. Рис. 3.2, связанная с этим показателем Ляпунова мода локализована в окрестности $x_2$), и при $u \neq 0$ вклад второго слагаемого вдали от центров возбуждения стремится к нулю. С другой стороны, теперь мода, связанная с $\gamma_2$, вносит вклад в амплитуду функции тока:

$$\psi(x_1 < x < x_2) \approx \Psi_3(x)e^{\gamma_3(x-x_1)} + \Psi_2(x)e^{\gamma_2(x-x_2)} + \Psi_1(x)e^{\gamma_1(x-x_2)} \tag{3.51}$$

В результате, при $\gamma_2 > 0$ ($u > 0$) течение локализовано по потоку (правый склон локализованного решения) с показателем $\gamma_3$: $\gamma_- = |\gamma_3|$. В левой же части окрестности области возбуждения в течении представлены две моды:

$$\psi(x < x_2) \approx \Psi_2(x)e^{\gamma_2(x-x_2)} + \Psi_1(x)e^{\gamma_1(x-x_2)}. \tag{3.52}$$

При умеренных значениях $u$, $\Psi_1(x)$ и $\Psi_2(x)$ соизмеримы, мода $\Psi_1(x)e^{\gamma_1(x-x_2)}$ при удалении от $x_2$ быстро "теряется" на фоне $\Psi_2(x)e^{\gamma_2(x-x_2)}$ и свойства лока-лизации определяются $\gamma_2$: $\gamma_+ = \gamma_2$. При $u = 0$ функция $\Psi_2(x) = 0$, и при малых $u$, по непрерывности, $\Psi_2(x)$ должно быть мало. Тогда течение (3.52) успевает существенно затухнуть в области, где еще доминирует мода, связанная с $\gamma_1$, и именно эта мода определяет характерную длину локализации: $\gamma_+ = \gamma_1$.

### 3.3.2. Показатели роста среднеквадратичных значений

Для приблизительной аналитической оценки ведущего показателя Ляпунова можно проанализировать поведение средних по реализациям шума значений полей (см. [62]; например, в работе [63] подобным образом оценивался показатель Ляпунова в стохастической системе, подобной (3.45)).

А именно, здесь полезен следующий результат из монографии [62], справедливый для систем линейных обыкновенных дифференциальных уравнений с шумом в коэффициентах:

$$\frac{dy_i}{dx} = L_{ij}(x)y_j + \eta(x)\varGamma_{ij}(x)y_j, \tag{3.53}$$

где $\eta(x)$ – $\delta$-коррелированный шум с кумулянтными функциями

$$K_n(x_1, x_2, ..., x_n) = K_n(x_1)\delta(x_1 - x_2)\delta(x_2 - x_3)...\delta(x_{n-1} - x_n)$$

(гауссовость, соответствующая $K_n(x) = 2D(x)\delta_{n,2}$, на данный момент не требуется). Динамика средних по реализациям шума $\langle y_i \rangle$ подчиняется системе уравнений:

$$\frac{d}{dx}\langle y_i \rangle = L_{ij}(x)\langle y_i \rangle + \sum_{n=1}^{\infty} \frac{1}{n!} K_n(x)[\mathbf{\Gamma}^n(x)]_{ij} \langle y_i \rangle. \tag{3.54}$$

Если матрица $\mathbf{\Gamma}$ разрежена (что довольно распространено в реальных ситуациях), может оказываться, что, начиная с некоторой степени, $\mathbf{\Gamma}^n = 0$, и система уравнений (3.54) замкнута относительно кумулянтных функций.

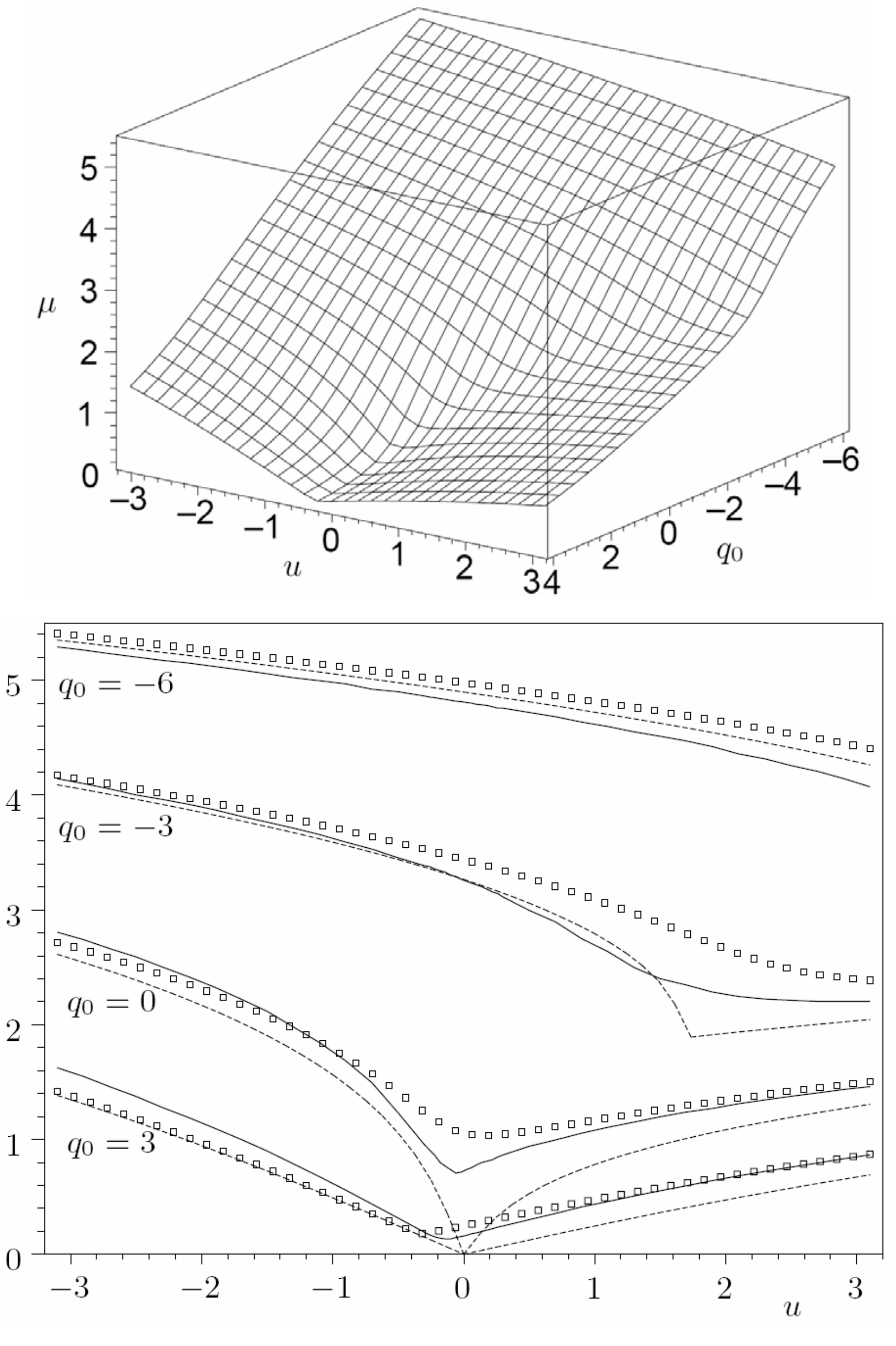

Рис. 3.3: Показатели роста в системе (3.45). На нижнем графике: сплошная линия – удвоенный максимальный показатель Ляпунова $2\gamma_1$, квадраты – показатель роста среднеквадратичных $\mu$, и пунктирная линия – удвоенный максимальный показатель Ляпунова системы (3.45) в отсутствии шума.

В рассматриваемом здесь случае стационарного (в статистическом смысле) гауссовского шума единичной интенсивности $K_n(x) = 2\delta_{n,2}$ и система (3.54) принимает вид

$$\frac{d}{dx}\langle y_i\rangle = [\mathrm{L} + \mathbf{\Gamma}^2]_{ij}\langle y_i\rangle. \tag{3.55}$$

Если рассматривать $\vec{y} = \{\theta, \psi, \phi\}$, то (см. (3.45))

$$\mathbf{\Gamma} = \begin{bmatrix} 0 & 0 & 0 \\ 0 & 0 & 0 \\ 0 & -1 & 0 \end{bmatrix}, \quad \mathbf{\Gamma}^2 = 0,$$

и шум себя не проявляет. Это связано с симметрией системы и мало говорит о ее поведении при конкретной реализации шума – для анализа того, что имеет место при конкретной реализации шума, следует рассматривать поведение среднеквадратичных значений (см. [62]).

Для $\vec{y} = \{\theta^2, \theta\psi, \theta\phi, \psi^2, \psi\phi, \phi^2\}$ уравнения (3.45) дают

$$\begin{aligned}
\frac{dy_1}{dx} &= 2\theta\frac{d\theta}{dx} = 2y_2, \\
\frac{dy_2}{dx} &= \frac{d\theta}{dx}\psi + \theta\frac{d\psi}{dx} = y_4 + y_3, \\
\frac{dy_3}{dx} &= \frac{d\theta}{dx}\phi + \theta\frac{d\phi}{dx} = y_5 - [q_0 + \xi(x)]y_2 - uy_1, \\
\frac{dy_4}{dx} &= 2\psi\frac{d\psi}{dx} = 2y_5, \\
\frac{dy_5}{dx} &= \frac{d\psi}{dx}\phi + \psi\frac{d\phi}{dx} = y_6 - [q_0 + \xi(x)]y_4 - uy_2, \\
\frac{dy_6}{dx} &= 2\phi\frac{d\phi}{dx} = -2[q_0 + \xi(x)]y_5 - 2uy_3.
\end{aligned}$$

Тогда

$$\mathbf{\Gamma} = \begin{bmatrix} 0 & 0 & 0 & 0 & 0 & 0 \\ 0 & 0 & 0 & 0 & 0 & 0 \\ 0 & -1 & 0 & 0 & 0 & 0 \\ 0 & 0 & 0 & 0 & 0 & 0 \\ 0 & 0 & 0 & -1 & 0 & 0 \\ 0 & 0 & 0 & 0 & -2 & 0 \end{bmatrix} \qquad (3.56)$$

и

$$\mathbf{A} \equiv \mathbf{L} + \mathbf{\Gamma}^2 = \begin{bmatrix} 0 & 2 & 0 & 0 & 0 & 0 \\ 0 & 0 & 1 & 1 & 0 & 0 \\ -u & -q_0 & 0 & 0 & 1 & 0 \\ 0 & 0 & 0 & 0 & 2 & 0 \\ 0 & -u & 0 & -q_0 & 0 & 1 \\ 0 & 0 & -2u & 2 & -2q_0 & 0 \end{bmatrix}. \qquad (3.57)$$

Показателем роста среднеквадратичных значений $\mu$ является максимальное вещественное (среднеквадратичная величина не может колебаться) собственное значение матрицы $\mathbf{A}$, имеющее физический смысл. Формальное собственное решение имеет физический смысл, если для него соблюдены условия $\left\langle\theta^2\right\rangle \geq 0$, $\left\langle\psi^2\right\rangle \geq 0$, $\left\langle\phi^2\right\rangle \geq 0$, $\left\langle\theta^2\right\rangle\left\langle\psi^2\right\rangle \geq \langle\theta\psi\rangle^2$, $\left\langle\theta^2\right\rangle\left\langle\phi^2\right\rangle \geq \langle\theta\phi\rangle^2$ и $\left\langle\psi^2\right\rangle\left\langle\phi^2\right\rangle \geq \langle\psi\phi\rangle^2$.

Хотя характеристическое уравнение матрицы $\mathbf{A}$ имеет шестую степень относительно собственных значений $\mu$, оно имеет лишь вторую степень относительно $u$, и поверхность максимального вещественного значения $\mu(q_0, u)$ может быть найдена в параметрическом виде – она состоит из двух частей:

$$u_{1,2} = \frac{1}{16}\Big(7\mu^3 + 4q_0\mu - 8 \pm \\ \pm\sqrt{81\mu^6 + 216q_0\mu^3 - 240\mu^3 + 144q_0^2m^2 - 192q_0\mu + 64}\Big), \quad (3.58)$$

определенных при $\mu > 0$, $q_0 \geq \frac{2\mu}{3} - \frac{3\mu^2}{4} + \sqrt{\frac{2\mu}{3}}$. На Рис. 3.3 можно видеть эту поверхность, а также оценить степень согласования между показателями роста $2\gamma_1$ и $\mu$ (которые не обязаны совпадать по определению).

Здесь уместно вновь акцентировать внимание на различии между показателями Ляпунова $\gamma_i$ и показателем роста среднеквадратичных $\mu$. А именно, стоит повторить, что показатель Ляпунова описывает асимптотическую эволюцию возмущений на больших расстояниях при конкретной реализации шума и, в этом плане, соответствует осреднению по состояниям системы (или вдоль ее траектории). В линейной задаче это явно не проявляется благодаря справедливости принципа суперпозиции для произвольных возмущений, но в нелинейной задаче это сказывалось бы более существенным образом. Именно такие характеристики (показатели Ляпунова) описывают локализацию течений в конкретной гидродинамической системе с заданным случайно неоднородным нагревом. В тоже время, показатели роста $\mu$ описывают поведение средних по реализациям шума значений полей, что не в точности соответствует наблюдаемым в конкретной задаче явлениям.

Также примечательно, что для матрицы $\mathbf{\Gamma}$, определяемой (3.56), оказывается $\mathbf{\Gamma}^3 = 0$ (это же справедливо и для всех последующих степеней матрицы $\mathbf{\Gamma}$). Таким образом, для среднеквадратичных по реализациям шума значений полей гауссовость шума не важна: в силу указанной вырожденности матрицы $\mathbf{\Gamma}$ и согласно формуле (3.54) поведение среднеквадратичных значений при любом $\delta$-коррелированном шуме определяется только интенсивностью этого шума.

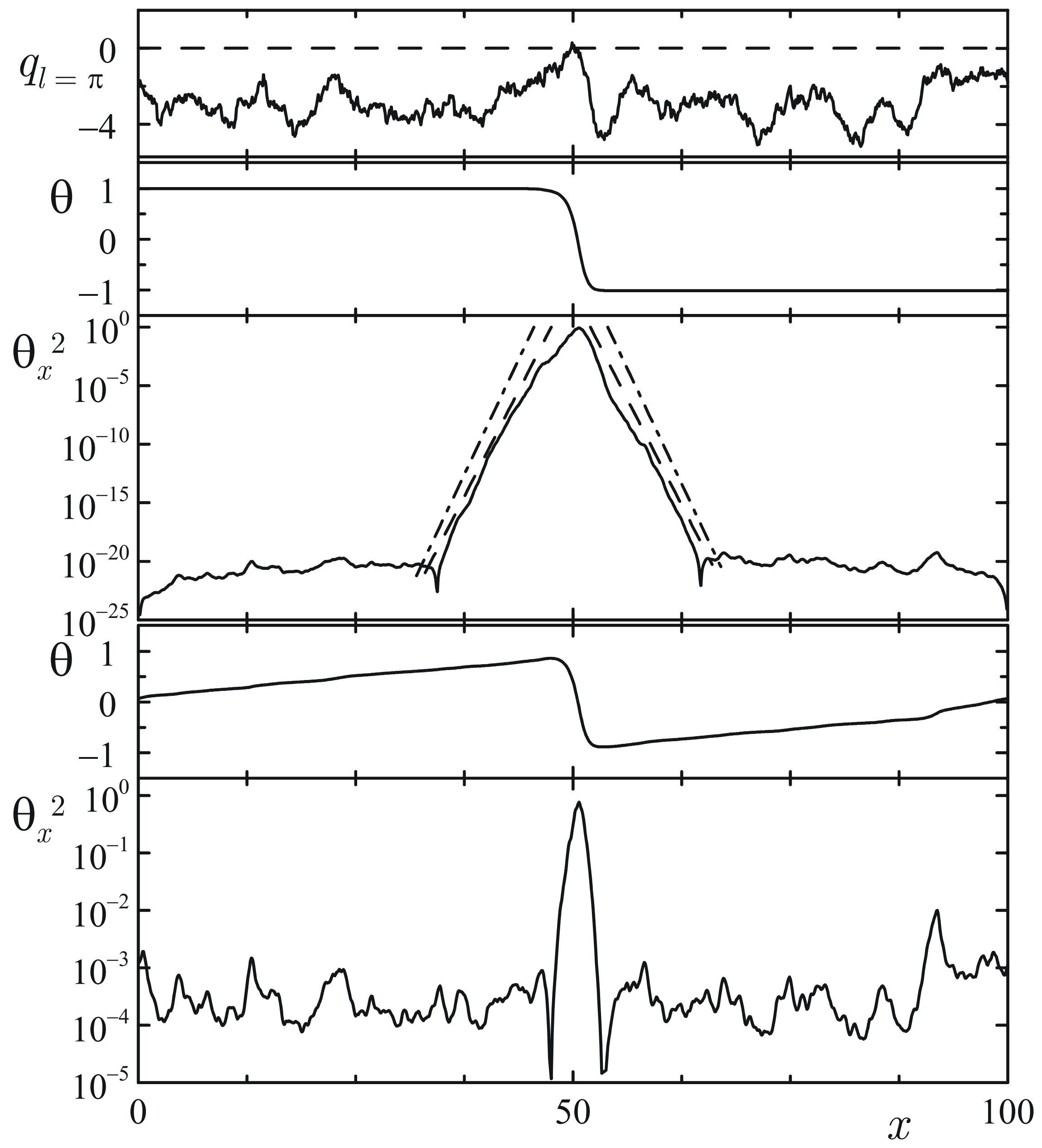

Рис. 3.4: Примеры стационарных решений уравнения (3.43) в отсутствии прокачивания при $q_0 = -3.1$ (реализации $q(x)$ представлены графиками $q_{l=\pi}(x)$). Боковые границы: адиабатические непроницаемые для решения, изображенного выше, и периодические – для нижнего. Штриховая линия отмечает наклон, соответствующий затуханию решения с показателем $2\gamma_1 = 3.332$, а штрихпунктирная – $\mu = 3.623$.

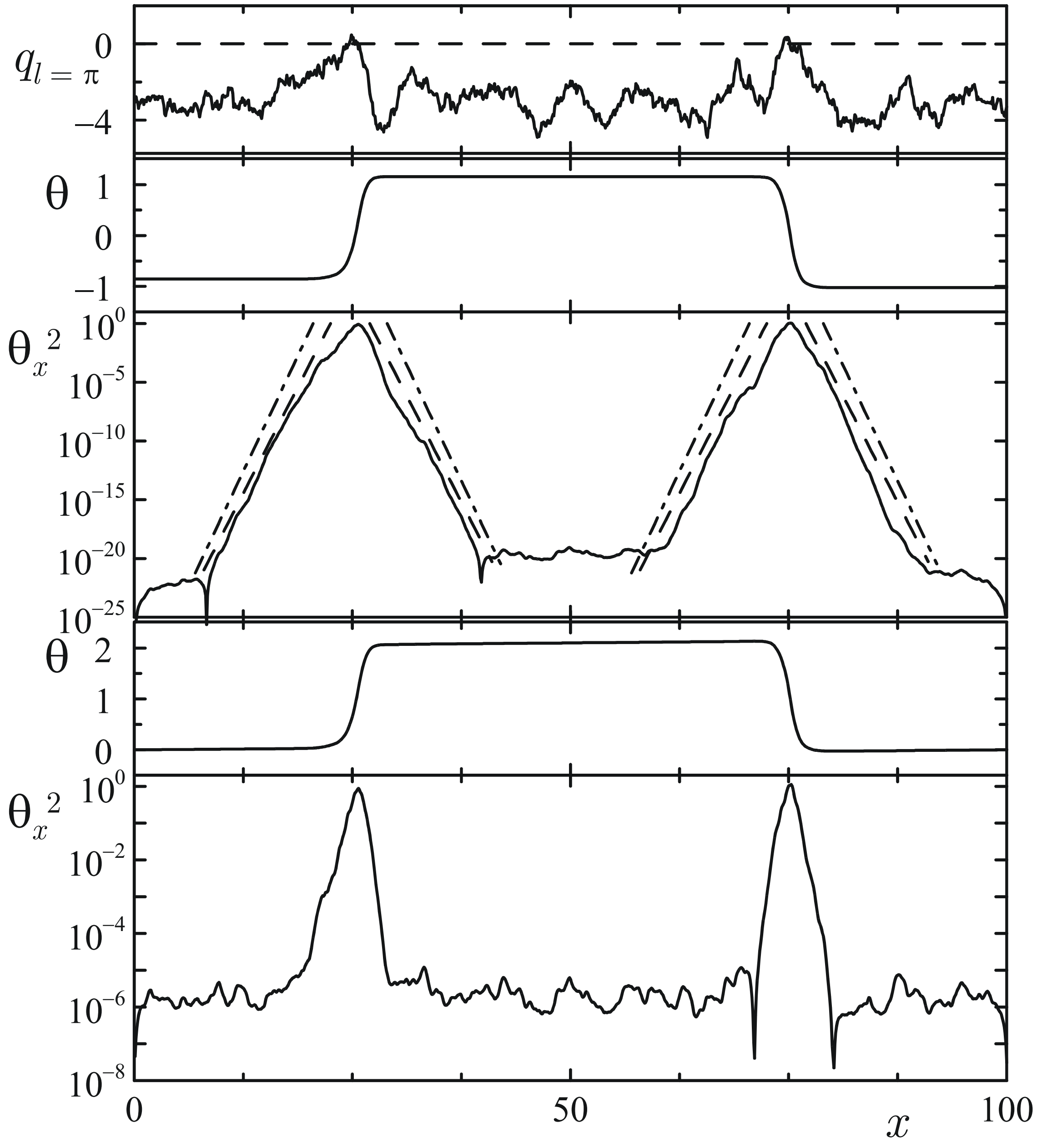

Рис. 3.5: Примеры стационарных решений уравнения (3.43) в отсутствии прокачивания при $q_0 = -3.1$ (реализации $q(x)$ представлены графиками $q_{l=\pi}(x)$). Боковые границы: адиабатические непроницаемые для решения, изображенного выше, и изотермические – для нижнего. Штриховая линия отмечает наклон, соответствующий затуханию решения с показателем $2\gamma_1 = 3.332$, а штрихпунктирная – $\mu = 3.623$.

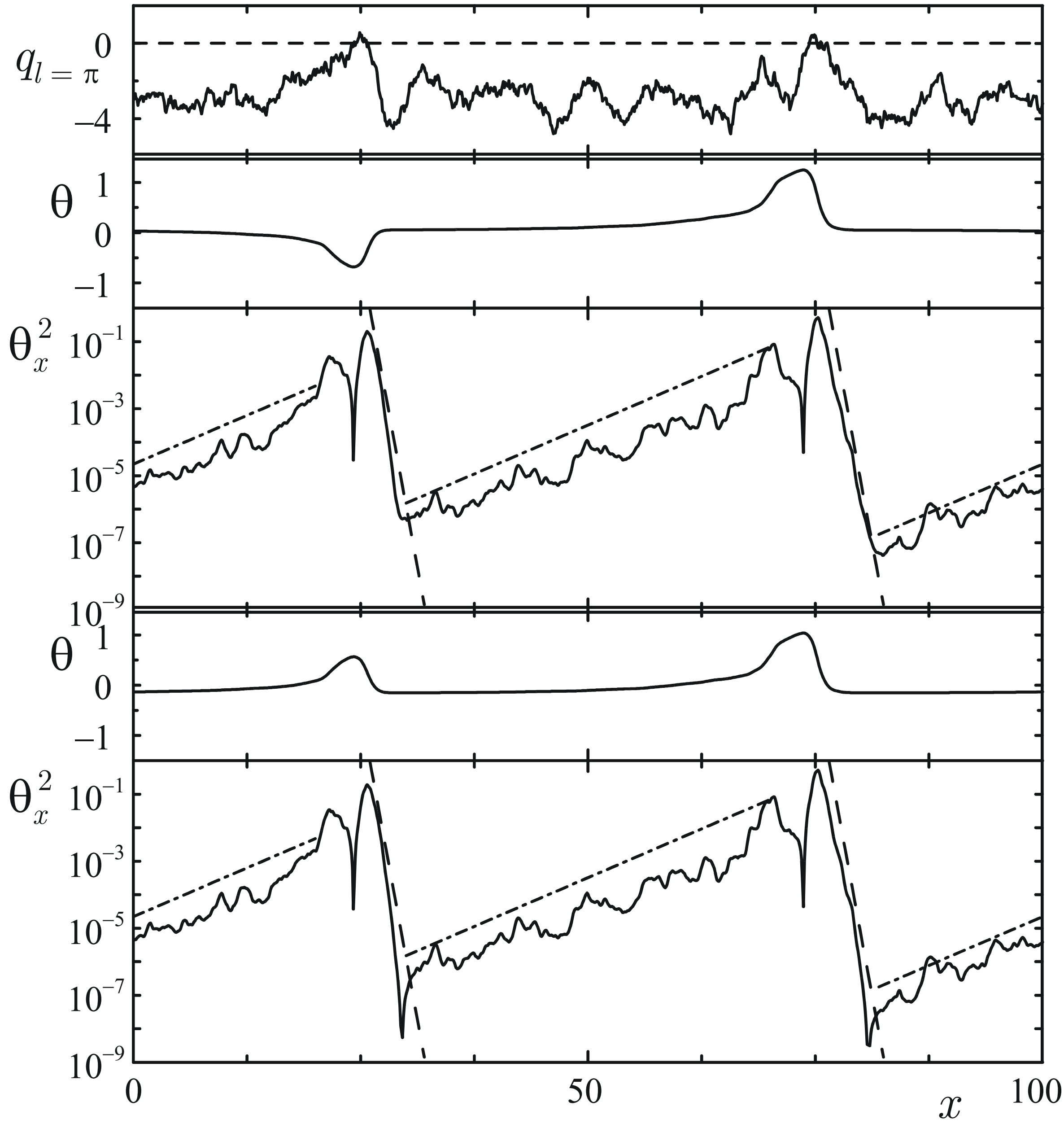

Рис. 3.6: Примеры стационарных решений уравнения (3.43) при $u = 0.3$, $q_0 = -3.0$ (реализации $q(x)$ представлены графиками $q_{l=\pi}(x)$) в слое с периодическими граничными условиями. Штриховая линия отмечает наклон, соответствующий затуханию решения с показателем $2\gamma_1 = 3.394$, а штрихпунктирная – $2\gamma_2 = 0.268$.

## 3.4. Решения нелинейной задачи

Хотя свойства локализации в стационарном линейном уравнении (3.44) при $u = 0$ хорошо известны, поведение гидродинамической задачи (3.43) в отсутствии прокачивания заслуживает внимания в силу упомянутых ранее принципиальных различий между данной системой и уравнением Шредингера.

В соответствии с ожиданиями относительно наблюдаемости локализованных решений при $q_0 < 0$, на Рис. 3.4 можно видеть примеры локализованных решений при $u = 0$. При адиабатических боковых границах (Рис. 3.4, верхнее решение) течение локализовано в окрестности области, где $q_{l=\pi}$ спонтанно принимает положительные значения, и демпфируется за ее пределами в соответствии с предсказаниями линейной теории (нерегулярный фон на уровне $\theta_x^2 \sim 10^{-25}$ связан с погрешностью численного счета). Адиабатические граничные условия для ограниченной в горизонтальном направлении области соответствуют большому (по сравнению с горизонтальными размерами области, где производится численный счет) расстоянию между областями возбуждения течений в бесконечном слое. При $q_0 = -3.1$, $P(q_{l=\pi} > 0) \approx 5 \times 10^{-5}$, и эти области, действительно, должны располагаться в пространстве достаточно редко. Следует отметить, что использованная оценка очень груба и не подразумевает достоверного определения характерных расстояний между центрами возбуждения течений, а лишь позволяет судить о степени их редкости.

При периодических граничных условиях (Рис. 3.4, нижнее решение) эффективно проявляется конечность расстояния между областями возбуждения. В отсутствии шума система (3.43) пропускает тривиальное решение в виде $\theta = q_0^{-1} S\,x$. При наличии шума с этой модой связано стационарное решение $\theta' = q_0^{-1} S + \psi_1$, $\langle \psi_1 \rangle = 0$, которое при малых $S$ подчиняется линейному уравнению

$$\frac{d^2\psi_1}{dx^2} + [q_0 + \xi(x)]\psi_1 = -q_0^{-1} S\,\xi(x),$$

(здесь $\langle \xi(x)\psi_1 \rangle = 0$), т.е. $\psi_1 \sim S$. При наличии шума это решение реализуется в области демпфирования течения ($q_l < 0$). В ситуации, изображенной на Рис. 3.4, $q_0^{-1}S = [\theta]/L$, где $[\theta] \approx 2$ – перепад температур, навязанный нелинейным течением в области, где $q_l > 0$, а $L \approx 100$ – период структуры, и $q_0^{-2}S^2 \approx 4 \times 10^{-4}$ – именно такой порядок интенсивности переходного течения и имеет место вдали от центра возбуждения и не позволяет наблюдать характер затухания "хвостов" интенсивного локализованного течения. Для стационарного течения поток тепла $S$ постоянен вдоль слоя, в силу чего среднее переходное течение $\langle \psi \rangle = \langle \theta' \rangle = q_0^{-1}S$ остается постоянным по всему слою даже при наличии нескольких областей возбуждения конвективного течения.

Примечательно, что связанный с переходным течением поток тепла $S$ при адиабатических граничных условиях должен затухать до нуля. Поэтому в слое с адиабатическими вертикальными границами стационарные локализованные решения реализуются без переходного течения при произвольном количестве областей возбуждения (Рис. 3.5, верхнее решение). В свою очередь, при изотермических вертикальных границах интенсивность фонового течения тем меньше, чем протяженней слой. Дело в том, что в системе имеется мультистабильность относительно знака скачка поля температуры при переходе через область интенсивного течения и при большом количестве скачков они в среднем друг друга компенсируют, что приводит к небольшому значению $q_0^{-1}S = L_{\text{layer}}^{-1}\sum_j [\theta]_j$, где просуммированы скачки температуры по всем областям возбуждения конвективных течений (например, на Рис. 3.5 фоновое течение существенно меньше, чем на Рис. 3.4).

Примеры течений при $u \neq 0$ можно видеть на Рис. 3.6. В соответствии с предсказаниями раздела 3.3.1 температура по обе стороны от области интенсивного течения одинакова: локализованы не только течения, но и возмущения поля температуры. И, что более примечательно, свойства локализации против потока определяются показателем $\gamma_2$ даже при довольно малых $u$ и, соответственно, $\gamma_2$, т.е. могут очень сильно изменяться.

Примечательно также, что горизонтальное прокачивание может приводить к подавлению конвективных течений. Причем, чем больше $u$, тем сильнее стабилизирующая роль прокачивания. Однако, этот эффект не связан с рассматриваемыми в работе эффектами локализации, и потому остается за рамками настоящего исследования.

# Глава 4: Синхронизация нелинейных систем общим шумом

В теории стохастических процессов широко распространены ситуации, требующие осреднения по реализациям шума. При этом имеется два классических круга задач, связанных с осреднением вдоль траектории системы при заданной реализации шума. Количественное описание явлений, наблюдаемых в этих задачах, в той или иной степени фактически связано с вычислением показателей Ляпунова. Для фигурирующих в таких задачах стохастических систем показатели Ляпунова определяются в том смысле, что они описывают эволюцию малых возмущений траектории системы при заданной реализации шума, т.е. описывают устойчивость отклика не к возмущениям шума, а к возмущениям состояния системы.

Первый круг задач связан с явлениями локализации в распределенных системах со случайной пространственной модуляцией параметров – задача такого типа рассматривается в предыдущей главе. Второй круг задач упомянутого типа связан с синхронизацией нелинейных систем общим шумом, и некоторые фундаментальные результаты относительно этого явления представлены в данной главе диссертации.

Основным эффектом при воздействии шума на периодические автоколебания является появление диффузии фазы: автоколебания становятся неидеальными [74]. Однако шум может играть и упорядочивающую роль, в частности синхронизовать автоколебания. Если на две одинаковые (или слабо отличающиеся) автоколебательные системы действует общий шум, то их состояния могут под действием этого шума синхронизоваться. Этот эффект определяется знаком максимального показателя Ляпунова, при периодических автоколебаниях он отвечает направлению вдоль предельного цикла. Для автономных систем он нулевой, и синхронизации в описанном выше смысле нет. Под действием шума показатель Ляпунова может стать отрицательным, что означает синхронизацию. Рассмотрению влияния белого гауссовского шума на динамические системы общего положения с предельным циклом и посвящена данная глава. Некоторые результаты воспроизведены для телеграфного шума.

Используемый для аналитического исследования [99,100] подход основан на сведении динамики автоколебательной системы к уравнению для фазы. Это оправдано при малом (в нужном смысле) внешнем воздействии. Выводится стохастическое уравнение для фазы автоколебаний, и находится стационарное распределение фазы. Показатель Ляпунова выражается интегралом по этому распределению. В частности, показывается, что при малой интенсивности шума показатель всегда отрицателен, что отвечает синхронизации.

При конечной интенсивности шума аналитическое исследование поведения систем в общем случае оказывается невозможно и требуется численный счет, который для некоторых систем обнаруживает возможность появления положительного показателя Ляпунова [101,103], т.е. десинхронизации колебаний.

В упомянутых случаях отдельное внимание уделяется неидеальным ситуациям [101,103]: когда имеется некоторая неидентичность в шуме, действующем на различные системы (внутренний шум), или в параметрах систем. При малом шуме и малых неидентичностях задачи допускают аналитическое исследование.

В основной части исследования рассматривается белый гауссовский шум [99,100,101,103]. По своей природе телеграфный шум существенно отличается от белого гауссовского и представляет интерес с точки зрения выяснения того, насколько существенны специфические свойства последнего, и будет ли ситуация аналогична для шумов с другими свойствами. Кроме того, телеграфный шум интересен также в связи с тем, что такому сигналу может быть очевидным образом сопоставлен ступенчатый периодический сигнал. Сравнение эффектов таких сигналов может помочь в понимании механизмов синхронизации общим шумом, и это сравнение произведено [102].

## 4.1. Системы с предельным циклом – показатель Ляпунова

Автоколебательные системы – это динамические системы, имеющие устойчивый предельный цикл, движению по которому и соответствуют автоколебания. Далее удобно говорить о таких системах именно как о системах с устойчивым предельным циклом.

Будем исходить из общих стохастических уравнений, описывающих динамику $N$-мерной колебательной системы $x_j$, $j = 1,2,\ldots,N$, в присутствии некоррелированных между собой шумовых воздействий $\xi_k(t)$ с амплитудами $Q_{jk}$, где $k = 1,2,\ldots,M \leq N$:

$$\dot{x}_j = F_j(\mathbf{x}) + \sum_{k=1}^{M} Q_{jk}(\mathbf{x})\,\xi_k(t). \qquad (4.1)$$

Свойства шумов будут уточнены ниже.

Если в динамической системе без шума существует цикл $\mathbf{x}^0(t) = \mathbf{x}^0(t + 2\pi/\omega_0)$, состояния на нем могут быть параметризованы посредством фазы $\varphi(\mathbf{x}^0)$ [92], линейно растущей со временем: $\dot{\varphi} = \omega_0$. Для предельного цикла фаза $\varphi$ может быть введена без неоднородной перенормировки времени (которая нежелательна в случае, если не ограничиваться $\delta$-коррелированными шумами) и в конечной его окрестности, где также будет справедливо соотношение $\dot{\varphi} = \omega_0$. С учетом шума уравнение эволюции фазы в малой окрестности цикла примет вид

$$\dot{\varphi} = \omega_0 + \sum_{j=1}^{N}\sum_{k=1}^{M} \frac{\partial\varphi(\mathbf{x})}{\partial x_j} Q_{jk}(\mathbf{x})\bigg|_{\mathbf{x}=\mathbf{x}^0(\varphi)} \xi_k(t). \qquad (4.2)$$

Эволюция системы будет происходить в малой окрестности цикла в двух случаях: либо при малой интенсивности шума, либо при большом по модулю отрицательном ведущем ляпуновском показателе цикла и конечных интенсивностях шума. Не изменяя характера шума (а лишь перенормируя его количественные характеристики), однородная нормировка времени $\omega_0 t \to t$ позволяет сделать частоту цикла равной $1$. Вид стохастического уравнения для фазы существенно зависит от того, как действует шум в исходной системе (4.1). Если в системе имеется единственный независимый шумовой сигнал, т.е. только для одного $k$ шумовое воздействие $\xi_k \neq 0$, получится

$$\dot{\varphi} = 1 + \varepsilon f(\varphi)\xi(t), \tag{4.3}$$

где $\xi(t)$ будет полагаться $\delta$-коррелированным гауссовским шумом с $\langle \xi(t) \rangle = 0$, нормированным так, что $\langle \xi(t)\xi(t+t') \rangle = 2\delta(t')$, параметр $\varepsilon$ описывает интенсивность шума, причем в результате нормировки времени $\varepsilon \sim \omega_0^{-1/2}$, а $f(\varphi)$ – нормированная периодическая функция фазы, описывающая чувствительность системы к шуму: $f(\varphi) = f(\varphi + 2\pi)$, $\int_0^{2\pi} f^2(\varphi) d\varphi = 2\pi$. Более сложное уравнение получается, если шумовой сигнал в исходной системе содержит несколько независимых компонент (далее упоминается как случай многокомпонентного шума):

$$\dot{\varphi} = 1 + \sum_{k=1}^{M} \varepsilon_k f_k(\varphi)\xi_k(t), \quad \langle \xi_i(t)\xi_k(t+t') \rangle = 2\delta(t')\delta_{ik}. \tag{4.4}$$

Целью исследования в данном разделе является аналитический анализ устойчивости решений стохастических уравнений (4.3),(4.4). Для этого можно рассматривать линеаризованное уравнение (4.3) для малого отклонения фазы $\alpha$:

$$\dot{\alpha} = \varepsilon \alpha f'(\varphi)\xi(t), \tag{4.5}$$

где штрих означает производную по аргументу. Ляпуновский показатель $\lambda$, задающий среднюю скорость экспоненциального роста отклонения $\alpha$, определяется усреднением соответствующей мгновенной скорости:

$$\lambda = \left\langle \frac{d \ln \alpha}{dt} \right\rangle = \left\langle \varepsilon f'(\varphi)\xi(t) \right\rangle. \tag{4.6}$$

Для многокомпонентного шума выражение для показателя Ляпунова имеет вид

$$\lambda = \sum_{k=1}^{M} \left\langle \varepsilon_k f_k'(\varphi)\, \xi_k(t) \right\rangle. \tag{4.7}$$

Следует отметить, что показатель Ляпунова определяет асимптотическое поведение малых возмущений решения и в данной задаче будет описывать, сближаются или расходятся близкие состояния системы со временем в ходе ее эволюции. При этом, конечно же, в отдельные моменты времени близкие состояния могут друг от друга удаляться при отрицательном показателе Ляпунова, что не препятствует их итоговому сближению, а при положительном показателе Ляпунова, наоборот, часть времени состояния могут расходиться.

### 4.1.1. Уравнение Фоккера–Планка и его стационарное решение

Уравнение Фоккера–Планка для стохастического уравнения (4.3), понимаемого в смысле Стратоновича, записывается стандартным образом [94,62]:

$$\frac{\partial W(\varphi,t)}{\partial t} + \frac{\partial}{\partial \varphi}\left[ W(\varphi,t) - \varepsilon^2 f(\varphi) \frac{\partial}{\partial \varphi}\left[ f(\varphi)\, W(\varphi,t) \right] \right] = 0. \tag{4.8}$$

В стационарном режиме поток вероятности $S$ постоянен:

$$W(\varphi) - \varepsilon^2 f(\varphi) \frac{d}{d\varphi}\left[ f(\varphi)\, W(\varphi) \right] = S. \tag{4.9}$$

Это позволяет найти решение в квадратурах, которое при периодических граничных условиях для плотности вероятности $W(\varphi)$ имеет вид

$$W(\varphi) = C\int_{\varphi}^{\varphi+2\pi} \frac{d\psi}{f(\varphi)f(\psi)} \exp\left(-\frac{1}{\varepsilon^2}\int_{\varphi}^{\psi} \frac{d\theta}{f^2(\theta)}\right), \tag{4.10}$$

где $C$ определяется из условия нормировки распределения:

$$C^{-1} = \int_0^{2\pi} d\varphi \int_{\varphi}^{\varphi+2\pi} \frac{d\psi}{f(\varphi)f(\psi)} \exp\left(-\frac{1}{\varepsilon^2}\int_{\varphi}^{\psi} \frac{d\theta}{f^2(\theta)}\right) \tag{4.11}$$

– а поток вероятности задается выражением

$$S = \left[1 - \exp\left(-\frac{1}{\varepsilon^2}\int_0^{2\pi} \frac{d\theta}{f^2(\theta)}\right)\right] C\,. \tag{4.12}$$

Для многокомпонентного шума аналогично можно получить

$$\frac{\partial W(\varphi,t)}{\partial t} + \frac{\partial}{\partial\varphi}\left[W(\varphi,t) - \sum_{k=1}^{M} \varepsilon_k^2 f_k(\varphi) \frac{\partial}{\partial\varphi}\Big[f_k(\varphi)\, W(\varphi,t)\Big]\right] = 0\,. \tag{4.13}$$

Интересно, что это уравнение эквивалентно его однокомпонентному варианту (4.8), если положить

$$f^2(\varphi) = \sum_{k=1}^{M} \varepsilon_k^2 f_k^2(\varphi) \Big/ \sum_{k=1}^{M} \varepsilon_k^2 \quad \text{и} \quad \varepsilon^2 = \sum_{k=1}^{M} \varepsilon_k^2\,. \tag{4.14}$$

Таким образом, выписанное выше стационарное решение (4.10) справедливо и в этом случае.

### 4.1.2. Показатель Ляпунова

Для вычисления показателя Ляпунова (4.6), (4.7) требуется найти средние вида $\left\langle F(\varphi)\xi(t)\right\rangle$. Такие средние для стохастических уравнений (4.3),(4.4) с $\delta$-коррелированным шумом вычисляются стандартным способом [62]:

$$\left\langle F(\varphi)\xi(t)\right\rangle = \varepsilon\left\langle F'(\varphi)f(\varphi)\right\rangle. \tag{4.15}$$

Записывая усреднение как интеграл по равновесной плотности распределения фазы, для показателя Ляпунова можно получить

$$\begin{aligned} \lambda &= \varepsilon^2\left\langle f''(\varphi)f(\varphi)\right\rangle = \\ &= \varepsilon^2 C\int\limits_0^{2\pi} d\varphi \int\limits_{\varphi}^{\varphi+2\pi} d\psi\, \frac{f''(\varphi)}{f(\psi)} \exp\left(-\frac{1}{\varepsilon^2}\int\limits_{\varphi}^{\psi}\frac{d\theta}{f^2(\theta)}\right). \end{aligned} \tag{4.16}$$

Для многокомпонентного шума аналогично получается

$$\lambda = \sum_{k=1}^{M}\varepsilon_k^2\int_0^{2\pi} f_k''(\varphi)f_k(\varphi)W(\varphi)\,d\varphi. \tag{4.17}$$

Прежде чем перейти к анализу полученных формул в конкретных ситуациях, стоит отметить, что в пределе слабого шума показатель Ляпунова всегда отрицателен: в ведущем порядке по $\varepsilon$

$$\lambda \approx -\sum_{k=1}^{M}\frac{\varepsilon_k^2}{2\pi}\int_0^{2\pi}\left[f_k'(\varphi)\right]^2 d\varphi, \tag{4.18}$$

где $M$ может принимать и значение $1$, что соответствует однокомпонентному шуму.

### 4.1.3. Пример: линейно поляризованный однородный шум

Если в исходной системе шум аддитивный и действует только на одну компоненту, а предельный цикл близок к окружности, то получается одноком-понентное стохастическое уравнение с $f(\varphi)=\sqrt{2}\sin\varphi$. Здесь также предположено, что скорость фазового потока на предельном цикле примерно постояннна. В этом случае

$$\int\limits_{\varphi}^{\varphi+2\pi}\frac{d\psi}{f(\psi)}\exp\left(-\frac{1}{\varepsilon^2}\int\limits_{\varphi}^{\psi}\frac{d\theta}{f^2(\theta)}\right)=\int\limits_{\varphi}^{\pi+\pi\left[\frac{\varphi}{\pi}\right]}\frac{d\psi}{\sqrt{2}\sin\psi}\exp\left(\frac{\operatorname{ctg}\psi-\operatorname{ctg}\varphi}{2\varepsilon^2}\right),$$

где квадратные скобки обозначает целую часть заключенного в них выражения. Для выбранного вида функции $f(\varphi)$ распределение имеет период $\pi$ и при $\varphi\in[0,\pi)$

$$W(\varphi)=\frac{C}{2}\int\limits_{\varphi}^{\pi}\frac{d\psi}{\sin\psi\sin\varphi}\exp\left(\frac{\operatorname{ctg}\psi-\operatorname{ctg}\varphi}{2\varepsilon^2}\right),$$

$$S=C=\left(\int\limits_{-\infty}^{+\infty}dy\int\limits_{-\infty}^{y}\frac{dx}{\sqrt{1+x^2}\sqrt{1+y^2}}\exp\left(\frac{x-y}{2\varepsilon^2}\right)\right)^{-1},$$

$$\lambda=-\frac{\varepsilon^2 C}{2}\left(\int\limits_{-\infty}^{+\infty}dy\int\limits_{-\infty}^{y}\frac{dx}{(1+x^2)^{1/2}(1+y^2)^{3/2}}\exp\left(\frac{x-y}{2\varepsilon^2}\right)\right),\qquad(4.19)$$

Для перехода к последним формулам введены переменные $x=\operatorname{ctg}\psi$ и $y=\operatorname{ctg}\varphi$ (переход нужен для того, чтобы сделать очевидной сходимость интегралов). Дальнейшее упрощение формул не представляется возможным, однако они уже достаточно удобны для построения численного решения. График зависимости искомого показателя Ляпунова от интенсивности шума представлен на Рис. 4.1.

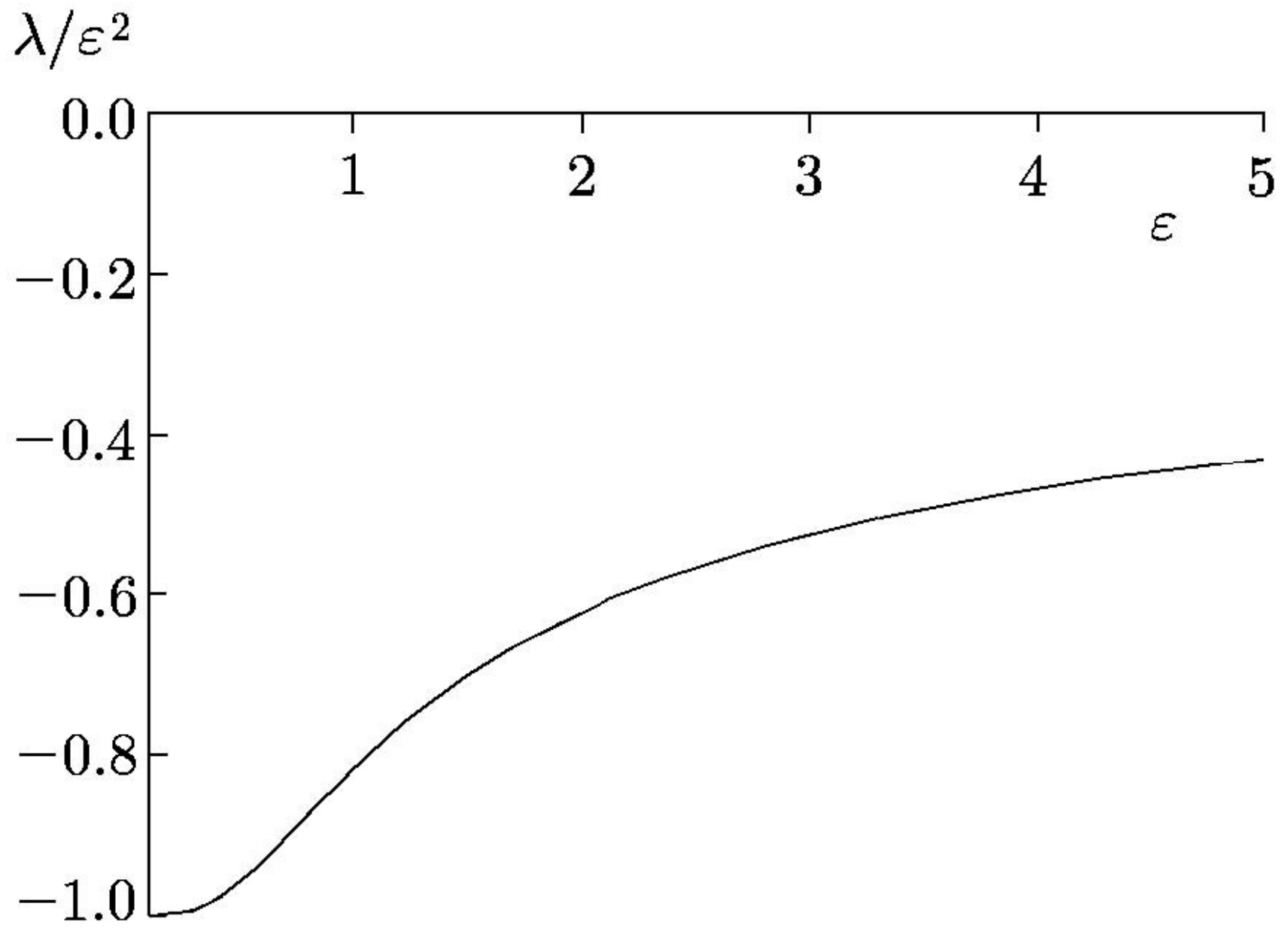

Рис. 4.1: Линейно поляризованный однородный шум в системе с круговым предельным циклом. Зависимость нормированного на $\varepsilon^2$ показателя Ляпунова от амплитуды шума $\varepsilon$. При слабых и сильных шумах для зависимости показателя $\lambda$ от амплитуды шума имеют место квадратичные асимптотики с разными коэффициентами.

### 4.1.4. Пример: суперпозиция двух независимых линейно поляризованных однородных шумов

В тех же предположениях, что и выше (предельный цикл в динамической системе близок к окружности и движение на нем равномерно), эффект действия шума на две компоненты, сдвинутые по фазе на $\pi/2$, естественно моделировать многокомпонентным шумом с $f_1(\varphi)=\sqrt{2}\sin\varphi$ и $f_2(\varphi)=\sqrt{2}\cos\varphi$. Для осредненной плотности вероятности имеют место эффективные функция $f(\varphi)=\sqrt{1+\Delta\cos(2\varphi)}$ и интенсивность шума $\varepsilon^2=\varepsilon_1^2+\varepsilon_2^2$, где $\Delta\equiv(\varepsilon_2^2-\varepsilon_1^2)/(\varepsilon_2^2+\varepsilon_1^2)$; очевидно, $\Delta\in[-1,1]$. Из вида функции $f(\varphi)$ можно сделать вывод о существовании в системе симметрии относительно замены $(\Delta, \varphi) \leftrightarrow (-\Delta, \varphi + \pi / 2)$. В этом случае

$$\int \frac{d\theta}{f^2(\theta)} = \frac{1}{\sqrt{1-\Delta^2}} \left( \pi \left[ \frac{\theta}{\pi} \right] - \operatorname{arctg} \left( \sqrt{\frac{1+\Delta}{1-\Delta}} \operatorname{ctg} \theta \right) \right),$$

и для осредненного распределения $W(\varphi)$ и потока вероятности $S$ можно получить:

$$W(\varphi) = C \int\limits_{\varphi}^{\varphi+\pi} d\psi \frac{e^{-\frac{1}{\varepsilon^2 \sqrt{1-\Delta^2}} \left( \pi \left[ \frac{\theta}{\pi} \right] - \operatorname{arctg} \left( \sqrt{\frac{1+\Delta}{1-\Delta}} \operatorname{ctg} \theta \right) \right) \Big|_{\varphi}^{\psi}}}{\sqrt{1+\Delta \cos(2\varphi)} \sqrt{1+\Delta \cos(2\psi)}},$$

$$C = \left( 2 \int\limits_0^{\pi} d\varphi \int\limits_{\varphi}^{\varphi+\pi} d\psi \frac{e^{-\frac{1}{\varepsilon^2 \sqrt{1-\Delta^2}} \left( \pi \left[ \frac{\theta}{\pi} \right] - \operatorname{arctg} \left( \sqrt{\frac{1+\Delta}{1-\Delta}} \operatorname{ctg} \theta \right) \right) \Big|_{\varphi}^{\psi}}}{\sqrt{1+\Delta \cos(2\varphi)} \sqrt{1+\Delta \cos(2\psi)}} \right)^{-1},$$

$$S = \left[ 1 - \exp \left( -\frac{\pi}{\varepsilon^2 \sqrt{1-\Delta^2}} \right) \right] C.$$

Окончательное выражение для показателя Ляпунова имеет вид

$$\lambda = -2\varepsilon^2 C \int\limits_0^{\pi} d\varphi \int\limits_{\varphi}^{\varphi+\pi} d\psi \frac{\sqrt{1+\Delta \cos(2\varphi)}}{\sqrt{1+\Delta \cos(2\psi)}} \times$$
$$\times \exp \left( -\frac{1}{\varepsilon^2 \sqrt{1-\Delta^2}} \left( \pi \left[ \frac{\theta}{\pi} \right] - \operatorname{arctg} \left( \sqrt{\frac{1+\Delta}{1-\Delta}} \operatorname{ctg} \theta \right) \right) \Big|_{\varphi}^{\psi} \right). \qquad (4.20)$$

Зависимость искомого показателя Ляпунова от амплитуды шума $\varepsilon$ и существенного параметра $\Delta$ представлена на Рис. 4.2.

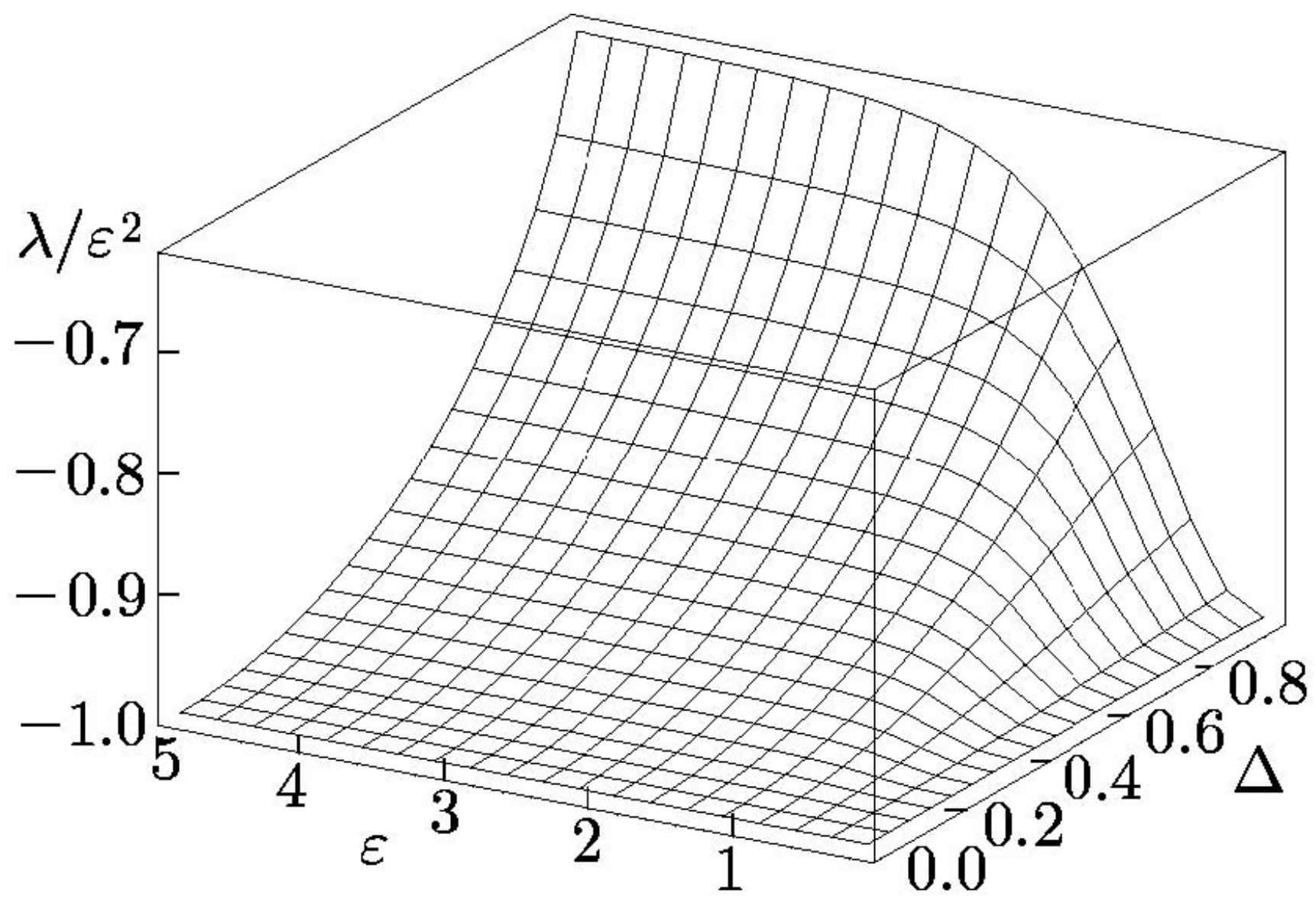

Рис. 4.2: Суперпозиция двух независимых линейно поляризованных однородных шумов. Зависимость нормированного на $\varepsilon^2$ показателя Ляпунова от $\varepsilon$ и $\Delta$.

### 4.1.5. Шум, не допускающий фазового приближения

Для того, чтобы описание состояния системы фазой было несправедливо, шум должен вызывать конечные отклонения от предельного цикла. Таким образом, как было отмечено выше (при введении фазового описания), фазовое приближение может быть справедливо как при малых интенсивностях шума в обычных системах, так и при конечных интенсивностях шума в системах, где предельный цикл имеет большой отрицательный максимальный показатель Ляпунова (имея в виду такие системы, рассматривались результаты построенной выше аналитической теории при конечных интенсивностях шума).

Случаи, не допускающие фазового описания, требуют численного анализа. В качестве одного из классических примеров системы с устойчивым предельным циклом можно использовать стандартный осциллятор Ван дер Поля–Дюффинга:

$$\ddot{x} - \mu(1 - x^2)\dot{x} + x + bx^3 = \varepsilon\xi(t), \tag{4.21}$$

где $\xi(t)$ – нормированный гауссовский шум, $\mu$ описывает близость системы к точке бифуркации Хопфа, а "параметр Дюффинга" $b$ характеризует степень неизохронности колебаний в окрестности предельного цикла.

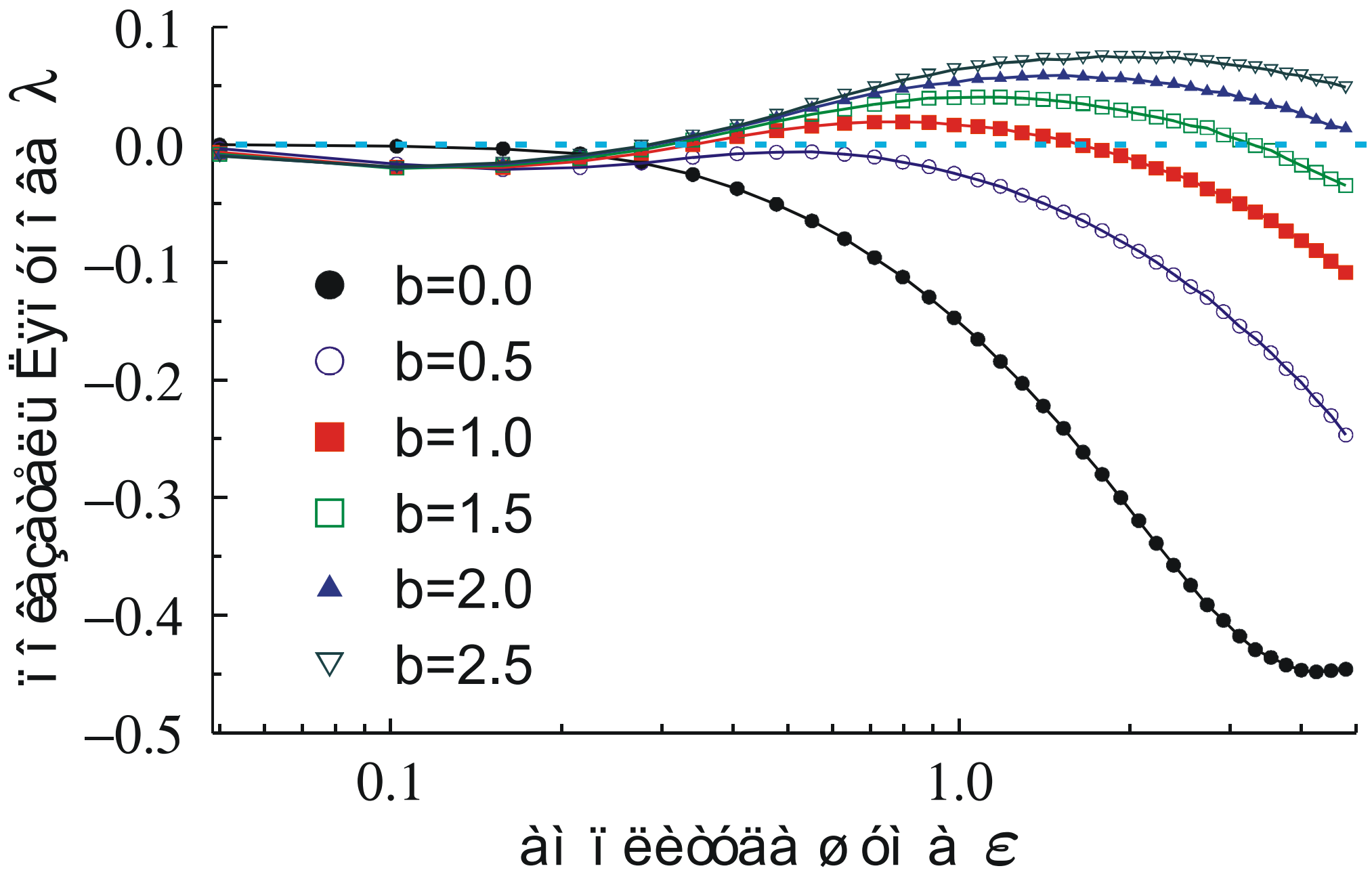

Рис. 4.3: Осциллятор Ван дер Поля–Дюффинга (4.21), подверженный воздействию белого гауссовского шума. Зависимости показателей Ляпунова от амплитуды шума представлены для $\mu = 0.2$ и различных значений $b$.

На Рис. 4.3 представлены зависимости показателя Ляпунова $\lambda$ от амплитуды шума $\varepsilon$ при $\mu = 0.2$ и различных значениях $b$. Можно видеть, что при $b \gtrsim 0.5$ (разумеется, критическое значение зависит от $\mu$) в некотором диапазоне амплитуд шума $\varepsilon$ наблюдаются положительные значения показателя Ляпунова $\lambda$, тогда как асимптотический закон

$$\lim_{\varepsilon \to 0} \lambda / \varepsilon^2 = const < 0$$

остается справедлив для всех значений $b$. По мере роста $b$ область положительных значений показателя Ляпунова расширяется.

## 4.2. Системы с предельным циклом – неидеальные ситуации

При обращении к вопросу о проявлениях синхронизации и десинхронизации осцилляторов общим шумом следует иметь в виду существенную роль возможных неидентичностей осцилляторов и дополнительных внутренних шумов в системах. Кроме того, в таких неидеальных ситуациях важную роль играет не только знак показателя Ляпунова $\lambda$, но и его абсолютная величина: при отрицательных $\lambda$ отклонения в состояниях систем, вызванные малой неидентичностью, тем меньше, чем больше $|\lambda|$. Поскольку в ситуации, когда шум ведет к десинхронизации, системы в любом случае ведут себя по-разному, имеет смысл рассматривать роль неидентичностей лишь при отрицательных $\lambda$. Аналитическую теорию удается построить в фазовом приближении, гарантированно справедливом при малых интенсивностях шума, которые ниже и предполагаются.

### 4.2.1. Слегка неидентичные осцилляторы

Эволюция $N$ слабо отличающихся осцилляторов может быть описана следующим обобщением фазового уравнения (4.3):

$$\dot{\varphi}_j = \omega + \sigma_j + \varepsilon f(\varphi_j)\xi(t), \qquad j = 1,2,\ldots,N, \tag{4.22}$$

где $\sigma_j$ – отклонения частот осцилляторов от средней частоты, $\sum_{j=1}^{N}\sigma_j = 0$. Отличиями в функциях $f$ можно пренебречь в силу малости $\varepsilon$. Можно ожидать, что состояния осцилляторов будут близки при малой по сравнению с показателем Ляпунова расстройке частот: $|\sigma_j| \ll |\lambda| \ll 1$. В этом случае удобно перейти к новым переменным $\varphi = \sum_{j=1}^{N} \varphi_j$ и $\theta_j = \varphi_j - \varphi$, $j = 1,2,...,N-1$. В свою очередь при малых $\theta_j$ система (4.22) может быть переписана в виде

$$\dot{\varphi} = \omega + \varepsilon f(\varphi)\xi(t), \tag{4.23}$$

$$\dot{\theta}_j = \sigma_j + \varepsilon f'(\varphi)\theta_j\xi(t). \tag{4.24}$$

Принимая во внимание, что отклонения $\theta_j$ с различными $j$ ведут себя независимо, можно рассматривать эволюцию каждого $\theta_j$ отдельно и опустить индекс $j$. Таким образом, эволюция $\varphi$ и $\theta$ оказывается такой же, как для двух слегка отличающихся осцилляторов.

На основе уравнений (4.23),(4.24) можно выписать уравнение Фоккера–Планка для плотности вероятности $W(\varphi,\theta,t)$:

$$\frac{\partial W}{\partial t} + \omega\frac{\partial W}{\partial \varphi} + \sigma\frac{\partial W}{\partial \theta} - \varepsilon^2\mathbf{Q}^2 W = 0, \tag{4.25}$$

где $\mathbf{Q}g \equiv \frac{\partial}{\partial\varphi}(f(\varphi)g) + \frac{\partial}{\partial\theta}(f'(\varphi)\theta g)$. Благодаря возможности ренормировки $\theta$ в (4.25) можно полагать $\sigma$ порядка $\varepsilon^2$. Разложение стационарного решения по степеням $\varepsilon^2$ обнаруживает, что в нулевом порядке $W_0 = w(\theta)$, а в первом порядке

$$\omega\frac{\partial W_1}{\partial \varphi} + \sigma\frac{\partial w}{\partial \theta} - \varepsilon^2\mathbf{Q}^2 w = 0. \tag{4.26}$$

Подставляя $\mathbf{Q}^2 w(\theta)$ и интегрируя уравнение (4.26) по $\varphi \in [0, 2\pi)$, можно получить (благодаря $2\pi$-периодической зависимости $W_1$ от $\varphi$)

$$\sigma \frac{dw}{d\theta} = \varepsilon^2 \overline{f'^2} \left( \theta^2 \frac{d^2 w}{d\theta^2} + 4\theta \frac{dw}{d\theta} + 2w \right) \tag{4.27}$$

(здесь $\overline{f'^2} \equiv (2\pi)^{-1} \int_0^{2\pi} f'^2(\varphi)\, d\varphi$). При $\sigma = 0$ решением последнего уравнения является $\delta$-функция. Если $\sigma \neq 0$, уравнение (4.27) может быть переписано в виде

$$x^2 \frac{d^2 w}{dx^2} + (4x - 1) \frac{dw}{dx} + 2w = 0, \tag{4.28}$$

где $x \equiv \varepsilon^2 \overline{f'^2} \sigma^{-1} \theta = |\lambda|\, \sigma^{-1} \theta$. Последнее дифференциальное уравнение удобно решать с помощью замены $w(x) = h(x) / x^2$. С учетом условия нормировки $\int_{-\infty}^{+\infty} w(\theta)\, d\theta = (2\pi)^{-1}$, стационарное решение уравнения (4.28) принимает вид:

$$w(\theta) = \begin{cases} \dfrac{|\sigma|}{2\pi |\lambda| \theta^2} \exp\left( -\dfrac{\sigma}{|\lambda| \theta} \right), & \sigma\theta > 0; \\ 0, & \sigma\theta \le 0. \end{cases} \tag{4.29}$$

Несмотря на такой способ задания, эта функция бесконечно гладка в $\theta = 0$.

Примечательно, что для пары осцилляторов, подверженных одному и тому же шумовому воздействию, фаза более быстрого осциллятора никогда не оказывается позади фазы медленного. Следует иметь в виду, что здесь не учитываются проскальзывания фазы на $2\pi n$.

Для плотности вероятности (4.29) могут быть вычислены моменты

$$\left\langle |\theta|^k \right\rangle = \left( |\sigma| \,/\, |\lambda| \right)^k \Gamma(1 - k) \tag{4.30}$$

(здесь $\Gamma$ – гамма функция) и наиболее вероятное значение $\theta_{\text{mp}} = \dfrac{\sigma}{2\,|\lambda|}$. Отсюда видно, что разность фаз $\theta$ мала при $|\sigma| \ll |\lambda|$. Формула (4.30) дает конечные моменты только при $k < 1$. Старшие моменты расходятся благодаря степенным асимптотам плотности вероятности $w(\theta)$. Для того, чтобы получить конечные моменты $\theta$, необходимо выходить за рамки линейного по $\theta$ приближения даже для сколь угодно малой расстройки частот $\sigma$.

В термодинамическом пределе $N \to \infty$ могут быть вычислены средние по ансамблю значения моментов отклонений состояний

$$\left\langle |\theta|^k \right\rangle_{\text{ens}} = \Gamma(1-k)\,|\lambda|^{-k} \int_{-\infty}^{+\infty} |\sigma|^k\, F(\sigma)\, d\sigma\,,$$

где $F(\sigma)$ – распределение $\sigma$.

### 4.2.2. Малый внутренний шум

Вполне обычна ситуация, когда $N$ идентичных систем, подверженных идентичному внешнему шумовому воздействию $\xi(t)$, испытывают также влияние дополнительных независимых (например, тепловых) шумов $\eta_j(t)$. Динамика фазы в этом случае определяется

$$\dot{\varphi}_j = \omega + \varepsilon f(\varphi_j)\xi(t) + \gamma_j g_j(\varphi_j)\eta_j(t), \quad j = 1,2,\ldots,N, \tag{4.31}$$

где функции $f$ и $g_j$ нормированы: $\overline{f^2} = \overline{g_j^2} = 1$ , $\varepsilon$ и $\gamma_j$ – амплитуды шума, и

$$\left\langle \xi(t)\xi(t+t') \right\rangle = 2\delta(t'),$$
$$\left\langle \eta_j(t)\eta_k(t+t') \right\rangle = 2\delta_{jk}\delta(t'),$$
$$\left\langle \xi(t)\eta_j(t+t') \right\rangle = 0.$$

Подобно случаю неидентичных осцилляторов можно ввести фазу $\varphi$, определяемую уравнением (4.23), и для малых отклонений $\theta_j$ получить

$$\dot{\theta} = \varepsilon f'(\varphi)\theta\xi(t) + \gamma g(\varphi)\eta(t), \tag{4.32}$$

где индекс $j$ вновь опущен, как и ранее. В этом случае соответствующее уравнение Фоккера–Планка принимает вид

$$\frac{\partial W}{\partial t} + \omega\frac{\partial W}{\partial \varphi} - \gamma^2 g^2(\varphi)\frac{\partial^2 W}{\partial \theta^2} - \varepsilon^2 \mathbf{Q}^2 W = 0. \tag{4.33}$$

Стационарное распределение плотности вероятности может быть найдено подобно тому, как это было сделано выше для случая неидентичных осцилляторов (4.25). Вместо уравнения (4.27) теперь получается

$$\gamma^2 \frac{d^2 w}{d\theta^2} + \varepsilon^2 \overline{f'^2}\left(\theta^2 \frac{d^2 w}{d\theta^2} + 4\theta\frac{dw}{d\theta} + 2w\right) = 0, \tag{4.34}$$

где благодаря условию нормировки $\overline{g^2} = 1$ зависимость от функций $g$ исчезает. После изменения масштаба координат $x \equiv \dfrac{\varepsilon\sqrt{\overline{f'^2}}\,\theta}{\gamma} = \dfrac{\sqrt{|\lambda|}\,\theta}{\gamma}$ последнее уравнение может быть переписано в виде

$$(x^2 + 1)\frac{d^2 w}{dx^2} + 4x\frac{dw}{dx} + 2w = 0, \tag{4.35}$$

и решено с помощью прежней замены $w(x) = h(x)/x^2$. С учетом условия нормировки решение принимает вид колоколообразного распределения

$$w(\theta) = \frac{\gamma}{2\pi^2\sqrt{|\lambda|}}\left[1 + \frac{|\lambda|}{\gamma^2}\theta^2\right]^{-1}. \tag{4.36}$$

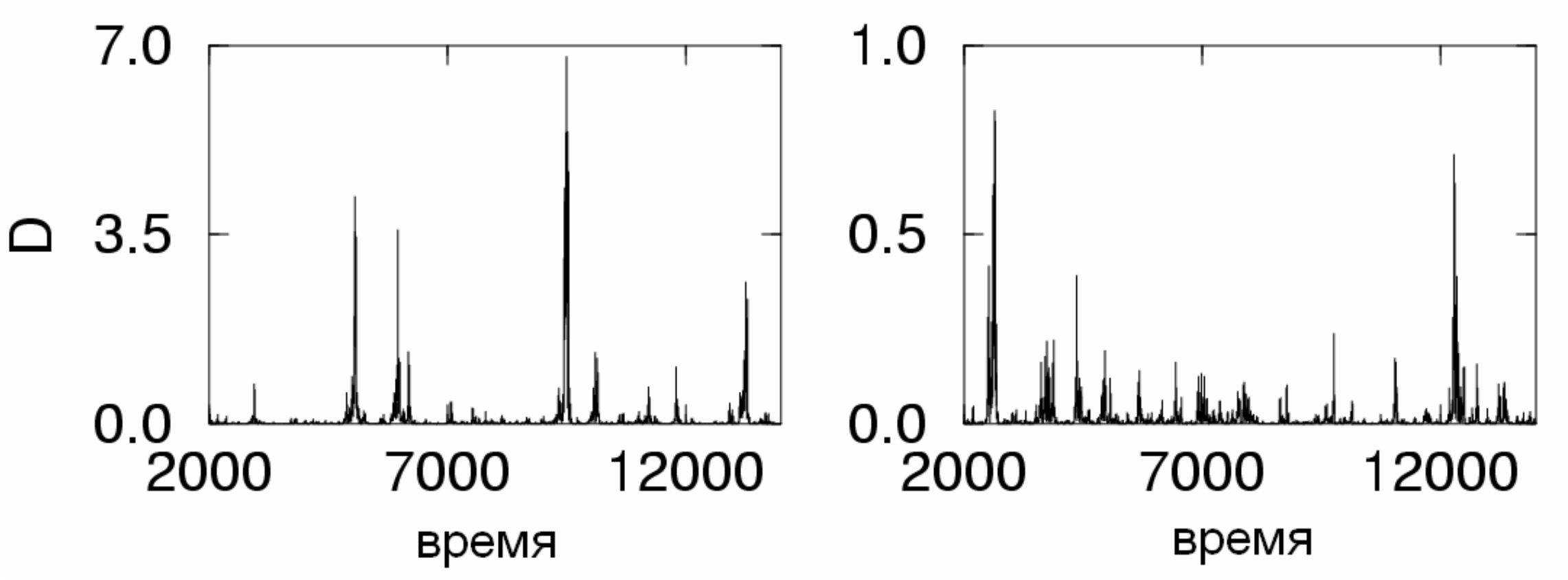

Рис. 4.4: Реализации разности $D \equiv \sqrt{(x_1 - x_2)^2 + (\dot{x}_1 - \dot{x}_2)^2}$ между двумя системами Ван дер Поля–Дюффинга (4.21). Левый график: общий гауссовский белый шум воздействует на осцилляторы с малой расстройкой частот (уравнения (4.37),(4.38) при $\sigma = 10^{-4}$). Правый график: идентичные осцилляторы под воздействием слегка отличающихся шумов (уравнения (4.39),(4.40) при $\gamma / \varepsilon = 2 \times 10^{-4}$). Параметры: $\mu = 0.2$, $b = 1$, $\varepsilon = 0.2$.

Подобно (4.29) распределение плотности вероятности (4.36) имеет степенные асимптоты, свидетельствующие о возможности больших флуктуаций для сколь угодно малых $\gamma$. В обоих случаях (малая расстройка и малая неидентичность шума) эти флуктуации имеют вид всплесков, локализованных во времени (Рис. 4.4), подобно тому, как это имеет место для других случаев неидеальной синхронизации [85].

В термодинамическом пределе $N \to \infty$ могут быть вычислены средние по ансамблю значения моментов отклонений состояний

$$\left\langle |\theta|^k \right\rangle_{\text{ens}} = \frac{1}{|\lambda|^{k/2} \cos(\pi k / 2)} \int_0^{+\infty} \gamma^k G(\gamma)\, d\gamma,$$

где $G(\gamma)$ – распределение $\gamma$.

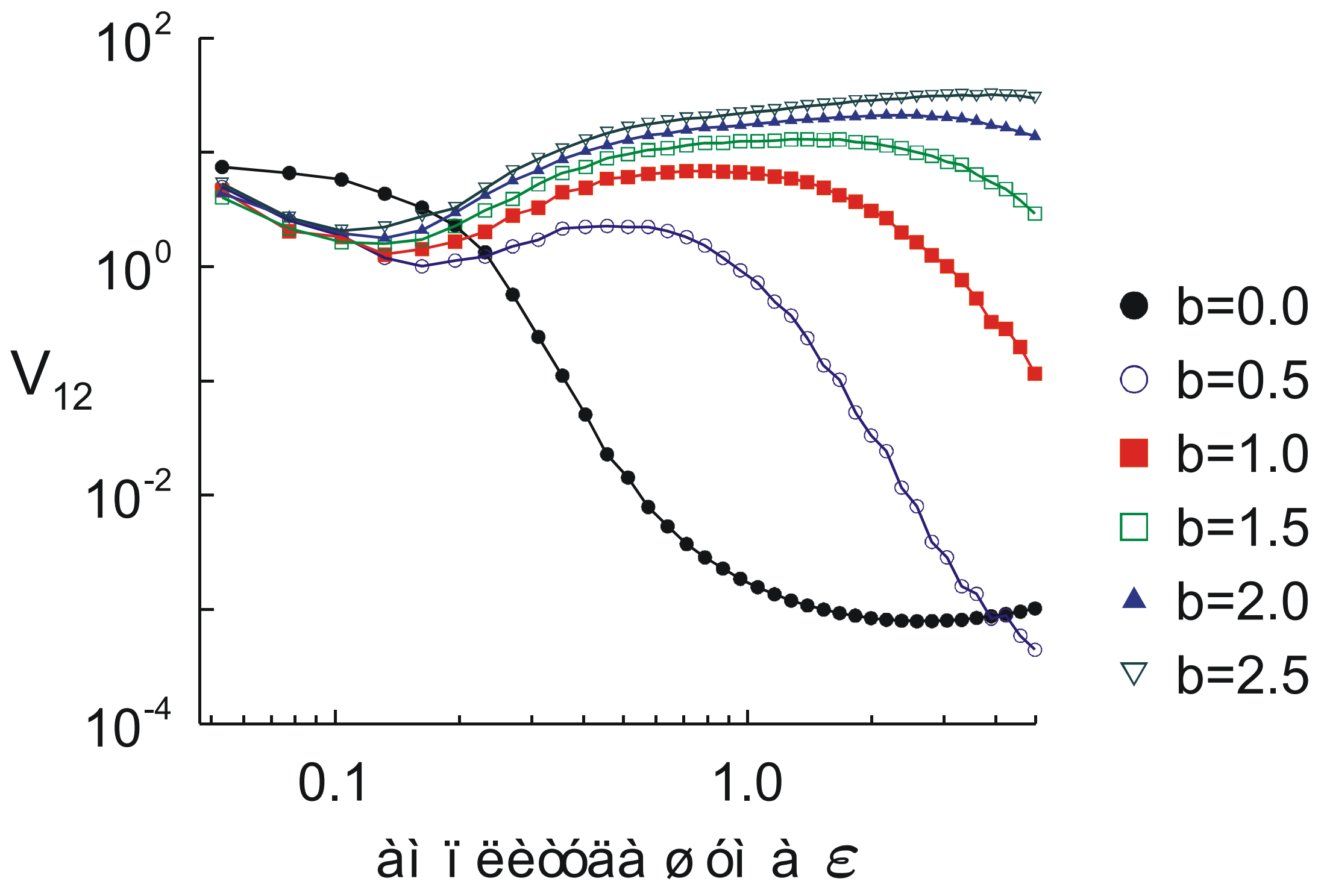

Рис. 4.5: Зависимости $V_{12}(\varepsilon)$ представлены при $\mu = 0.2$ и $\sigma = 0.002$ для пары осцилляторов Ван дер Поля–Дюффинга подверженных воздействию белого гауссовского шума (4.37),(4.38).

### 4.2.3. Неидеальные ситуации – численные результаты

Чтобы количественно характеризовать переход синхронизация–десинхронизация в системе (4.21) в неидеальных ситуациях, было выполнено численное моделирование двух слегка неидентичных осцилляторов Ван дер Поля–Дюффинга, подверженных одному шумовому воздействию

$$\ddot{x}_1 - \mu(1 - x_1^2)\dot{x}_1 + (1+\sigma)x_1 + bx_1^3 = \varepsilon\xi(t), \tag{4.37}$$

$$\ddot{x}_2 - \mu(1 - x_2^2)\dot{x}_2 + (1-\sigma)x_2 + bx_2^3 = \varepsilon\xi(t), \tag{4.38}$$

и двух идентичных осцилляторов Ван дер Поля–Дюффинга, подверженных слегка отличающемуся шумовому воздействию

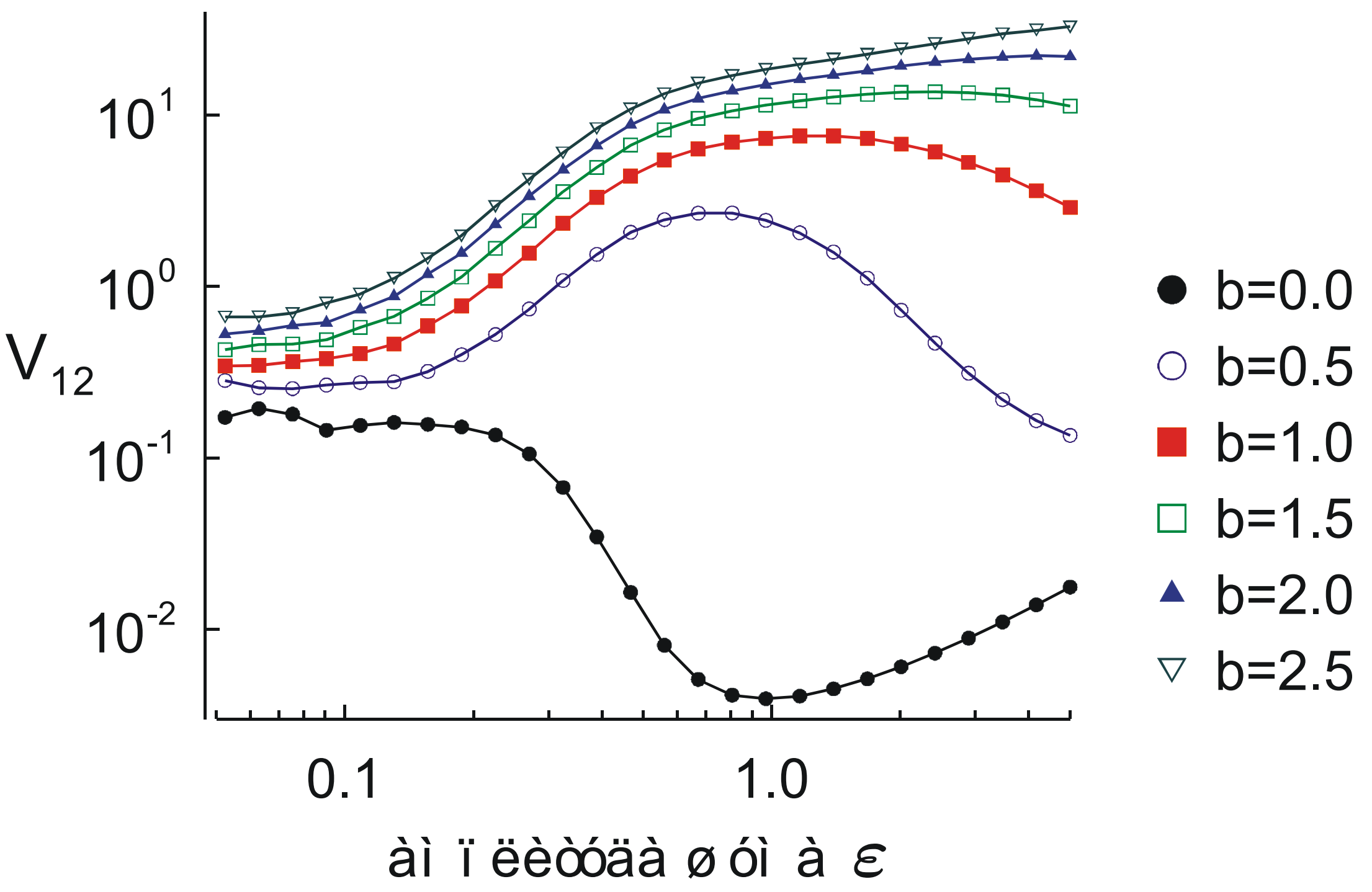

Рис. 4.6: Зависимости $V_{12}(\varepsilon)$ представлены при $\mu = 0.2$ и $\gamma / \varepsilon = 0.01$ для пары идентичных осцилляторов Ван дер Поля–Дюффинга подверженных воздействию различных белых гауссовских шумов (4.39),(4.40).

$$\ddot{x}_1 - \mu(1 - x_1^2)\dot{x}_1 + x_1 + b x_1^3 = \varepsilon\xi(t) + \gamma\eta(t), \tag{4.39}$$

$$\ddot{x}_2 - \mu(1 - x_2^2)\dot{x}_2 + x_2 + b x_2^3 = \varepsilon\xi(t) - \gamma\eta(t). \tag{4.40}$$

Степень синхронности характеризуется средней разностью $V_{12} = \left\langle (x_1 - x_2)^2 + (\dot{x}_1 - \dot{x}_2)^2 \right\rangle$. Зависимость $V_{12}(\varepsilon)$ имеет максимум в области положительных значений показателя Ляпунова (см. Рис. 4.5, 4.6, ср. Рис. 4.3).

Также было выполнено численное интегрирование большого ансамбля слегка отличающихся осцилляторов, находящихся под воздействием одного и того же шума. Распределение состояний системы на плоскости $(x, \dot{x})$ в некото-рый момент времени локализовано для отрицательных значений показателя Ляпунова и имеет широко раскинувшуюся фрактальную структуру – для положительных (Рис. 4.7). Эти распределения отвечают различным типам мгновенных аттракторов системы (4.21) (см. [95]).

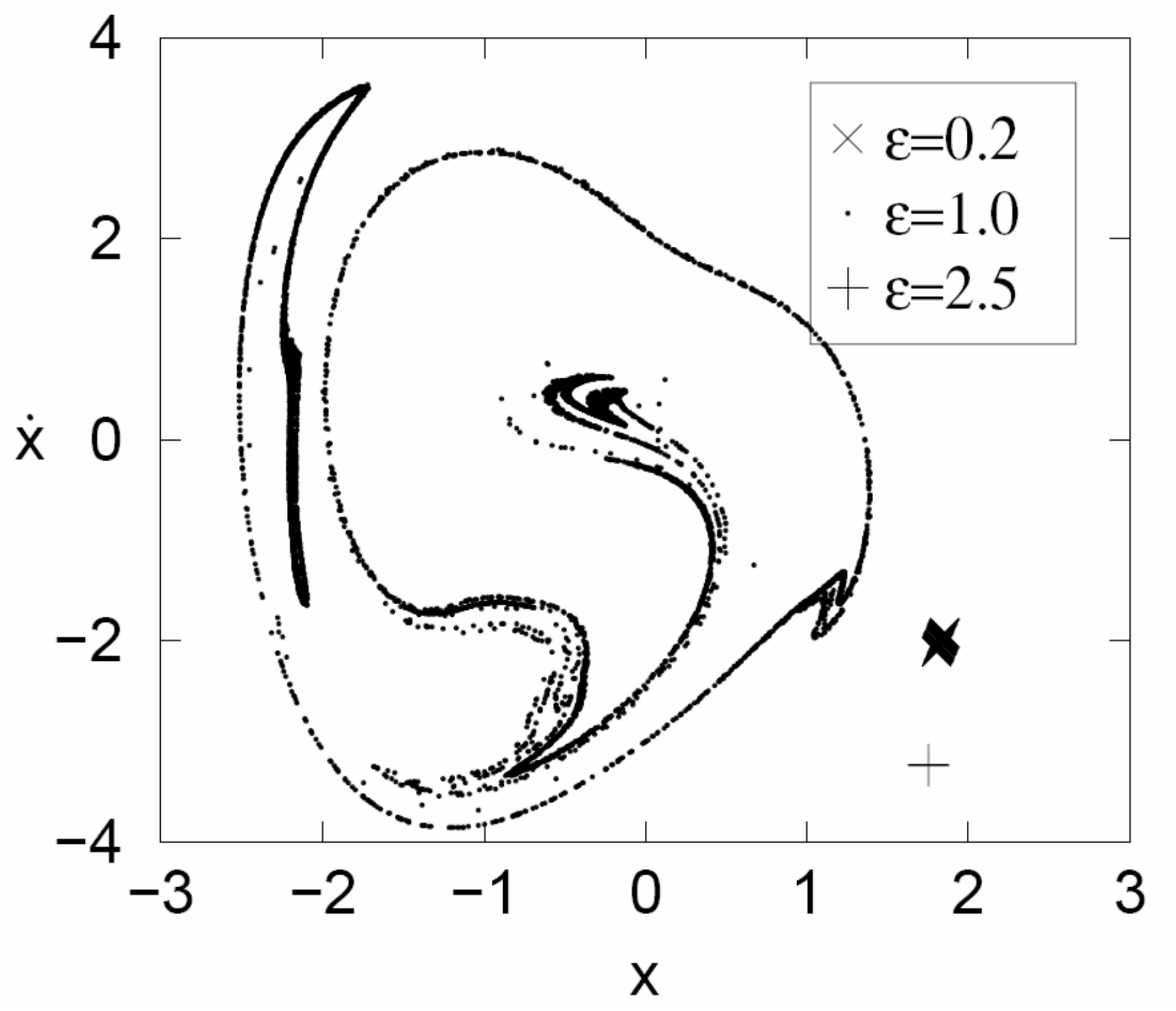

Рис. 4.7: Моментальное распределение состояний в ансамбле $10\,000$ осцилляторов Ван дер Поля–Дюффинга с равномерным распределением $\sigma_j$ в интервале $[-0.01; 0.01]$, подверженных воздействию общего белого гауссовского шума при $\mu = 0.2$ и $b = 1$. Три выбранных значений амплитуды шума $\varepsilon$ соответствуют отрицательным ($\varepsilon = 0.2$, состояния в окрестности точки $(1.82;-2.07)$; и $\varepsilon = 2.5$, состояния в окрестности точки $(1.76;-3.29)$) и положительным ($\varepsilon = 1$) значениям показателя Ляпунова.

## 4.3. Системы с предельным циклом – телеграфный шум

По своей природе телеграфный шум существенно отличается от рассматривавшегося выше белого гауссовского шума и представляет интерес с точки зрения выяснения того, насколько существенны специфические свойства последнего, и будет ли ситуация аналогична для шумов с другими свойствами. Кроме того, телеграфный шум интересен также в связи с тем, что такому сигналу может быть очевидным образом сопоставлен ступенчатый периодический сигнал. Сопоставление эффектов таких сигналов может помочь в понимании механизмов синхронизации общим шумом.

Вновь, исходим из фазового уравнения

$$\dot{\varphi} = \omega + \varepsilon f(\varphi)\xi(t), \tag{4.41}$$

где $2\pi / \omega$ – период предельного цикла системы без шума, $\varepsilon$ – амплитуда шума, $f(\varphi)$ – нормированная восприимчивость системы к шуму, $\int_0^{2\pi} f^2(\varphi)\,d\varphi = 2\pi$, а $\xi(t)$ – нормированный телеграфный шум. Под нормированным телеграфным шумом подразумевается сигнал, принимающий значения $\pm 1$ и мгновенно переключающийся между ними с постоянной вероятностью переключения, не зависящей от предыстории; среднее время между переключениями $\tau$.

### 4.3.1. Телеграфный шум – мастер–уравнение

Для описания статистических свойств рассматриваемой системы удобно ввести два распределения плотности вероятности $W_{\pm}(\varphi, t)$, определяющие вероятность обнаружить систему в окрестности точки $\varphi$ с $\xi = \pm 1$, соответственно, в момент времени $t$. Для этих распределений мастер уравнение принимает вид:

$$\frac{\partial W_+(\varphi,t)}{\partial t}+\frac{\partial}{\partial\varphi}\big[(\omega+\varepsilon f(\varphi))W_+(\varphi,t)\big]=\frac{1}{\tau}W_-(\varphi,t)-\frac{1}{\tau}W_+(\varphi,t), \quad (4.42)$$

$$\frac{\partial W_-(\varphi,t)}{\partial t}+\frac{\partial}{\partial\varphi}\big[(\omega-\varepsilon f(\varphi))W_-(\varphi,t)\big]=\frac{1}{\tau}W_+(\varphi,t)-\frac{1}{\tau}W_-(\varphi,t). \quad (4.43)$$

В терминах $W=W_+ + W_-$ и $V=W_+ - W_-$ последняя система может быть переписана как

$$\dot W=-\omega W_\varphi-\varepsilon(fV)_\varphi, \quad (4.44)$$

$$\dot V=-\omega V_\varphi-\varepsilon(fW)_\varphi-2\tau^{-1}V. \quad (4.45)$$

Для стационарных распределений поток вероятности $S$ постоянен:

$$S=\omega W(\varphi)+\varepsilon f(\varphi)V(\varphi);$$

и система (4.44),(4.45) с периодическими граничными условиями имеет решение

$$V(\varphi)=-\frac{\varepsilon\omega C}{\omega^2-\varepsilon^2 f^2(\varphi)}\int\limits_{\varphi}^{\varphi+2\pi}d\psi\; f'(\psi)\exp\left(\frac{2}{\tau}\int\limits_{\varphi}^{\psi}\frac{d\theta}{\omega^2-\varepsilon^2 f^2(\theta)}\right), \quad (4.46)$$

где константа $C$ определяется условием нормировки $\int_0^{2\pi}W(\varphi)d\varphi=2\pi$:

$$C^{-1}=2\pi\left(\exp\left(\frac{2}{\tau}\int\limits_0^{2\pi}\frac{d\theta}{\omega^2-\varepsilon^2 f^2(\theta)}\right)-1\right)+ \\ +\varepsilon^2\int\limits_0^{2\pi}d\varphi\int\limits_{\varphi}^{\varphi+2\pi}d\psi\frac{f(\varphi)f'(\psi)}{\omega^2-\varepsilon^2 f^2(\varphi)}\exp\left(\frac{2}{\tau}\int\limits_{\varphi}^{\psi}\frac{d\theta}{\omega^2-\varepsilon^2 f^2(\theta)}\right). \quad (4.47)$$

В свою очередь, поток вероятности

$$S=\omega\left(\exp\left(\frac{2}{\tau}\int_0^{2\pi}\frac{d\theta}{\omega^2-\varepsilon^2 f^2(\theta)}\right)-1\right)C\,.$$

### 4.3.2. Телеграфный шум – показатель Ляпунова

Малые возмущения $\alpha$ решений уравнения (4.41), как и прежде, подчиняются уравнению

$$\dot{\alpha}=\varepsilon\,\alpha\, f'(\varphi)\xi(t),$$

и показатель Ляпунова

$$\lambda=\left\langle\frac{d\ln\alpha}{dt}\right\rangle=\left\langle\varepsilon f'(\varphi)\xi(t)\right\rangle=\varepsilon\int_0^{2\pi} f'(\varphi)V(\varphi)\,d\varphi=$$
$$=-\varepsilon^2\omega C\int_0^{2\pi}d\varphi\int_{\varphi}^{\varphi+2\pi}d\psi\,\frac{f'(\varphi)f'(\psi)}{\omega^2-\varepsilon^2 f^2(\varphi)}\exp\left(\frac{2}{\tau}\int_{\varphi}^{\psi}\frac{d\theta}{\omega^2-\varepsilon^2 f^2(\theta)}\right). \quad (4.48)$$

Выражение (4.48) может быть существенно упрощено при $\tau\ll 1$ или $\varepsilon\ll\omega$:

$$\lambda_{\text{app}}=-\frac{\varepsilon^2}{2\pi\omega}\frac{\int_0^{2\pi}d\varphi\int_0^{2\pi}d\psi\, f'(\varphi)f'(\psi+\varphi)\, e^{\frac{2\psi}{\tau\omega^2}}}{e^{\frac{4\pi}{\tau\omega^2}}-1}. \quad (4.49)$$

Последнее выражение всегда отрицательно. В самом деле, в Фурье пространстве оно принимает вид:

$$\lambda_{\text{app}}=-\omega\tau\varepsilon^2\sum_{k=1}^{\infty}\left|C_k\right|^2\frac{k^2}{1+(k\tau\omega^2/2)^2}, \quad (4.50)$$

где $C_k=(2\pi)^{-1}\int_0^{2\pi}f(\varphi)\,e^{-ik\varphi}d\varphi$.

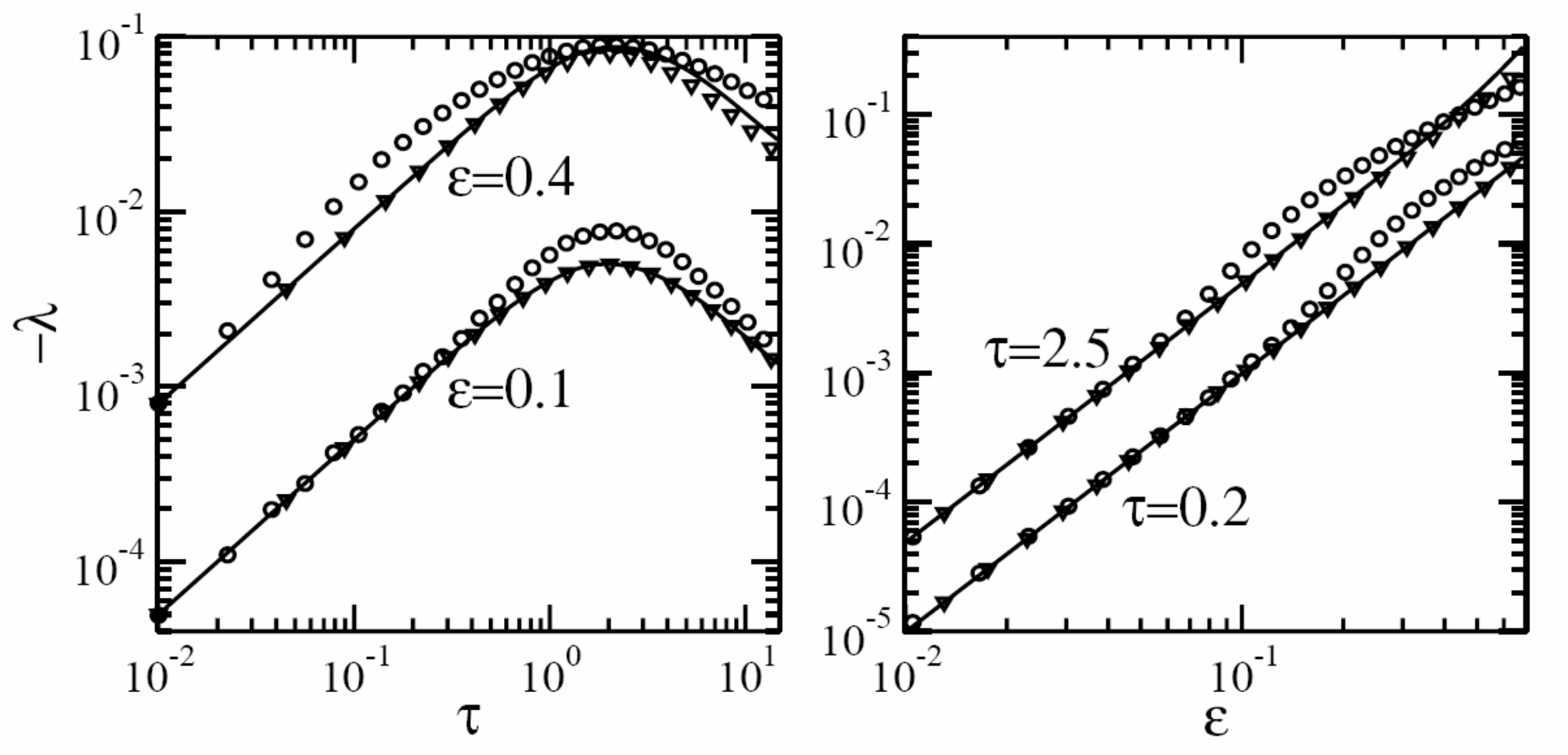

Рис. 4.8: Примеры зависимостей $\lambda(\varepsilon, \tau)$ для модифицировано осциллятора Ван дер Поля (4.51) при $\mu = 0.1$. Линии представляют аналитические результаты фазового описания, треугольники – приближения (4.49), а круги отвечают численному моделированию системы (4.51).

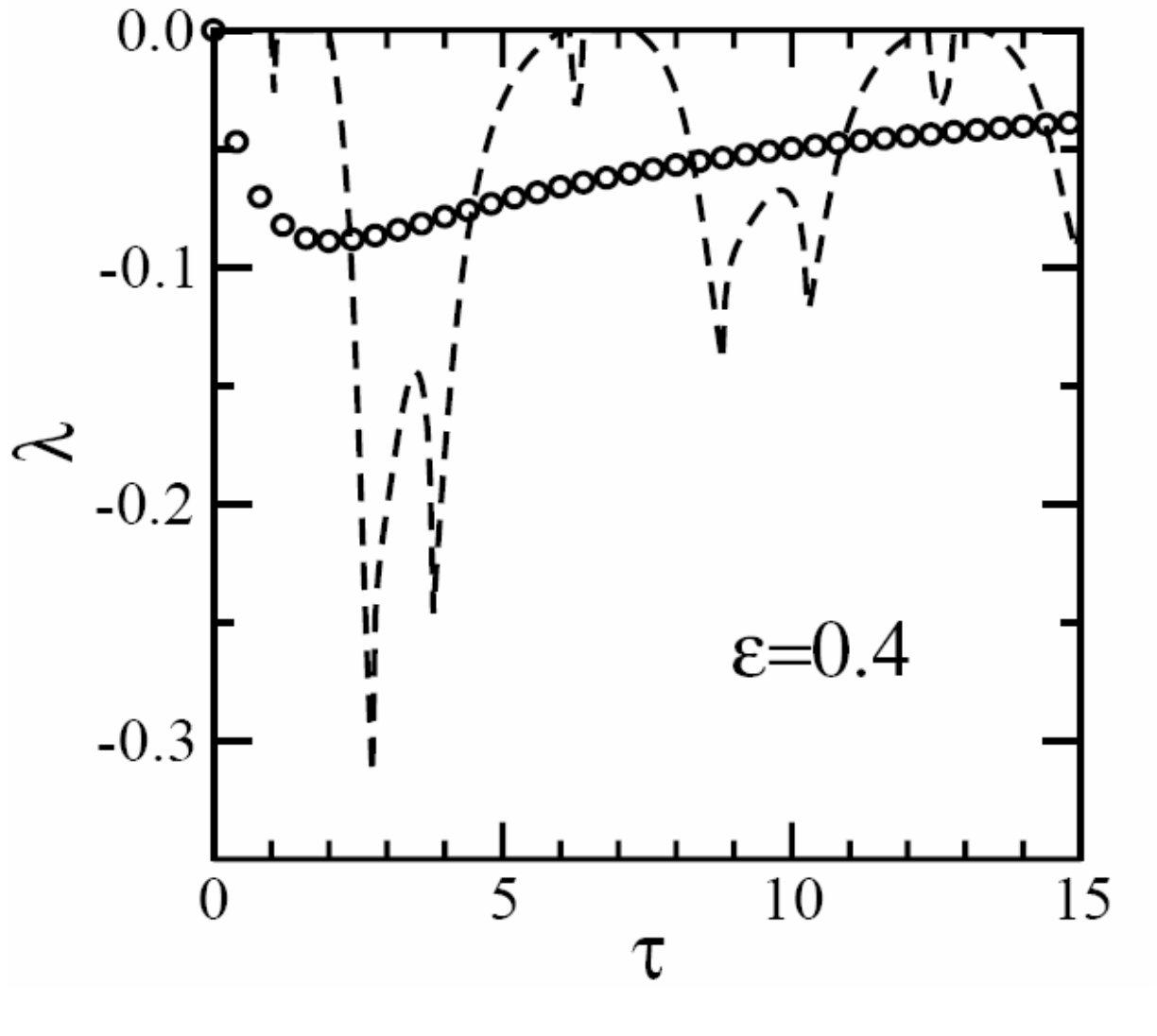

Рис. 4.9: Примеры зависимостей $\lambda(\varepsilon, \tau)$ для модифицированого осциллятора Ван дер Поля (4.51) при $\mu = 0.1$ при шумовом (круги) и периодическом (пунктирная линия) воздействии.

### 4.3.2. Телеграфный шум – численные результаты

Как видно из формулы (4.50), при малой интенсивности или частых переключениях в рамках фазового приближения телеграфный шум ведет к синхронизации при любых гладких функциях $f$. Аналогичная ситуация имела место ранее для белого гауссовского шума: слабый шум ведет к синхронизации любых гладких автоколебательных систем – однако, в некоторых случаях умеренный шум мог вести к десинхронизации. В свете этого представляют интерес следующие вопросы:

- Какова область применимости построенной аналитической теории?
- Обнаруживает ли стохастическая синхронизация следы синхронизации периодическим воздействием?
- Может ли телеграфный шум вести к десинхронизации?

Для ответа на первые два вопроса произведено численное интегрирование модифицированого осциллятора Ван дер Поля:

$$\ddot{x} - \mu\left(1 - x^2 - \dot{x}^2\right)\dot{x} + x = \varepsilon\sqrt{2}\,\xi(t), \qquad (4.51)$$

где $\xi(t)$ – либо нормированный телеграфный шум со средним временем переключений $\tau$, либо $\xi(t) = \operatorname{sign}\left[\cos(\pi t / \tau)\right]$, т.е. периодический сигнал с постоянным временем переключений $\tau$. Примечательной особенностью этой системы является то, что в отсутствии воздействия при $\mu > 0$ она имеет круговой устойчивый предельный цикл единичного радиуса. Однако, фазовое уравнение (4.41) с $\omega = 1$ и простой функцией $f(\varphi) = \sqrt{2}\cos\varphi$ может быть использовано для описания системы, только если модуль фазового потока близок к постоянному по всему циклу, что справедливо только при малых $\mu$.

На Рис. 4.8 можно видеть, что построенная в фазовом приближении аналитическая теория находится в хорошем согласии с результатами численного счета для системы (4.51) не только при малых интенсивностях шума. Так же заслуживает внимание тот факт, что зависимость $\lambda(\varepsilon, \tau)$ для шумового воздействия не обнаруживает никаких следов таковой для периодического воздействия (Рис. 4.9).

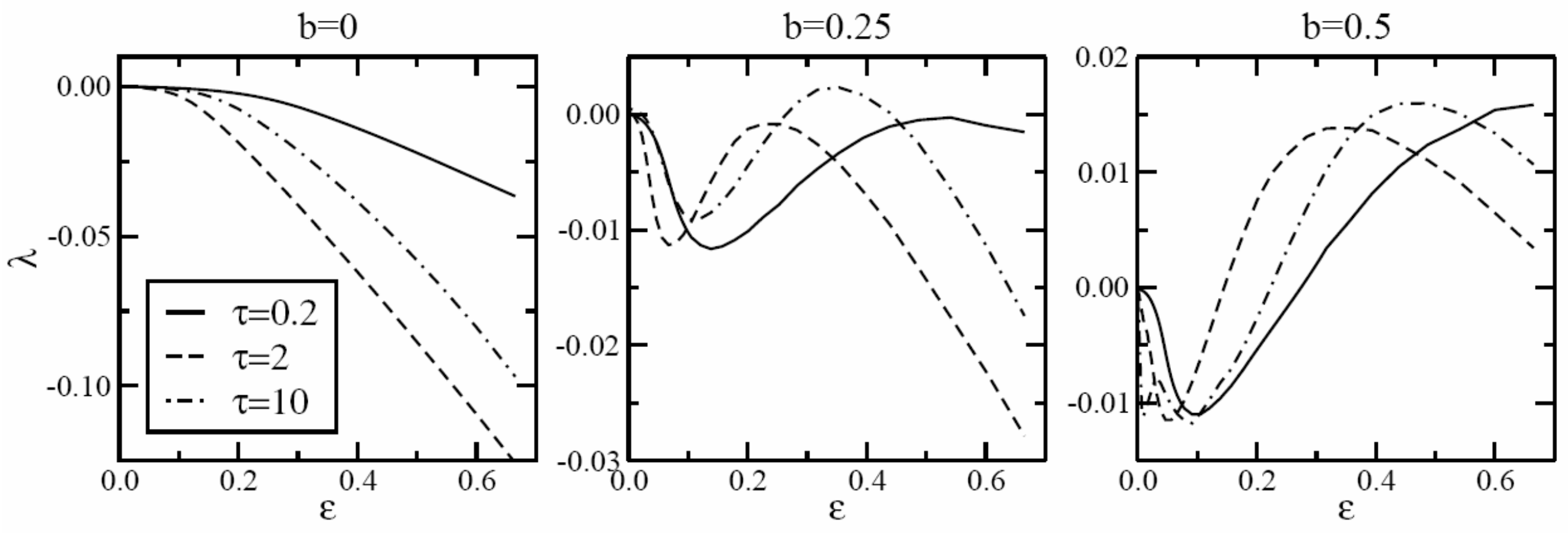

Рис. 4.10: Примеры зависимостей $\lambda(\varepsilon, \tau)$ для осциллятора Ван дер Поля–Дюффинга (4.52) при $\mu = 0.1$.

Хотя система (4.51) не обнаруживает положительных показателей Ляпунова при любых интенсивностях шума и значениях параметра $\mu$, они могут наблюдаться для модели Ван дер Поля–Дюффинга (то же самое имело место для гауссовского белого шума):

$$\ddot{x} - \mu\left(1 - 2x^2\right)\dot{x} + x + 2bx^3 = \varepsilon\sqrt{2}\,\xi(t), \qquad (4.52)$$

где "параметр Дюффинга" $b$ описывает неизохронность траекторий. На Рис. 4.10 можно видеть, что при достаточно больших $b$ при некоторых значениях параметров воздействия наблюдаются положительные значения показателя Ляпунова.

Здесь не представляются результаты для слабо-неидентичных систем, подобно тому, как это было сделано для белого гауссовского шума (разделы 4.2.1, 4.2.2), поскольку, несмотря на более сложный вывод, все окончательные результаты в точности воспроизводятся лишь с заменой показателя Ляпунова на $\lambda_{\mathrm{app}}$, определяемое формулой (4.49).

# Заключение

## Глава 1: Параметрическое возбуждение термоконцентрационной конвекции в слое пористой среды

В первой главе исследована термоконцентрационная конвекция бинарной смеси в подогреваемом снизу горизонтальном слое пористой среды при наличии модуляции поля тяжести. Для случая фиксированного теплового потока через границы с учетом эффекта Соре выведены уравнения, описывающие нелинейную динамику длинноволновых структур при конечной надкритичности.

Аналитически исследована линейная устойчивость состояния механического равновесия по отношению к длинноволновым возмущениям в отсутствии модуляции: найдены границы монотонной и колебательной неустойчивостей. Для проверки обоснованности использования длинноволнового приближения исследована линейная устойчивость системы по отношению к возмущениям конечной длины. Это исследование показало, что длинноволновые возмущения, действительно, являются самыми опасными.

Построенная в линейном приближении для длинноволновых возмущений теория показала, что длина волны возмущения в отсутствии модуляции определяет лишь характерный масштаб временной эволюции возмущения, а не ее характер. В результате, задачи о конечной области, когда спектр волновых чисел дискретен, и бесконечной, когда спектр непрерывен, оказываются существенно различными. Влияние модуляции исследовано аналитически для малых амплитуд модуляции: рассмотрены изменение монотонного уровня и первые три резонанса – и численно для конечных амплитуд. То и другое выполнено как для дискретного, так и для непрерывного спектров волновых чисел.

Оказалось, что параметрическое воздействие не оказывает влияния на модули комплексных мультипликаторов возмущения с фиксированной длиной волны (центральное многообразие системы двумерно), а влияет лишь на грани-цы области параметров, где эти мультипликаторы остаются комплексными. В результате, устойчивость системы по отношению к квазипериодическим колебаниям возмущения с данной длиной волны не зависит от амплитуды модуляции – дестабилизация связана исключительно с резонансами и монотонной неустойчивостью немодулированной системы.

При дискретном спектре граница устойчивости системы может определяться для сколь угодно малых амплитуд модуляции резонансами как первого порядка, так и старших, а при непрерывном спектре и малой модуляции параметрическая дестабилизация колебательного уровня определяется резонансом исключительно первого порядка.

Для некоторых значений параметров системы монотонный уровень неустойчивости может дестабилизироваться модуляцией. При этой дестабилизации наиболее опасными оказываются однородные возмущения при непрерывном спектре, а при дискретном – возмущения с наибольшей возможной длиной волны. Стабилизация монотонного уровня невозможна ни при дискретном, ни при непрерывном спектрах волновых чисел.

## Глава 2: Термоконцентрационная конвекция в слое пористой среды от источников тепла или примеси

Во второй главе исследованы нелинейные режимы стационарной термоконцентрационной конвекция бинарной смеси в тонком слое пористой среды при наличии внутреннего источника примеси или тепла.

Обнаружено, что решения определяются локальными характеристиками потоков (тепла и примеси), а структура источника и то, как он организован, не влияют на течение в конкретной точке пространства: это течение полностью определяется интенсивностью источника. Найденные решения справедливы, начиная с некоторого конечного удаления от источника, которое при слабом источнике может быть малым – порядка толщины слоя.

Выяснено, что в случаях, когда стационарное течение устойчиво, области длинноволнового течения могут быть разделены одним или несколькими кольцами переходного течения.

Кроме того, оказалось, что в случае локализованного источника примеси, ее вынос из непосредственной окрестности источника осуществляется преимущественно конвективным образом. В тоже время, для локализованного источника тепла возможны два типа режимов: с преимущественно конвективным и преимущественно диффузионным механизмами выноса тепла из окрестности источника. В некоторых случаях между этими режимами с разными механизмами выноса тепла возможна мультистабильность.

## Глава 3: Локализация течений в горизонтальном слое при случайно неоднородном нагреве

В третьей главе рассмотрена двухмерная тепловая конвекция жидкости в тонком горизонтальном слое со случайной стационарной неоднородностью нагрева, обеспечиваемой фиксированным потоком тепла поперек слоя.

Выведены уравнения, описывающие нелинейные режимы длинноволновой конвекции при неоднородном нагреве.

При неоднородности, моделируемой белым гауссовским шумом, в системе при средней величине потока тепла ниже критического значения для однородного слоя возможны локальные превышения критического значения теплового потока, приводящие, как показано, к возбуждению локализованных течений, изучению свойств которых и посвящена данная глава диссертации.

Исследованы свойства локализации нелинейных течений и влияние на них прокачивания жидкости в горизонтальном направлении. Обнаружено, что прокачивание приводит к локализации не только течений, но и возмущений поля температуры. Вычислены показатели локализации в направлении по потоку прокачивания и против. Аналитически определен показатель роста среднеквад-ратичных значений, дающий оценку показателей локализации, определяемых численно.

В соответствии с предсказаниями теории, численное интегрирование полной нелинейной системы выявило радикальное влияние прокачивания на свойства локализации в направлении против потока прокачивания: длина локализации при малых конечных скоростях прокачивания может увеличиваться на порядок.

## Глава 4: Синхронизация нелинейных систем общим шумом

В четвертой главе исследована возможность синхронизации одинаковых автоколебательных динамических систем посредством воздействия общим внешним шумом (там, где это не оговорено, идет речь о белом гауссовском шуме).

Количественной характеристикой способности систем синхронизоваться является показатель Ляпунова, соответствующий возмущениям вдоль фазовой траектории динамической системы. В фазовом приближении задача его отыскания решена аналитически в квадратурах для случаев одного и нескольких независимых шумовых сигналов.

В качестве примеров систем, соответствующих первому и второму случаям, рассмотрены задачи о влиянии одного и двух независимых линейно поляризованных однородных шумов на синхронизацию систем с предельным циклом, имеющим близкую к окружности форму, и примерно постоянный на этом цикле модуль скорости фазового потока.

Для обоих рассмотренных примеров шум может приводить только к синхронизации, поскольку показатель Ляпунова отрицателен. Аналогичный вывод следует и из общего асимптотического выражения, справедливого для любых систем в пределе малого шума. Причем в этом пределе предположение о гауссовости является излишним, поскольку вклад старших кумулянтов шумового сигнала (гауссовость используется для их устранения) в значения интересующих нас величин пропорционален старшим степеням интенсивности шума. Сами же асимптотические значения показателя Ляпунова в этом пределе прямо пропорциональны интенсивности шума и обратно пропорциональны частоте автоколебаний.

В ситуациях, когда фазовое приближение не справедливо (при конечной интенсивности шума и небольшой устойчивости предельного цикла) аналитическое исследование поведения систем в общем случае оказывается невозможно и требуется численный счет, который для некоторых систем (например, осциллятора Ван дер Поля–Дюффинга) обнаружил возможность появления положительного показателя Ляпунова, т.е. десинхронизацию колебаний.

В упомянутых случаях отдельное внимание уделено неидеальным ситуациям: когда имеется ансамбль идентичных осцилляторов, находящихся под воздействием слегка отличающихся шумов (внутренний шум), или ансамбль слегка отличающихся осцилляторов под действием идентичного шума. При малом шуме, когда показатель Ляпунова отрицателен, и малых неидентичностях (в шуме или параметрах) задачи допускают аналитическое исследование в рамках фазового приближения.

Оказывается, что при неидентичности в параметрах, характерные отклонения состояний систем пропорциональны расстройке частот и обратно пропорциональны модулю показателя Ляпунова, а при дополнительном индивидуальном шуме – пропорциональны амплитуде индивидуального шума и обратно пропорциональны корню из модуля показателя Ляпунова.

Найдены распределения разностей фаз осцилляторов. В обоих случаях эти распределения имеют степенные асимптоты, свидетельствующие о перемежаемости между эпохами синхронного и асинхронного поведения при сколь угодно малых неидентичностях.

Результаты численного счета для конкретных систем при конечных интенсивностях шума обнаружили тот же характер соотношения между степенью синхронности поведения систем с малой неидентичностью и показателем Ляпунова, что и в аналитической теории.

Для выяснения существенности роли специфических свойств рассматривавшегося шума (гауссовость и $\delta$-коррелированность) рассмотрен случай существенно иного шума: телеграфного.

Вновь оказалось, что при малом шуме или частых переключениях показатель Ляпунова отрицателен, а в некоторых системах при умеренной интенсивности шума может происходить десинхронизация колебаний.

В тоже время, для некоторых систем фазовое описание дает адекватные результаты даже при умеренных значениях интенсивности шума и среднего времени переключений. Зависимость показателя Ляпунова от параметров шума (амплитуды и среднего времени переключений) не обнаруживает каких-либо следов таких зависимостей, соответствующих периодическому воздействию.

Поведение отклонений в состояниях систем при малых неидентичностях и амплитудах шума для телеграфного шума не отличается от такового для белого гауссовского шума.

## Список литературы